\documentclass[11pt,aps,prd,showpacs,showkeys,superscriptaddress,preprintnumbers,amsmath,amssymb,nofootinbib]{revtex4-1}

\pdfoutput=1 

\usepackage{graphicx}
\usepackage{dcolumn}
\usepackage{bm}
\usepackage{color}
\usepackage{epsfig}
\usepackage{amssymb}
\usepackage{amsmath}
\usepackage{amsfonts}

\usepackage{natbib}
\usepackage{amsmath,amssymb,enumitem}
\usepackage{braket}
\usepackage{fontenc}[T1]
\usepackage[utf8]{inputenc}
\usepackage{multirow}
\usepackage{placeins}
\usepackage{slashed}
\usepackage{braket}
\usepackage{xspace}
 
\usepackage[usenames, dvipsnames]{xcolor}
\usepackage{url}
\usepackage[linktocpage=true]{hyperref}
\hypersetup{
   colorlinks   = true,
    linkcolor= blue,
    citecolor= blue,
    urlcolor= blue}

\newcommand{\Azero}[1]{\ensuremath{{\tilde A}_0^{B\to #1}}}
\newcommand{\Aone}[1]{\ensuremath{{\tilde A}_1^{B\to #1}}}
\newcommand{\Atwo}[1]{\ensuremath{{\tilde A}_2^{B\to #1}}}

\newcommand{\V}[1]{\ensuremath{{\tilde V}^{B\to #1}}}
\newcommand{\Tone}[1]{\ensuremath{{\tilde T}_1^{B\to #1}}}
\newcommand{\Ttwo}[1]{\ensuremath{{\tilde T}_2^{B\to #1}}}
\newcommand{\Tthree}[1]{\ensuremath{{\tilde T}_3^{B\to #1}}}

\newcommand{\Athree}[1]{\ensuremath{{\tilde A}_3^{B\to #1}}}
\newcommand{\Athreezero}[1]{\ensuremath{{\tilde A}_{30}^{B\to #1}}}
\newcommand{\TtwothreeA}[1]{\ensuremath{{\tilde T}_{23A}^{B\to #1}}}
\newcommand{\TtwothreeB}[1]{\ensuremath{{\tilde T}_{23B}^{B\to #1}}}
\newcommand{\F}[1]{\ensuremath{{ F}^{B\to #1}}}
\newcommand{\MS}{\ensuremath{\overline{\text{MS}}}}
\newcommand{\nn}{\nonumber}
\newcommand{\be}{\begin{equation}}
\newcommand{\ee}{\end{equation}}
\newcommand{\GeV}{\ensuremath{\text{GeV}}}
\newcommand{\MeV}{\ensuremath{\text{MeV}}}
\newcommand{\refapp}[1]{Appendix~\ref{app:#1}}
\newcommand{\refeq}[1]{Eq.~(\ref{eq:#1})}
\newcommand{\reffig}[1]{Fig.~\ref{fig:#1}}
\newcommand{\refeqs}[2]{Eqs.~(\ref{eq:#1})--(\ref{eq:#2})}
\newcommand{\refsec}[1]{Sec.~\ref{sec:#1}}
\newcommand{\reftab}[1]{Table~\ref{tab:#1}}
\newcommand{\Jint}[1]{\ensuremath{j_\text{int}^{#1}}}
\newcommand{\Jweak}[1]{\ensuremath{j_\text{weak}^{#1}}}
\makeatletter

\def\ak{\@ifstar\@@ak\@ak}
\newcommand{\@ak}[1]{\textcolor{ForestGreen}{[\textbf{AK:} #1]}}
\newcommand{\@@ak}[1]{\textcolor{ForestGreen}{#1}}
\makeatother

\begin{document}

\title{\boldmath $B \rightarrow T$ transition form factors in light-cone sum rules}

\author{T.~M.~Aliev}
\email[Electronic address:~]{taliev@metu.edu.tr}
\affiliation{Physics Department, Middle East Technical University, 06531 Ankara, Turkey}
\author{H.~Dag}
\email[Electronic address:~]{huseyindagph@gmail.com}
\affiliation{ \"Ozye\u{g}in University, Department of Natural and Mathematical Sciences \\ \c{C}ekmek\"{o}y, \.{I}stanbul, Turkey}
\affiliation{Physik Department, Technische Universit\"at M\"unchen, James-Franck-Stra\ss{}e 1, D-85748 Garching, Germany}
\author{A. Kokulu}
\email[Electronic address:~]{ahmetkokulu@gmail.com}
 \affiliation{Physik Department, Technische Universit\"at M\"unchen, James-Franck-Stra\ss{}e 1, D-85748 Garching, Germany} 
\author{A.~Ozpineci}
\email[Electronic address:~]{ozpineci@metu.edu.tr}
\affiliation{Physics Department, Middle East Technical University, 06531 Ankara, Turkey}


\begin{abstract}
\vspace{1cm}
We present a new calculation of the semileptonic tree-level and flavor-changing neutral current form factors describing $B$-meson transitions to tensor mesons $T=D_2^*,K_2^*,a_2,f_2$ ($J^{P}=2^{+}$). We employ the QCD Light-Cone Sum Rules approach with $B$-meson distribution amplitudes. We go beyond the leading-twist accuracy and provide analytically, for the first time, higher-twist corrections for the two-particle contributions up to twist four terms. We observe that the impact of higher twist terms to the sum rules is noticeable. We study the phenomenological implications of our results on the radiative $ { B} \to K_2^{*}\gamma$ and semileptonic ${ B} \to D_2^* \ell {\bar \nu}_\ell$, $ { B} \to K_2^{*}\ell^+\ell^-$ decays.

\end{abstract}


\preprint{TUM-HEP-1207/19}

\maketitle

\section{Introduction}
\label{sec:intro}

$B$-meson decays represent a promising area for checking the gauge structure of the Standard Model (SM), looking for physics beyond it, as well as precise determination of the elements of the Cabibbo-Kobayashi-Maskawa (CKM) matrix. 

Interest to the $ B $ meson decays increases considerably after a number of measurements that deviate from the respective Standard Model predictions. These results are observed in two types of decays:
\begin{enumerate}
[label={(\arabic*)}]
\item Decays due to the flavor-changing neutral currents: $ b\to s\mu^+\mu^- $. Discrepancies with the SM predictions are obtained in several observables in $ B\to K^*\mu^+\mu^- $ \cite{Aaij:2013qta,Aaij:2015oid,Abdesselam:2016llu,Aaboud:2018krd,Sirunyan:2017dhj,Khodjamirian:2010vf,Bobeth:2017vxj} and $ B_{s} \to \phi\mu^+\mu^- $ \cite{Aaij:2013aln,Geng:2003su,Erkol:2002hc,Yilmaz:2008pa,Chang:2011jka} as well as in the measurements of $ R _{K(K^*)} = B(B\to K^{(*)}\mu^+\mu^-)/ B(B\to K^{(*)}e^+e^-)$ \cite{Aaij:2017vbb,Aaij:2014ora,Bordone:2016gaq}.

\item The charged current $b\to cl\nu $ transitions that take place at tree-level in SM. Tension between theory predictions and experimental data has been observed in the ratios $ R _{D^{(*)}} = B(B\to D^{(*)}\tau\bar\nu _\tau)/B(B\to D^{(*)}\ell {\bar\nu}_{\ell}) $ ($ \ell=e,\mu $) \cite{Huschle:2015rga,Lees:2013uzd,Abdesselam:2016xqt,Amhis:2016xyh} as well as $R _{J\psi} = B(B _c\to J/\psi \tau \bar \nu _\tau)/B(B _c\to J/\psi \mu\bar\nu _\mu)$ \cite{Aaij:2017tyk,Dutta:2017xmj}. 
\end{enumerate}
If these results are confirmed by the forthcoming experiments, it will be an unambiguous discovery of existence of new physics (NP).

With respect to these experimental observations one expects that if NP exists at the quark-level $b\to c$ transition, then such discrepancies should also be seen in $B$-meson transitions to tensor mesons in addition to $B$ decays to pseudo-scalar or vector-mesons\footnote{Tests of lepton flavor universality (LFU) regarding the quark-level $b\to c$ transition are not restricted to mesonic decays. For a very recent analysis of the baryonic counter part decay $\Lambda_b \to \Lambda_c (\to \Lambda \pi)\ell\bar\nu$, we refer the reader to see \cite{Boer:2019oyr}.}.

In regard to seeking NP effects, the $B$-meson decays to tensor mesons have the following advantage: tensor mesons have additional polarizations compared to the vector mesons and therefore this could provide additional kinematical quantities that are sensitive to the existence of NP. As a result, $B$-meson decays to tensor mesons could provide a complementary platform to search for new helicity structures, that deviate from the SM ones.

The main ingredients of $B\to T$ transitions are the relevant form factors. In this work, the form factors of $B \to T$ transitions are calculated within the light-cone QCD sum rules (LCSRs) \cite{Balitsky:1989ry,Chernyak:1990ag} (for reviews see e.g. \cite{Colangelo:2000dp}) using $B$-meson Light-Cone Distribution Amplitudes (LCDAs). Note that the light-cone sum rules have successfully been applied to a wide range of problems of hadron physics. The recent applications of LCSRs with heavy meson and heavy baryon distribution amplitudes are discussed in detail in many works (see for example \cite{Khodjamirian:2006st,Faller:2008tr,Wang:2015vgv,Wang:2015ndk,Cheng:2017smj,Gubernari:2018wyi,Gao:2019lta,Cheng:2019tgh,Descotes-Genon:2019bud} and references therein).

It should be noted that the $B \to T$($J^{P}=2^+$) form factors have previously been calculated by several groups using various methods \cite{Khosravi:2015jfa,Bernlochner:2017jxt,Azizi:2013aua,Emmerich:2018rug,Yang:2010qd,Wang:2010ni,Wang:2010tz,Sharma:2010yx,Cheng:2010sn,Cheng:2003sm,Cheng:2009ms,Safir:2001cd,Scora:1995ty,Charles:1999gy,Ebert:2001pc,Datta:2007yk}. For example, the $B \to f _2(1270)$ form factors have recently been calculated in \cite{Emmerich:2018rug} within the LCSRs framework using the $f _2(1270)$ meson DAs. Also, for the light tensor meson final states $B\to f_2,a_2,K_2^*,f_2^{'}$ the form factor calculation has been carried out previously by \cite{Yang:2010qd} within LCSRs employing tensor-meson DAs, and in \cite{Wang:2010ni} using perturbative QCD approach. Within three-point QCD sum-rule approach, a sub-set of the relevant form factors under consideration was estimated in Ref.~\cite{Khosravi:2015jfa} for $B\to f_2,a_2,K_2^*$ transitions, and in Ref.~\cite{Azizi:2013aua} for $B\to D_2^*$. The LCSRs analysis carried out in Ref.~\cite{Wang:2010tz} computes the relevant form factors for $B\to f_2,a_2,K_2^*$ transitions considering only the $\phi_+,\bar{\phi}$ $B$-LCDAs with vanishing virtual quark masses regarding the $f_2,a_2$ final states. Our analysis here extends previous works by providing {\it new} results for the {\it full} set of $B\to T$($D_2^*,K_2^*,a_2, f_2$) transition form factors up to and including twist-four accuracy of $B$-LCSRs as well as takes into account the finite virtual quark mass effects in the results of the form factors. Moreover, we provided results for the tensor form factors in $B\to D_2^{*}$ transitions for the first time.

We should further mention that the tensor isosinglet final state $f_2(1270)$ considered in this study, in principle, possesses a mixing pattern with the other isosinglet tensor meson of the same quantum numbers $f_2'(1525)$ in the form
\be
f_2 \equiv \frac{1}{\sqrt{2}} (u\bar{u} + d\bar{d}) \cos{\delta}  + s\bar{s} \sin{\delta}\, , \quad  f_2' \equiv  -  s\bar{s} \cos{\delta} +  \frac{1}{\sqrt{2}} (u\bar{u} + d\bar{d}) \sin{\delta}  \, ,
\ee
where the mixing angle $\delta$ has been found to be small indicating that $f_2$ could be considered nearly as a pure $\frac{1}{\sqrt{2}} (u\bar{u} + d\bar{d})$ state ($\sim98.2\%$) while $f_2'$ is nearly a pure $s\bar{s}$ state (for details, see Refs.~\cite{Li:2000zb} and \cite{Agashe:2014kda}). We will therefore assume no mixing between $f_2$ with $f_2'$ when studying the $B\to f_2$ form factors in this paper (see e.g. \cite{Emmerich:2018rug,Khosravi:2015jfa} for similar assumptions in regard to analyses of $B\to f_2$ form factors).

The outline of our paper is as follows. In \refsec{detailsBT}, the LCSRs for the relevant form factors are derived. \refsec{results} is devoted to the numerical analysis of the sum rules obtained in \refsec{detailsBT}. In \refsec{phenoAnalysis}, we study the phenomenological implications of our form factor results on the radiative $ { B} \to K_2^{*}\gamma$ and semileptonic ${ B} \to D_2^* \ell {\bar \nu}_\ell$, $ { B} \to K_2^{*}\ell^+\ell^-$ decays within the context of SM. \refsec{summary} contains a summary of our findings. Last, in \refapp{LCDAs} we collect the two-particle DAs of the $B$-meson and in \refapp{coefficients} we present analytical expressions for the coefficient functions needed for the determination of the relevant form factors.

\section{Theoretical Framework and Analytical Results for the Form Factors}
\label{sec:detailsBT}

In general, the $B\to T$ transitions, where $T=D_2^*,K_2^*,a_2, f_2$, can be described by seven form factors $\Azero{T}$,
$\Aone{T}$, $\Atwo{T}$, $\V{T}$, $\Tone{T}$, $\Ttwo{T}$, and $\Tthree{T}$,
which are defined in analogy to the $B\to V$ case\footnote{Note that for the case of semileptonic tree-level transitions ${ B} \to D_2^* \ell {\bar \nu}_\ell$, \Tone{D_2^*},\Ttwo{D_2^*} and \Tthree{D_2^*} form factors would only be induced by possible NP tensor type operators, which are absent in the SM.}:
\begin{align}
    \label{eq:BtoT:vector}
    \bra{T(k, \varepsilon)} \bar{q}_1 \gamma^\rho b \ket{B(p)}
        & =   {-} 2 \epsilon^{\rho\beta\delta\sigma} \varepsilon^*_\beta p_\delta k_\sigma \frac{\V{T}}{m_B + m_T}\,,\\
    \label{eq:BtoT:axial}
      \bra{T(k, \varepsilon)} \bar{q}_1 \gamma^\rho \gamma_5 b \ket{B(p)}
        & = i \varepsilon^*_\beta\ \bigg[g^{\rho\beta} (m_B + m_T) \Aone{T} - \frac{(p + k)^\rho q^\beta}{m_B + m_T}\, \Atwo{T}\\
    \nonumber
        & \qquad\qquad  - q^\rho q^\beta \frac{2 m_T}{q^2} \left( {\tilde A}_3 - {\tilde A}_0\right)\bigg]\,,\\
    \label{eq:BtoT:tensor}
      \bra{T(k, \varepsilon)} \bar{q}_1  \sigma^{\rho\alpha}\, q_\alpha b \ket{B(p)}
        & = -  2  i \epsilon^{\rho\beta\delta\sigma} \varepsilon^*_\beta p_\delta k_\sigma \Tone{T}\,,\\
    \label{eq:BtoT:tensor5}
      \bra{T(k, \varepsilon)} \bar{q}_1  \sigma^{\rho\alpha}\, q_\alpha \gamma_5 b \ket{B(p)}
        & =  \varepsilon^*_\beta \bigg[\left( g^{\rho\beta} (m_B^2 - m_T^2) - (p + k)^\rho q^\beta\right) \Ttwo{T}\\
    \nonumber
        & \qquad\qquad +  q^\beta \left(q^\rho - \frac{q^2}{m_B^2 - m_T^2} (p + k)^\rho\right) \Tthree{T}\bigg]\,,
\end{align}
where $\varepsilon$ represents the polarization of the final state tensor meson with shorthand notation $\varepsilon_\beta=\varepsilon_{\beta\alpha} q^\alpha/m_B$, and we use $\epsilon_{0123} = +1$. The polarization tensor $\varepsilon_{\beta\alpha}$ is symmetric in its indices and satisfies $\varepsilon_{\beta\alpha}(k)  k^\alpha=0$. Throughout, $k$ and $p$ represent the final-state tensor meson's and the $B$-meson's momentum, respectively, with $q^2 = (p - k)^2$ being the momentum transfer squared.

$\Athree{T}$ is superfluous, because it is correlated with \Aone{T} and \Atwo{T} form factors as
\begin{equation}
    \Athree{T}   =  \frac{m_B + m_T}{2 m_T} \Aone{T} - \frac{m_B - m_T}{2 m_T} \Atwo{T}\,.
    \label{eq::A3rel}
\end{equation}

The unphysical
singularities of the matrix elements defined in \refeq{BtoT:axial} at $q^2 = 0$  are removed by
\begin{align}
    \Azero{T}(q^2 = 0)
        & = \Athree{T}(q^2 = 0)\,.
        \label{eq::A3A0rel}
\end{align}

Besides, one has the following identity using algebraic relations between $\sigma^{\mu\nu}$ and $\sigma^{\mu\nu}\gamma_5$: 
\begin{equation}
\label{eq:T1eqT2atq2zero}
    \Tone{T}(q^2 = 0)
        = \Ttwo{T}(q^2 = 0)\,.
\end{equation}

\begin{table}[t]
    \renewcommand{\arraystretch}{1.25}
    \centering
\begin{tabular}{|c|c|c|c|}
    \hline
      Transition  & $\Jint{\mu\nu}$ & $\Jweak{\rho}$ & Form factor \\ \hline\hline
       \multirow{ 4 }{*}{$\bar{B}^0 \to {D_2^*}^{+}$}       &\multirow{ 4 }{*} {$ \frac{i}{2} \, \bar{d} \left[ \gamma^\mu \overleftrightarrow{D^\nu}    + \mu \leftrightarrow \nu    \right] c$}
                                                       & $\bar{c} \gamma^\rho h_v$                 & $\V{D_2^*}$                                  \\
                                                      && $\bar{c} \gamma^\rho \gamma_5 h_v$        & $\Azero{D_2^*},\,\Aone{D_2^*},\,\Atwo{D_2^*}$    \\
                                                      && $\bar{c} \sigma^{\rho\{q\}} h_v$          & $\Tone{D_2^*}$                               \\
                                                      && $\bar{c} \sigma^{\rho\{q\}} \gamma_5 h_v$ & $\Ttwo{D_2^*}, \Tthree{D_2^*}$  \\ \hline
                                                             \multirow{ 4 }{*}{$\bar{B}^0 \to {K_2^*}^{0}$} &\multirow{ 4 }{*}    {$ \frac{i}{2} \, \bar{d} \left[ \gamma^\mu   \overleftrightarrow{D^\nu}   + \mu \leftrightarrow \nu    \right] s$}
                                                       & $\bar{s} \gamma^\rho h_v$                 & $\V{K_2^*}$                                  \\
                                                      && $\bar{s} \gamma^\rho \gamma_5 h_v$        & $\Azero{K_2^*},\,\Aone{K_2^*},\,\Atwo{K_2^*}$    \\
                                                      && $\bar{s} \sigma^{\rho\{q\}} h_v$          & $\Tone{K_2^*}$                               \\
                                                      && $\bar{s} \sigma^{\rho\{q\}} \gamma_5 h_v$ & $\Ttwo{K_2^*}, \Tthree{K_2^*}$  \\ \hline
       \multirow{ 4 }{*}{$\bar{B}^0 \to {a_2^+}$}       &\multirow{ 4 }{*} {$ \frac{i}{2} \, \bar{d} \left[ \gamma^\mu \overleftrightarrow{D^\nu}    + \mu \leftrightarrow \nu    \right] u $}
                                                       & $\bar{u} \gamma^\rho h_v$                 & $\V{a_2}$                                  \\
                                                      && $\bar{u} \gamma^\rho \gamma_5 h_v$        & $\Azero{a_2},\,\Aone{a_2},\,\Atwo{a_2}$    \\
                                                      && $\bar{u} \sigma^{\rho\{q\}} h_v$          & $\Tone{a_2}$                               \\
                                                      && $\bar{u} \sigma^{\rho\{q\}} \gamma_5 h_v$ & $\Ttwo{a_2}, \Tthree{a_2}$      \\ \hline
       \multirow{4}{*}{${B} \to {f_2^0}$}       &\multirow{4 }{*}      {$ \frac{i}{2 \sqrt{2}}  \, \bar{u} \left[ \gamma^\mu  \overleftrightarrow{D^\nu} + \mu \leftrightarrow \nu  \right] u $ + $u \leftrightarrow d$  }  
                                                       & $\bar{u}  (\bar{d}) \gamma^\rho h_v$                 & $\V{f_2}$                                 \\
                                                      && $\bar{u}  (\bar{d})  \gamma^\rho \gamma_5 h_v$        & $\Azero{f_2},\,\Aone{f_2},\,\Atwo{f_2}$ \\
                                                      && $\bar{u}   (\bar{d})  \sigma^{\rho\{q\}} h_v$          & $\Tone{f_2}$                              \\
                                                      && $\bar{u}  (\bar{d}) \sigma^{\rho\{q\}} \gamma_5 h_v$ & $\Ttwo{f_2}, \Tthree{f_2}$  
                \\ \hline\hline
\end{tabular}
    \caption{
In this table,
    $\sigma^{\rho\lbrace q\rbrace} \equiv \sigma^{\rho\alpha} q_\alpha$, and  the position-space covariant derivative $D^\nu$ is defined in \refeq{JintTensor}.
    }
    \label{tab:listcurrents}
\end{table}

Our starting point is the correlation function
\begin{align}
    \label{eq:correlator}
    \Pi^{\mu\nu\rho}(q, k)
        \equiv i \int \text{d}^4 x\, e^{i k\cdot x}\,
        \bra{0} \mathcal{T}\lbrace \Jint{\mu\nu}(x), \Jweak{\rho}(0)\rbrace \ket{\bar{B}_{q_2}(q + k)}
\end{align}
of two quark currents $\Jint{\mu\nu} = \bar{q}_2(x) \Gamma_2^{\mu\nu} q_1(x)$ and $\Jweak{\rho}(0) = \bar{q}_1(0) \Gamma_1^\rho h_v(0)$, where $h_v$ denotes the Heavy Quark Effective Theory field instead of a $b$-quark.
The spin structures of $\Gamma_{1,2}$ together with various choices of quark flavors $q_1$ and $q_2$ for the form factors extracted
in this paper are given in \reftab{listcurrents}.

The interpolating current for tensor mesons (with valence quark content $q_1{\bar q}_2$) is, in general, given by
\be
 \label{eq:JintTensor}
 \Jint{\mu\nu} = \frac{i}{2} \, \bar{q}_2(x) \left[ \gamma^\mu  \overleftrightarrow{D^\nu}    + \mu \leftrightarrow \nu    \right] q_1(x)\, ; \quad  \overleftrightarrow{D^\nu} = \frac{1}{2} \left[  \overrightarrow{D^\nu} -  \overleftarrow{D^\nu}   \right] \,, \\ \vspace{0.15cm}
\ee
where $\overrightarrow{D^\nu} = \overrightarrow{\partial^\nu} - i g \frac{\lambda^a}{2} A_a^\nu(x)$ and $\overleftarrow{D^\nu} = \overleftarrow{\partial^\nu} + i g \frac{\lambda^a}{2} A_a^\nu(x) $ with $\partial^\nu=\frac{\partial}{\partial x_\nu}$. When regard to two-particle contributions, which we are interested in this work, it suffices to take the first terms in the covariant derivatives, because the second terms involving the fields $A^\nu(x)$ only contribute to three-particle effects\footnote{In a recent comprehensive work with $B$-LCDAs \cite{Gubernari:2018wyi} for $B\to P,V$ transitions it has been shown that compared to two-particle contributions, the relative impact of the three-particle contributions to the form-factors is only at percent level or less (for details see \cite{Gubernari:2018wyi}). The same conclusion was also drawn e.g. in Ref.~\cite{Wang:2010tz} for $B\to T$ transitions. We therefore feel safe to neglect such effects in the present analysis.}.

The higher Fock state contributions to the correlation function arise when expanding the position-space virtual-quark $q_1$ propagator in $x^2$ near the light-cone $x^2 \simeq 0$. In the present work, we focus on the two-particle contributions, while higher Fock state contributions are beyond our current scope. We summarize the two-particle Operator-Product-Expansion (OPE) contributions as
\begin{multline}
    \label{eq:correlatorOPE2pt}
    \Pi_{\text{OPE}}^{\mu\nu\rho}(q, k)
        =  \int \text{d}^4 x\,\int  \frac{ \text{d}^4 p'}{(2 \pi)^4} \, e^{i (k - p')\cdot x}\,
        \left[  \Gamma_2^{\mu\nu}\, \frac{\slashed{p}'+m_{q_1}}{m_{q_1}^2 - p'^2}  \Gamma_1^\rho \right]_{\alpha\beta} \times \, 
        \bra{0} \bar{q}_{2}^{\alpha}(x) h_{v}^{\beta}(0) \ket{\bar{B}_{q_2}(v)},
\end{multline}
where $p'=k-l$, and $l$ describes the momentum of the spectator quark inside the $B$-meson with $\alpha$,$\beta$ being spinor indices. In \refeq{correlatorOPE2pt}, the $B$-meson to vacuum matrix elements are non-perturbative objects that are expressed in terms of $B$-meson LCDAs, whose explicit definitions are relegated to \refapp{LCDAs}.

The hadronic correlator $\Pi^{\mu\nu\rho}$ reads:
\begin{align}
    \label{eq:hadronicdisp}
    \Pi_{\text{had}}^{\mu\nu\rho}(q, k) = \frac{\bra{0}  \Jint{\mu\nu}(x) \ket{T(k)} \bra{T(k)} \Jweak{\rho}(0) \ket{\bar{B}_{q_2}(q + k)}}{m_T^2-k^2}
    + \int^{\infty}_{s_{\text{thr.}}^h} \text{d}s\frac{{\rho}_\text{had}^{\mu\nu\rho}(s)}{s-k^2}\, .
\end{align}

The decay constant $f_T$ is defined via\footnote{Note that this definition implies $f_T$ to be dimensionless.}:
\begin{align}
\label{eq:fTdef}
      \bra{0} \Jint{\mu\nu}  \ket{T(k, \varepsilon)}  =  \varepsilon^{\mu\nu} m_T^3 f_T\,.
\end{align}

The spin sum for the tensor mesons is given by:
\be
\varepsilon_{\mu\nu}(k) \varepsilon^{*}_{\alpha\beta}(k) = \frac{1}{2} \kappa_{\mu\alpha} \kappa_{\nu\beta} + \frac{1}{2} \kappa_{\mu\beta} \kappa_{\nu\alpha} - \frac{1}{3} \kappa_{\mu\nu} \kappa_{\alpha\beta}; \quad    \kappa_{\mu\nu} = -g_{\mu\nu} + \frac{k_\mu k_\nu}{m^2_T} \, .
\ee

\vspace{0.3cm}
The form-factors are extracted by matching independent Lorentz structures appearing in both correlators $\Pi_{\text{OPE}}^{\mu\nu\rho}(q, k)$ and $\Pi_{\text{had}}^{\mu\nu\rho}(q, k)$.

For the particular choice of the weak currents as in \reftab{listcurrents}, the correlator $\Pi_{\text{OPE}}^{\mu\nu\rho}(q, k)$ can be split as:
\begin{align}
\label{eq:DecompCorrOPElo-Tensor-VmAC}
\nn
\Pi^{\mu\nu\rho, BT}_{\text{(V-A)}}(q, k)  &= q^\rho q^\mu q^\nu \, \Pi^{(1,BT)}_{\rm OPE}(q^2, k^2)  +    k^\rho q^\mu q^\nu  \, \Pi^{(2,BT)}_{\rm OPE}(q^2, k^2)
\\  
& + \epsilon_{\mu\rho\alpha\beta} q^\nu q^\beta k^\alpha \, \Pi^{(3,BT)}_{\rm OPE}(q^2, k^2)   + q^\nu g^{\mu\rho}  \, \Pi^{(4,BT)}_{\rm OPE}(q^2, k^2)  + \text{...} \,  ,  \\
\nn
\Pi^{\mu\nu\rho, BT}_{\text{Tensor}}(q, k)  &= q^\rho q^\mu q^\nu \, \Pi^{(5,BT)}_{\rm OPE}(q^2, k^2)  +    k^\rho q^\mu q^\nu  \, \Pi^{(6,BT)}_{\rm OPE}(q^2, k^2)
\\   
& + \epsilon_{\mu\rho\alpha\beta} q^\nu q^\beta k^\alpha \, \Pi^{(7,BT)}_{\rm OPE}(q^2, k^2)  \,   + \text{...} \, ,
\label{eq:DecompCorrOPElo-Tensor-tensorC}
\end{align}
where the ellipsis stand for terms involving other Lorentz structures. The extraction of the $B \rightarrow T$ form factors is then achieved as follows: 
\begin{itemize}
\item ${\tilde V}$: we considered terms
with Lorentz-structure $\epsilon_{\mu\rho\alpha\beta} q^\nu q^\beta k^\alpha$ in \refeq{DecompCorrOPElo-Tensor-VmAC}.
\item ${\tilde A}_1$: we considered terms
with Lorentz-structure $q^\nu g^{\mu\rho}$ in \refeq{DecompCorrOPElo-Tensor-VmAC}.
\item ${\tilde A}_2$: we considered terms
with Lorentz-structure $k^\rho q^\mu q^\nu$ in \refeq{DecompCorrOPElo-Tensor-VmAC}.
\item $({\tilde A}_3-{\tilde A}_0)$: in this case, the form factors ${\tilde A}_0,{\tilde A}_2$ and ${\tilde A}_3$ possess some common Lorentz structures. Hence, we define a combined term as ${\tilde A}_{023}=\frac{{\tilde A}_2}{m_B+m_T} + \frac{2 m_T ({\tilde A}_3-{\tilde A}_0)}{q^2}$, and then extract ${\tilde A}_{023}$ by considering terms
with Lorentz-structure $q^\rho q^\mu q^\nu$ in \refeq{DecompCorrOPElo-Tensor-VmAC}.
\item ${\tilde T}_1$: we considered terms
with Lorentz-structure $\epsilon_{\mu\rho\alpha\beta} q^\nu q^\beta k^\alpha$ in \refeq{DecompCorrOPElo-Tensor-tensorC}.
\item ${\tilde T}_2$ and ${\tilde T}_3$: in this case, the form factors ${\tilde T}_2$ and ${\tilde T}_3$ possess some common Lorentz structures. We, therefore, define the combination terms ${\tilde T}_{23A},{\tilde T}_{23B}$ as in \refeq{defTA}-\refeq{defTB}, from which we then extract ${\tilde T}_{23A}({\tilde T}_{23B})$ by considering terms
with Lorentz-structure $k^\rho q^\mu q^\nu(q^\rho q^\mu q^\nu)$ in \refeq{DecompCorrOPElo-Tensor-tensorC}.
\end{itemize}
The choice of these structures is dictated by the fact that they contain
contributions coming {\it purely} from tensor mesons.

Following the formulation introduced in Ref.~\cite{Gubernari:2018wyi}, we write down the sum rule for all the $B \to T$ form factors in a form of a master-formula\footnote{Our results in this work provide additional ingredients to the master-formula introduced in Ref.~\cite{Gubernari:2018wyi} to also include the $B \to T$ form factors at the same twist accuracy of the $B$-LCDAs. For details on the derivation of this formula we refer the reader to see Ref.~\cite{Gubernari:2018wyi}. } as:

\begin{align}
 F^{B \to T}= 
    &\frac{f_B M_B\, }{ \chi  \, K^{(F)}} \sum_{n=1}^{\infty}\Bigg\{(-1)^{n}\int_{0}^{\sigma_0} d \sigma \;e^{(-s(\sigma,q^2)+m^2_{T})/M^2} \frac{1}{(n-1)!(M^2)^{n-1}}I_n^{(F)}\nonumber\\
        & - \Bigg[\frac{(-1)^{n-1}}{(n-1)!}e^{(-s(\sigma,q^2)+m^2_{T})/M^2}\sum_{j=1}^{n-1}\frac{1}{(M^2)^{n-j-1}}\frac{1}{s'}
        \left(\frac{\text{d}}{\text{d}\sigma}\frac{1}{s'}\right)^{j-1}I_n^{(F)}\Bigg]_{\sigma=\sigma_0}\Bigg\rbrace\,,
        \label{eq:masterformula}
\end{align}
where
\begin{align}
\label{eq:defs}
    s(\sigma,q^2)=\sigma m^2_B +\frac{m_{q_1}^2-\sigma q^2}{\bar{\sigma}}\,,
    \qquad
    s'(\sigma,q^2)=\frac{\text{d} s(\sigma,q^2)}{\text{d} \sigma}\,, \quad \text{with} \quad \bar{\sigma} = 1 -\sigma \, .
\end{align}
In \refeq{masterformula}, $\chi = \sqrt{2} \, \left(\chi=1\right)$ for the light unflavored states $f_2^0,a_2^0$ (for other states), and the differential operator is understood to act as

\begin{equation*}
    \left(\frac{\text{d}}{\text{d}\sigma}\frac{1}{s'}\right)^{n} I(\sigma) \to 
    \left(\frac{\text{d}}{\text{d}\sigma}\frac{1}{s'}\left(\frac{\text{d}}{\text{d}\sigma}\frac{1}{s'}\dots I(\sigma)\right)\right)\,.
\end{equation*}
Using the first relation in \refeq{defs} one obtains

\be
\sigma_0 = \frac{s_0+m_B^2-q^2- \sqrt{4 (m_{q_1}^2-s_0) m_B^2 + (m_B^2 +s_0 -q^2)^2 }}{2 m_B^2}\, ,
\ee
where $s_0$ is an effective threshold parameter to be determined and supplied as an input.

The two-particle LCDAs appear in the definitions of the functions $I_n^{(F)}$ \cite{Gubernari:2018wyi}:
\begin{align}
    I_n^{(F,\,\text{2p})}(\sigma,q^2)
     &= \frac{1}{\bar{\sigma}^n} \sum_{\psi_\text{2p}}  C^{(F,\psi_\text{2p})}_n(\sigma,q^2)\, \psi_\text{2p} (\sigma m_B),  \, \hspace{1.cm}\psi_\text{2p}=\phi_+,\bar{\phi},g_+,\bar{g}
    \label{eq:CoeffFuncs2pt}  \, ;
\end{align}
with $\sigma=\omega/m_B$ in \refeq{CoeffFuncs2pt}. The analytical expressions for the normalization factors $K^{(F)}$ together with the matching coefficients $C^{(F,\psi_{\text{2p}})}$ for the considered $B \to T$ transition form factors are relegated to \refapp{coefficients}.

We provide results for $F =  \V{T}, \Aone{T}, \Atwo{T}, \Athreezero{T}, \Tone{T}, \TtwothreeA{T}$, and $\TtwothreeB{T}$. The remainder of the form factors $\Azero{T},\Ttwo{T}$ and $\Tthree{T}$ are then simply obtained using
\begin{align}
  \Azero{T} 
        & = \Athree{T} -  \Athreezero{T}\,,\\
    \label{eq:defTA}
    \Ttwo{T}
        & = \frac{2 q^2}{m_B^2-m_T^2}  \TtwothreeB{T}  +   \frac{(m_B^2-m_T^2- q^2)}{m_B^2-m_T^2}  \TtwothreeA{T}  \,,\\
    \label{eq:defTB}
    \Tthree{T}
        & =   \TtwothreeA{T} -2 \TtwothreeB{T}    \,.
\end{align}

Further, we give our results for generic final state tensor meson $T(q_1 \bar{q}_2)$, where $q_1=c,s,u$\, ($q_1=u(d)$) for ${D_2^*}^{+},{K_2^*}^{0},a_2^+$ ($f_2^0$), respectively\footnote{The theoretical approach presented in this work, together with our form factor results, is generic and can also be readily applied to other tensor mesons with $J^{P}=2^{+}$ by making obvious replacements.}.

The analytical results presented in this work for $C^{(F^{B\to T},\psi_\text{2p})}_n$ coefficients of \refeq{CoeffFuncs2pt} constitute the first complete results up to twist four accuracy of $B$-LCDAs for the two-particle Fock state contributions to the correlation function. As a result, our present $B \to T$ form factor results improve upon previous results in the literature.

At this stage, a remark on our form factor results is in order. We compared our analytical results related to the two-particle contributions at the leading-twist limit to those of Ref.~\cite{Wang:2010tz}. We observe the followings: first, we see that the surface-term contributions\footnote{Surface-terms arise after performing continuum subtraction. We observed and stress that the numerical impact of the surface terms on the form factor results could be sizable.} given in \refeq{masterformula} of our paper have not been taken into account in the work of \cite{Wang:2010tz}. Nonetheless, when we still compare our analytical results to \cite{Wang:2010tz}, after also dropping the mentioned surface-term effects in our results, we then reproduce the analytical results for their form factors called $V, A_1, T_1, {\tilde A}_3$ and ${\tilde T}_3$. Next, for ${\tilde A}_2$ of Ref.~\cite{Wang:2010tz} we reproduce their results for $\phi_+$ terms, while for the ${\bar \phi}$ terms we have a disagreement. Last, for the ${T}_2$ form factor of Ref.~\cite{Wang:2010tz} we have a complete disagreement. The disagreement in the ${T}_2$ form factor of Ref.~\cite{Wang:2010tz} is particularly interesting because while in our case the condition $\Tone{T}(0)=\Ttwo{T}(0)$ is {\it exactly} fulfilled (as required by equation-of-motion conditions), the analytical results given in Eqs.~20--21 of Ref.~\cite{Wang:2010tz}  (arXiv v6) for these two form factors seem {not} to satisfy this condition.

\section{Numerical Illustrations}
\label{sec:results}

\subsection{Input}
\label{sec:results:input}

In this section we collect the input parameters used in our numerical estimates. We use up-to-date input parameters. 

The meson masses entering our numerics are quoted from the latest PDG averages \cite{Tanabashi:2018oca}:
\begin{align}
\nn
&m_{f_2} = 1275.5 \pm 0.8 \, \MeV, \quad  m_{a_2} = 1318.3 ^{+0.5}_{-0.6} \, \MeV  \, ,      \nn \\
& m_{K_2^{*}} = 1425.6 \pm 1.5 \, \MeV, \quad m_{D_2^{*}} = 2465.4 \pm 1.3  \, \MeV  \, , \nn \\
&   m_B = 5279.55 \pm 0.26 \, \MeV   \,. \nn
\end{align}
Moreover, the quark masses $m_{q_1}$($q_1=c,s,u(d)$) appearing in the $C^{(F,\psi_\text{2p})}_n$ coefficients of \refapp{coefficients} together with ${b}$-quark mass are defined  in $\MS$ scheme, for which we use \cite{Tanabashi:2018oca}
\begin{align}
\nn
m_{b}(m_b) &= 4.18  \, \GeV, \quad m_{c}(m_c) = 1.275  \, \GeV,  \nn \\  
  m_{s}(1\GeV) &= 0.128\, \GeV \, , \quad m_{u(d)}(1\GeV) = 0.005\,  \GeV \, .      \nn
\end{align}
$B$-meson decay constant is taken from the currently most precise lattice-QCD analysis $f_B = 189.4 \pm 1.4\,\MeV$   \cite{Bazavov:2017lyh}, on the other side the tensor mesons' decay constants used in our numerical results are quoted in \reftab{fTdecayconstants}.

\begin{table}[t]
    \centering
    \renewcommand{\arraystretch}{1.25}
    \begin{tabular}{|c|c|c|c|c|}
        \hline
        Tensor meson  &  $f_2^0$  &  $a_2$  &  $K_2^*$ & $D_2^*$  \\\hline\hline
        $f_{T}$  \,&\,  $0.040$ \cite{Aliev:1981ju}  \,&\, $0.0406\pm 0.0023$ \cite{Cheng:2010hn} \,&\, $0.050 \pm 0.002$ \cite{Aliev:2009nn}  \,&\,     $0.0185 \pm 0.0020$ \cite{Wang:2014yza}\,              
\\ \hline \hline
    \end{tabular}
        \caption{Tensor mesons' decay constants used in our numerical results.}
          \label{tab:fTdecayconstants}
\end{table}

For the non-perturbative parameters entering the explicit expressions of $B$-LCDAs we use:
\be
\lambda_B = 460 \pm 110 \, \MeV \,  \text{\cite{Braun:2003wx}}, \quad  \lambda_E^2= 0.03\pm 0.02\,\GeV^2 \, \text{\cite{Nishikawa:2011qk}}  \,, \quad  \lambda_H^2 = 0.06 \pm 0.03\, \GeV^2  \text{\cite{Nishikawa:2011qk}}  \,.
\ee
%

\subsection{LCSRs Results}
\label{sec:results:LCSRs}

We obtained results for the full set of $B\to T$ form factors within LCSRs up to $q^2 =  0$ $\GeV^2$. Our LCSRs results involve, besides other input, free parameters introduced by the method; the continuum threshold $s_0$ and Borel mass parameter $M^2$, which we determine by fulfilling some physical criteria. First, the working interval of the Borel parameter $M^2$ is determined following a standard criteria, i.e demanding that both the power corrections and the continuum contributions in the sum rules should be suppressed. Next, the working region of the continuum threshold is determined by defining so-called first-moments for each form factor and respective final state by differentiating the OPE correlator with respect to $-1/M^2$ and normalizing it to itself. These first-moments are then expected to give the mass squares $m_T^2$ of the respective final state mesons. Imposing $\pm5\%$ uncertainty on the mass of each final state tensor meson, we were then able to find validity window for $s_0$ too.

Based on these discussions, we determined the following working regions for $s_0$ and $M^2$ for the considered transitions:
\begin{align}
     7.2 \,  \GeV^2  &<  s_0^{D_2^*}  < 8.3 \,   \GeV^2 \,, \quad     5.0  \,  \GeV^2 <  M^2_{D_2^*}  < 7.0 \,   \GeV^2  \nn \, ,\\
          2.7 \,  \GeV^2  &<  s_0^{K_2^*}  < 3.3 \,   \GeV^2 \,, \quad     1.5  \,  \GeV^2 <  M^2_{K_2^*}  < 2.5 \,   \GeV^2  \nn \, ,\\
     2.3 \,  \GeV^2  &<  s_0^{a_2}  < 2.7 \,   \GeV^2 \,, \quad     1.5  \,  \GeV^2 <  M^2_{a_2}  < 2.0 \,   \GeV^2  \nn \, ,\\
          2.15 \,  \GeV^2  &<  s_0^{f_2}  < 2.45 \,   \GeV^2 \,, \quad     1.5  \,  \GeV^2 <  M^2_{f_2}  < 1.7 \,   \GeV^2   \, .
          \label{eq:s0windows}
\end{align}

With these working regions for $M^2$ and $s_0$'s, the smallness of the sub-leading twist-4 contributions compared to the leading twist-2 ones as well as the suppression of higher state contributions are satisfied simultaneously.

\begin{figure}[htbp]
\begin{center}
\includegraphics[width=0.65\textwidth]{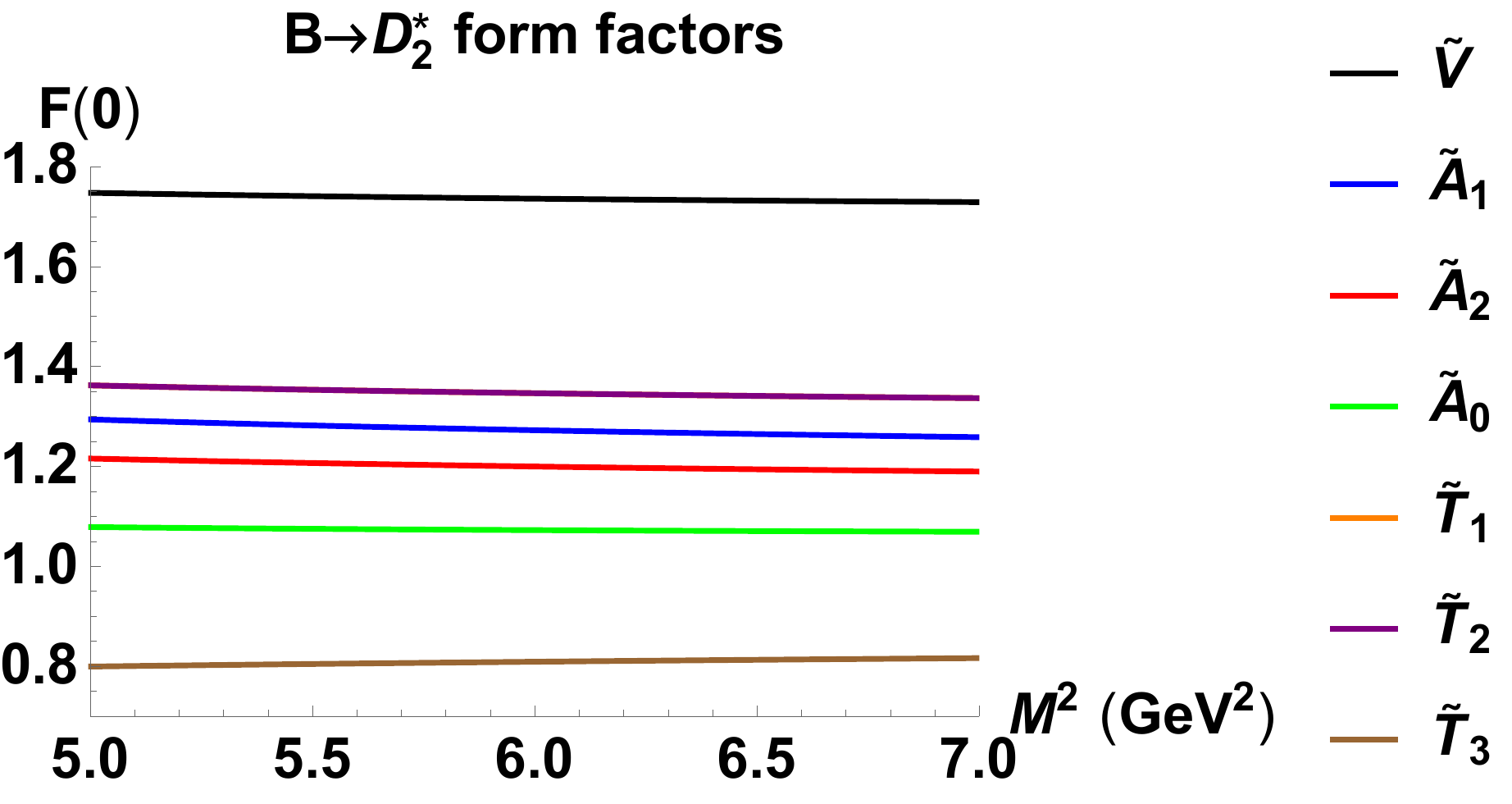}
\caption{Borel parameter dependence of the $B \to D_2^*$ form factors at $q^2=0$ based on our LCSRs results. The rest of the parameters are fixed to their central input values. Similar stability behavior also holds at other negative $q^2$ values and for the other final states ($ K_2^*, a_2, f_2$) too in their respective $M^2$ ranges, we therefore skip giving them here for brevity.}
\label{fig:M2depofBtoD2FFs}
\end{center}
\end{figure}

In \reffig{M2depofBtoD2FFs}, we illustrate the Borel parameter dependence of all form factors for $B \to D_2^*$ transition at $q^2=0$ based on our LCSRs results including higher twist contributions. Within the chosen interval for $M^2$, it is seen that the form factors posses a very mild dependence on $M^2$. Similar behavior holds at other negative $q^2$ values and for the other final states ($K_2^*, a_2, f_2$) too in their respective $M^2$ ranges.

\subsection{Parametrization of the form-factors}
\label{sec:results:paramfits}

After determining the best fit intervals of the threshold and the Borel parameters from \refsec{results:LCSRs}, we are now in a position to extrapolate our LCSRs results to the physical region where the phenomenology of the considered $B\to T$ transitions take place. As mentioned in \refsec{results:LCSRs} we truncate our LCSRs results at $q^2 = 0$ $\GeV^2$ for all the tensor meson final states. The extrapolation from the calculated LCSRs input points ($q^2 \lesssim 0\,  \GeV^2$) to larger $q^2$ values\footnote{For instance, the upper $q^2$ limit in the case of semi-leptonic decays is $q_{\text{max}}^2=(m_B-m_T)^2$.} is then achieved by parametrizing each of the form factors in a simple pole form with $z$-expansion\footnote{We observed that the $B \to T$ transition form factors under consideration, are well fitted by the fit-function of \refeq{fitform} to first order in $z$.} as \cite{Straub:2015ica}:
\begin{equation}
    \F{T}(q^2) \equiv \frac{1}{1 - q^2 / m_{R,F}^2}\, \sum_{n=0}^{1} \alpha_n^{F} \left[z(q^2) - z(0)\right]^n\,,
        \label{eq:fitform}
\end{equation}
where  $z(s) \equiv \frac{\sqrt{t_+ - s} - \sqrt{t_+ - t_0}}{\sqrt{t_+ - s} + \sqrt{t_+ - t_0}}$, $t_\pm = (m_B \pm m_{T})^2$ and $t_0 \equiv t_+ \left(1 - \sqrt{1 - t_- / t_+}\right)$. In \refeq{fitform}, $\alpha_{0,1}^{F}$ are the fit parameters that are constrained and presented in \reftab{Fitparams} for each form factor and final state transition separately. Beside, $m_{R,F}$ quantities describe the mass of the resonances associated with the quantum numbers of the
respective form factor $F$, whose values can be found in Ref.~\cite{Gubernari:2018wyi} (for details see Table~5 of \cite{Gubernari:2018wyi} and references therein). Note that the kinematical conditions given in \refeqs{:A3A0rel}{T1eqT2atq2zero} impose the following relations among the fit parameters
\be
\alpha_0^{{\tilde T}_1} = \alpha_0^{{\tilde T}_2},\quad   \alpha_0^{{\tilde A}_0}  = \frac{m_B + m_T}{2 m_T}  \alpha_0^{{\tilde A}_1}  -  \frac{m_B - m_T}{2 m_T}  \alpha_0^{{\tilde A}_2}  \, ,
\ee
which are respected in our numerical results presented in \reftab{Fitparams}. 

The uncertainties in the values of the form factors of \reftab{Fitparams} are due to the variation of various input parameters involved in the LCSR calculation. In particular, the non-perturbative parameters $\lambda_B, \, \lambda_H^2, \, \lambda_E^2$ of $B$-LCDAs together with the continuum threshold $s_0$ are mostly responsible for these errors.

In order to estimate the uncertainties of the results presented in this work, such as the form factors, decay rates etc., we followed a Monte Carlo based analysis as performed e.g. in Refs. \cite{Leinweber:1995fn,Erkol:2008gp}. For this analysis, randomly selected data sets of thousands of data points are generated for any input parameter and its given uncertainty. This led us to determine the mean and corresponding standard deviations of our results.

\begin{widetext}
\renewcommand{\arraystretch}{0.52}
\begin{table}[ht]
    \begin{tabular}{| c | c | c |}
      \hline
  Form Factor &~ $\alpha_0$~&~$\alpha_1$ \\
\hline \hline
                    \multicolumn{3}{c}{$B\to D_2^{*}$} \\ \hline
 $\V{D_2^*} $ ~&~ $1.45 ^{+0.31}_{-0.35}$   ~&~  $-8.63^{+1.99}_{-1.26}$  \\ \hline
 $\Azero{D_2^*} $ ~&~ $1.13 ^{+0.11}_{-0.17}$   ~&~  $-6.25^{+0.77}_{-0.17}$ \\\hline  
 $\Aone{D_2^*} $ ~&~ $1.10 ^{+0.21}_{-0.23}$   ~&~  $-4.57^{+0.75}_{-0.16}$ \\ \hline
 $\Atwo{D_2^*}  $ ~&~ $1.03 ^{+0.38}_{-0.33}$   ~&~  $-6.16^{+2.13}_{-2.17}$ \\ \hline  
 $\Tone{D_2^*}  $ ~&~ $1.15 ^{+0.23}_{-0.25}$   ~&~  $-6.67^{+1.34}_{-0.83}$ \\ \hline
 $\Ttwo{D_2^*}  $ ~&~ $1.15 ^{+0.23}_{-0.25}$   ~&~  $-3.86^{+0.55}_{-0.22}$ \\ \hline
 $\Tthree{D_2^*}$ ~&~ $0.44 ^{+0.23}_{-0.18}$   ~&~  $-2.18^{+0.98}_{-1.27}$ \\ \hline \hline
              \multicolumn{3}{c}{$B\to K_2^{*}$} \\ \hline
  $\V{K_2^{*}} $ ~&~ $0.22 ^{+0.11}_{-0.08}$   ~&~  $-0.90^{+0.37}_{-0.50}$\\ \hline
  $\Azero{K_2^{*}} $ ~&~ $ 0.30 ^{+0.06}_{-0.05}$   ~&~  $-1.23^{+0.23}_{-0.23}$ \\ \hline
  $\Aone{K_2^{*}} $ ~&~ $ 0.19 ^{+0.09}_{-0.07}$   ~&~  $-0.46^{+0.19}_{-0.25}$ \\ \hline
  $\Atwo{K_2^{*}} $ ~&~ $ 0.11 ^{+0.05}_{-0.06}$   ~&~  $-0.40^{+0.23}_{-0.16}$ \\ \hline
  $\Tone{K_2^{*}} $ ~&~ $ 0.19 ^{+0.09}_{-0.06}$   ~&~  $-0.75^{+0.28}_{-0.38}$  \\ \hline
  $\Ttwo{K_2^{*}} $ ~&~ $ 0.19 ^{+0.09}_{-0.06}$   ~&~  $-0.17^{+0.09}_{-0.12}$ \\ \hline
  $\Tthree{K_2^{*}}$ ~&~ $ 0.09 ^{+0.06}_{-0.04}$   ~&~  $-0.27^{+0.15}_{-0.22}$\\ \hline
\hline
              \multicolumn{3}{c}{$B \to a_2^+$} \\ \hline
  $\V{a_2^+} $ ~&~ $ 0.18 ^{+0.12}_{-0.07}$   ~&~  $-0.70^{+0.31}_{-0.52}$ \\ \hline
  $\Azero{a_2^+} $ ~&~ $ 0.30 ^{+0.06}_{-0.05}$   ~&~  $-1.21^{+0.22}_{-0.26}$  \\ \hline
  $\Aone{a_2^+} $ ~&~ $ 0.16 ^{+0.09}_{-0.05}$   ~&~  $-0.33^{+0.15}_{-0.26}$ \\ \hline
  $\Atwo{a_2^+} $ ~&~ $ 0.07 ^{+0.08}_{-0.03}$   ~&~  $-0.15^{+0.06}_{-0.31}$  \\ \hline
  $\Tone{a_2^+} $ ~&~ $ 0.15 ^{+0.09}_{-0.05}$   ~&~  $-0.59^{+0.23}_{-0.39}$  \\ \hline
  $\Ttwo{a_2^+} $ ~&~ $ 0.15 ^{+0.09}_{-0.05}$   ~&~  $-0.10^{+0.07}_{-0.12}$\\ \hline
  $\Tthree{a_2^+}$ ~&~ $ 0.07 ^{+0.06}_{-0.03}$   ~&~  $-0.19^{+0.12}_{-0.22}$ \\ \hline
\hline
              \multicolumn{3}{c}{$B \to f_2$} \\ \hline
  $\V{f_2} $ ~&~ $ 0.11 ^{+0.07}_{-0.05}$   ~&~  $-0.42^{+0.20}_{-0.32}$ \\ \hline
  $\Azero{f_2} $ ~&~ $ 0.20 ^{+0.04}_{-0.04}$   ~&~  $-0.80^{+0.15}_{-0.17}$  \\ \hline
  $\Aone{f_2} $ ~&~ $ 0.10 ^{+0.06}_{-0.04}$   ~&~  $-0.20^{+0.10}_{-0.16}$   \\ \hline
  $\Atwo{f_2} $ ~&~ $ 0.04 ^{+0.05}_{-0.01}$   ~&~  $-0.07^{+0.02}_{-0.20}$  \\ \hline
  $\Tone{f_2} $ ~&~ $ 0.10 ^{+0.05}_{-0.04}$   ~&~  $-0.36^{+0.15}_{-0.24}$ \\ \hline
  $\Ttwo{f_2} $ ~&~ $ 0.10 ^{+0.05}_{-0.04}$   ~&~  $-0.06^{+0.05}_{-0.08}$   \\ \hline
  $\Tthree{f_2}$ ~&~ $ 0.04 ^{+0.04}_{-0.02}$   ~&~  $-0.11^{+0.08}_{-0.13}$ \\ \hline
\hline
\end{tabular}
\caption{Results for the fit parameters $\alpha_n^{F}$ by fitting our LCSRs results for $B\to T$ form factors to \refeq{fitform}. }
\label{tab:Fitparams}
\end{table}
\end{widetext}
\FloatBarrier

\subsection{Illustrations}
\label{sec:results:illustrations}

The $q^2$ dependence of the complete set of $B \to T$ form factors is depicted in Figs.~\ref{fig:q2DepofBtoD2FFs},\ref{fig:q2DepofBtoK2FFs},\ref{fig:q2DepofBtoa2FFs} and \ref{fig:q2DepofBtof2FFs}. In these plots, comparing the leading-twist central results (empty red-circles) with the corresponding new results including twist-four terms (dotted-blue curves) we see that the relative impact of the calculated higher twist terms for the two-particle contributions could be sizable\footnote{For the $B \to T$ form factors under consideration, in the charmed case the relative impact of the calculated higher-twist terms is observed to be relatively less significant when compared to light final state transitions. In our opinion, this could mainly be related to the presence of the heavy mass scale $m_c$ in the problem.} (in particular for light tensor meson transitions) and therefore should be included in the estimations of the form factors. The magnitude of the central values of the form factors based on the leading-twist terms, is observed to decrease due to the calculated higher twist terms.

In \reftab{allFFcomparison}, we have also compared our present results for the $B\to T$ form factors at $q^2=0$ with existing results in literature. Regarding the comparison of $B\to D_2^{*}$ form factors with Ref. \cite{Azizi:2013aua}, we normalized their results to obtain dimensionless form factors (as in our case), and extracted their value for $\Azero{D_2^{*}}(0)$ using Eq. (\ref{eq::A3rel}) and Eq. (\ref{eq::A3A0rel}). We observe that our numerical results for $\V{D^*_2},\Azero{D^*_2},\Aone{D^*_2},\Atwo{D^*_2}$ at $q^2=0$, given in the top-left pane of \reftab{allFFcomparison}, severely differ\footnote{A remark on this point is in order. Our definition for the ${\tilde A}_3 - {\tilde A}_0$ form factor is related to the form factor $b_{-}$ of Ref.~\cite{Azizi:2013aua} (arXiv v3) in the following way: $(2 m_T/q^2)({\tilde A}_3 - {\tilde A}_0)/m_B = - b_{-}$. At $q^2=0$, ${\tilde A}_3 - {\tilde A}_0$ should exactly be zero according to the equation-of-motion condition given in \refeq{:A3A0rel} of our paper. However, the $b_{-} (q^2=0)$ form factor of Ref.~\cite{Azizi:2013aua} is seen to differ from zero (see Table 2 of the mentioned reference), in explicit violation of this condition.} from the corresponding values quoted in Ref. \cite{Azizi:2013aua}, which use three-point sum rules. On the other side, concerning the light tensor transition form factors, our numerical results are in agreement with some of the existing results in the literature, which use various calculation methods.

\begin{widetext}
    \renewcommand{\arraystretch}{.35}
    \begin{table}[h]
    \begin{minipage}{3in}
        \centering
        \begin{tabular}{| c | c | c |}
            \hline
            Form Factor  ~&~ This work  ~&~  Literature   \\
            \hline \hline
             $\V{D_2^{*}}(0)$    	 &  $1.45 ^{+0.31}_{-0.35}$ &  $-0.41\pm0.12$ \cite{Azizi:2013aua}          \\ \hline
$\Azero{D_2^{*}}(0)$     	 &  $1.13 ^{+0.11}_{-0.17}$   &  $-0.12\pm0.33$  \cite{Azizi:2013aua} \\ \hline
$\Aone{D_2^{*}}(0)$  	 &  $1.10 ^{+0.21}_{-0.23}$ &   $0.37\pm0.10$ \cite{Azizi:2013aua}          \\ \hline
$\Atwo{D_2^{*}}(0)$   	 &  $1.03 ^{+0.38}_{-0.33}$  &  $1.23\pm0.41$  \cite{Azizi:2013aua}           \\ \hline
$\Tone{D_2^{*}}(0)$     	 &  $1.15 ^{+0.23}_{-0.25}$  &  ---                  \\ \hline
$\Ttwo{D_2^{*}}(0)$     	 &  $1.15 ^{+0.23}_{-0.25}$   &  ---        \\ \hline
$\Tthree{D_2^{*}}(0)$    	 &  $0.44 ^{+0.23}_{-0.18}$  &  ---    \\ \hline
            \hline 
        \end{tabular}
         \end{minipage}
          \hspace{0.8cm}
                    \vspace{-1.6cm}
            %
            \begin{minipage}{3in}
        \centering
        \begin{tabular}{| c | c | c |}
            \hline
            Form Factor  ~&~ This work  ~&~  Literature    \\        \hline\hline  
             \multirow{ 3 }{*}{$\V{K_2^{*}}(0)$}     	 &  \multirow{ 3 }{*}{$0.22 ^{+0.11}_{-0.08}$}  &   $0.16 \pm 0.02$  \cite{Yang:2010qd} \\ && $0.21^{+0.06}_{-0.05}$ \cite{Wang:2010ni} \\  && $ 0.71^{+0.29}_{-0.20}$ \cite{Wang:2010tz}          \\ \hline
\multirow{ 3 }{*}{$\Azero{K_2^{*}}(0)$}     	 &  \multirow{ 3 }{*}{$ 0.30 ^{+0.06}_{-0.05}$}  &   $0.25 \pm 0.04$  \cite{Yang:2010qd} \\ && $0.18^{+0.05}_{-0.04}$ \cite{Wang:2010ni} \\  && $0.40^{+0.57}_{-0.37}$ \cite{Wang:2010tz}        \\ \hline
\multirow{ 3 }{*}{$\Aone{K_2^{*}}(0)$}     	 &  \multirow{ 3 }{*}{$ 0.19 ^{+0.09}_{-0.07}$ }  &   $0.14 \pm 0.02$  \cite{Yang:2010qd} \\ && $0.13^{+0.04}_{-0.03}$ \cite{Wang:2010ni} \\  && $0.43^{+0.19}_{-0.12}$ \cite{Wang:2010tz}         \\ \hline
\multirow{ 3 }{*}{$\Atwo{K_2^{*}}(0)$}     	 &  \multirow{ 3 }{*}{$ 0.11 ^{+0.05}_{-0.06}$}  &   $0.05 \pm 0.02$  \cite{Yang:2010qd} \\ && $0.08^{+0.03}_{-0.02}$ \cite{Wang:2010ni} \\  && $0.45^{+0.26}_{-0.18}$ \cite{Wang:2010tz}          \\ \hline
\multirow{ 3 }{*}{$\Tone{K_2^{*}}(0)$}     	 &  \multirow{ 3 }{*}{$ 0.19 ^{+0.09}_{-0.06}$}  &   $0.14 \pm 0.02$  \cite{Yang:2010qd} \\   && $0.17^{+0.05}_{-0.04}$ \cite{Wang:2010ni}  \\  && $0.54^{+0.22}_{-0.15}$ \cite{Wang:2010tz}        \\ \hline
\multirow{ 3 }{*}{$\Ttwo{K_2^{*}}(0)$}     	 &  \multirow{ 3 }{*}{$ 0.19 ^{+0.09}_{-0.06}$ }  &   $0.14 \pm 0.02$ \cite{Yang:2010qd} \\   && $0.17^{+0.05}_{-0.04}$ \cite{Wang:2010ni}  \\  && $0.54^{+0.23}_{-0.15}$ \cite{Wang:2010tz}        \\ \hline
\multirow{ 3 }{*}{$\Tthree{K_2^{*}}(0)$}     	 &  \multirow{ 3 }{*}{$ 0.09 ^{+0.06}_{-0.04}$}  &  $0.01^{+0.02}_{-0.01}$ \cite{Yang:2010qd} \\ && $0.14^{+0.05}_{-0.03}$ \cite{Wang:2010ni} \\  && $0.45^{+0.23}_{-0.15}$ \cite{Wang:2010tz}        \\  \hline 
            \hline 
        \end{tabular}
         \end{minipage}
         %
                              \begin{minipage}{3in}
        \centering
        \begin{tabular}{| c | c | c |}
            \hline
            Form Factor  ~&~ This work  ~&~  Literature    \\        \hline\hline  
                         \multirow{ 5 }{*}{$\V{f_2}(0)$}     	 &  \multirow{ 5 }{*}{$ 0.11 ^{+0.07}_{-0.05}$ }  &   $0.12 \pm 0.04$ \cite{Khosravi:2015jfa} \\  &&  $0.30 \pm 0.03$ \cite{Emmerich:2018rug} \\ && $0.18 \pm 0.02$  \cite{Yang:2010qd} \\ && $0.12^{+0.03}_{-0.03}$ \cite{Wang:2010ni} \\  && $0.57^{+0.26}_{-0.16}$ \cite{Wang:2010tz}           \\ \hline
\multirow{ 5 }{*}{$\Azero{f_2}(0)$}     	 &  \multirow{ 5 }{*}{$ 0.20 ^{+0.04}_{-0.04}$}  &  $0.24 \pm 0.06$ \cite{Khosravi:2015jfa} \\  && $0.22 \pm 0.02$  \cite{Emmerich:2018rug} \\ && $0.20 \pm 0.04$ \cite{Yang:2010qd} \\ && $0.13^{+0.04}_{-0.03}$ \cite{Wang:2010ni} \\  && $0.32^{+0.59}_{-0.37}$ \cite{Wang:2010tz}     \\ \hline
\multirow{ 5 }{*}{$\Aone{f_2}(0)$}     	 &  \multirow{ 5 }{*}{$ 0.10 ^{+0.06}_{-0.04}$}  &  $0.10 \pm 0.02$ \cite{Khosravi:2015jfa} \\  && $0.17 \pm 0.01$   \cite{Emmerich:2018rug} \\ && $0.14 \pm 0.02$ \cite{Yang:2010qd} \\ && $0.08^{+0.02}_{-0.02}$ \cite{Wang:2010ni} \\  && $0.35^{+0.17}_{-0.10}$ \cite{Wang:2010tz}       \\ \hline
\multirow{ 5}{*}{$\Atwo{f_2}(0)$}     	 &  \multirow{ 5 }{*}{$ 0.04 ^{+0.05}_{-0.01}$}  & $0.09 \pm 0.02$ \cite{Khosravi:2015jfa} \\  && $0.11 \pm 0.02$  \cite{Emmerich:2018rug} \\ && $0.10 \pm 0.02$ \cite{Yang:2010qd} \\ && $0.04^{+0.01}_{-0.01}$ \cite{Wang:2010ni} \\  && $0.37^{+0.25}_{-0.17}$ \cite{Wang:2010tz}         \\ \hline
\multirow{ 3 }{*}{$\Tone{f_2}(0)$}     	 &  \multirow{ 3 }{*}{$ 0.10 ^{+0.05}_{-0.04}$}  &   $0.11 \pm 0.02$  \cite{Emmerich:2018rug} \\ && $0.15 \pm 0.02$ \cite{Yang:2010qd} \\ && $0.10^{+0.03}_{-0.02}$ \cite{Wang:2010ni} \\  && $0.44^{+0.20}_{-0.13}$ \cite{Wang:2010tz}   \\ \hline
\multirow{ 3 }{*}{$\Ttwo{f_2}(0)$}     	 &  \multirow{ 3 }{*}{$ 0.10 ^{+0.05}_{-0.04}$}  & $0.12 \pm 0.01$  \cite{Emmerich:2018rug} \\ &&  $0.14 \pm 0.02$ \cite{Yang:2010qd} \\ && $0.10^{+0.03}_{-0.02}$ \cite{Wang:2010ni}   \\  && $0.44^{+0.20}_{-0.13}$ \cite{Wang:2010tz}    \\ \hline
\multirow{ 4 }{*}{$\Tthree{f_2}(0)$}     	 &  \multirow{ 4 }{*}{$ 0.04 ^{+0.04}_{-0.02}$}  &  $- 0.02 \pm 0.04$  \cite{Emmerich:2018rug} \\ && $0.06 \pm 0.02$ \cite{Yang:2010qd} \\ && $0.09^{+0.03}_{-0.02}$ \cite{Wang:2010ni} \\  && $0.38^{+0.20}_{-0.13}$ \cite{Wang:2010tz}    
\\ \hline \hline  
        \end{tabular}
                                      \vspace{0.8cm}
         \end{minipage}
          \hspace{.8cm}
         %
             \begin{minipage}{3in}
        \centering
        \begin{tabular}{| c | c | c |}
            \hline
            Form Factor  ~&~ This work  ~&~  Literature    \\        \hline\hline  
             \multirow{ 4 }{*}{$\V{a_2^{+}}(0)$}     	 &  \multirow{ 4 }{*}{$ 0.18 ^{+0.12}_{-0.07}$}  &    $0.13 \pm 0.03$ \cite{Khosravi:2015jfa}  \\ &&$0.18 \pm 0.02$ \cite{Yang:2010qd} \\ && $0.18^{+0.05}_{-0.04}$ \cite{Wang:2010ni} \\  && $0.60^{+0.28}_{-0.17}$  \cite{Wang:2010tz}      \\ \hline
\multirow{ 4 }{*}{$\Azero{a_2^{+}}(0)$}     	 &  \multirow{ 4 }{*}{$ 0.30 ^{+0.06}_{-0.05}$}  &   $0.26 \pm 0.07$ \cite{Khosravi:2015jfa} \\ && $0.21 \pm 0.04$  \cite{Yang:2010qd} \\ && $0.18^{+0.06}_{-0.04}$ \cite{Wang:2010ni} \\  && $0.35^{+0.56}_{-0.39}$  \cite{Wang:2010tz}        \\ \hline
\multirow{ 4 }{*}{$\Aone{a_2^{+}}(0)$}     	 &  \multirow{ 4 }{*}{$ 0.16 ^{+0.09}_{-0.05}$ }  &   $0.11 \pm 0.04$ \cite{Khosravi:2015jfa} \\ && $0.14 \pm 0.02$ \cite{Yang:2010qd} \\ && $0.11^{+0.03}_{-0.03}$ \cite{Wang:2010ni} \\  && $0.37^{+0.16}_{-0.11}$  \cite{Wang:2010tz}         \\ \hline
\multirow{ 4 }{*}{$\Atwo{a_2^{+}}(0)$}     	 &  \multirow{ 4 }{*}{$ 0.07 ^{+0.08}_{-0.03}$}  &   $0.09 \pm 0.02$  \cite{Khosravi:2015jfa} \\ && $0.09 \pm 0.02$  \cite{Yang:2010qd} \\ && $0.06^{+0.02}_{-0.01}$ \cite{Wang:2010ni} \\  && $0.38^{+0.26}_{-0.18}$ \cite{Wang:2010tz}         \\ \hline
\multirow{ 3 }{*}{$\Tone{a_2^{+}}(0)$}     	 &  \multirow{ 3 }{*}{$ 0.15 ^{+0.09}_{-0.05}$ }  &  $0.15 \pm 0.02$  \cite{Yang:2010qd} \\ && $0.15^{+0.04}_{-0.03}$ \cite{Wang:2010ni}     \\  && $0.46^{+0.21}_{-0.14}$ \cite{Wang:2010tz}   \\ \hline
\multirow{ 3 }{*}{$\Ttwo{a_2^{+}}(0)$}     	 &  \multirow{ 3 }{*}{$ 0.15 ^{+0.09}_{-0.05}$ }  &    $0.15 \pm 0.02$  \cite{Yang:2010qd} \\ && $0.15^{+0.04}_{-0.03}$ \cite{Wang:2010ni}     \\  && $0.46^{+0.21}_{-0.14}$ \cite{Wang:2010tz}    \\ \hline
\multirow{ 3 }{*}{$\Tthree{a_2^{+}}(0)$}     	 &  \multirow{ 3 }{*}{$ 0.07 ^{+0.06}_{-0.03}$}  &   $0.04 \pm 0.02$  \cite{Yang:2010qd} \\ && $0.13^{+0.04}_{-0.03}$ \cite{Wang:2010ni} \\  && $0.39^{+0.21}_{-0.14}$ \cite{Wang:2010tz}      \\ \hline
            \hline 
        \end{tabular}
         \end{minipage}  
        \caption{%
            Comparison of our form factor results at $q^2=0$ with existing results in the literature. }
        \label{tab:allFFcomparison}
    \end{table}
\end{widetext}
\FloatBarrier

\begin{widetext}
\renewcommand{\arraystretch}{3.}
\begin{figure}[htbp]
\centering
\begin{tabular}{c p{0.01\textwidth} c}
\includegraphics[width=0.34\textwidth]{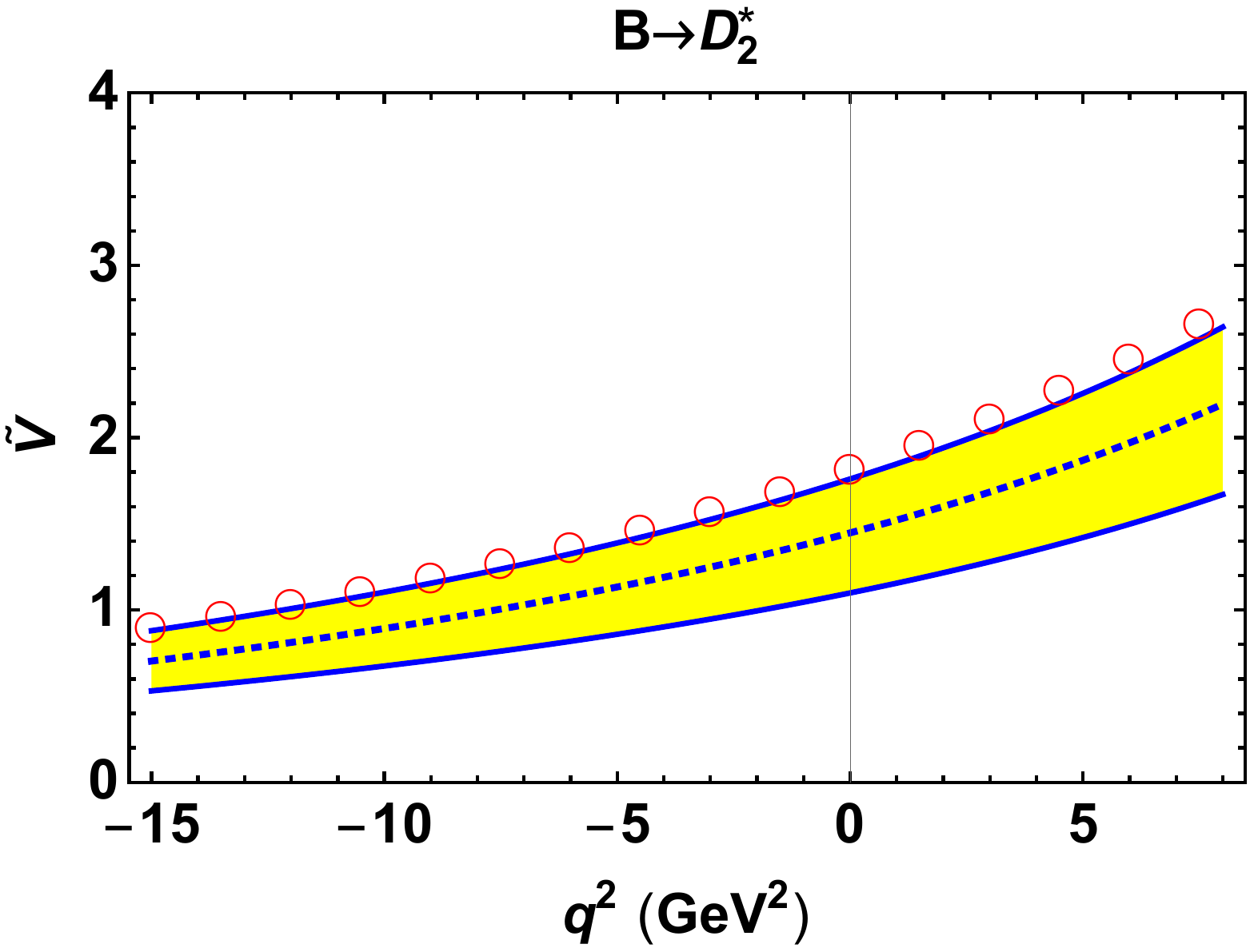}   \quad  &&
\includegraphics[width=0.34\textwidth]{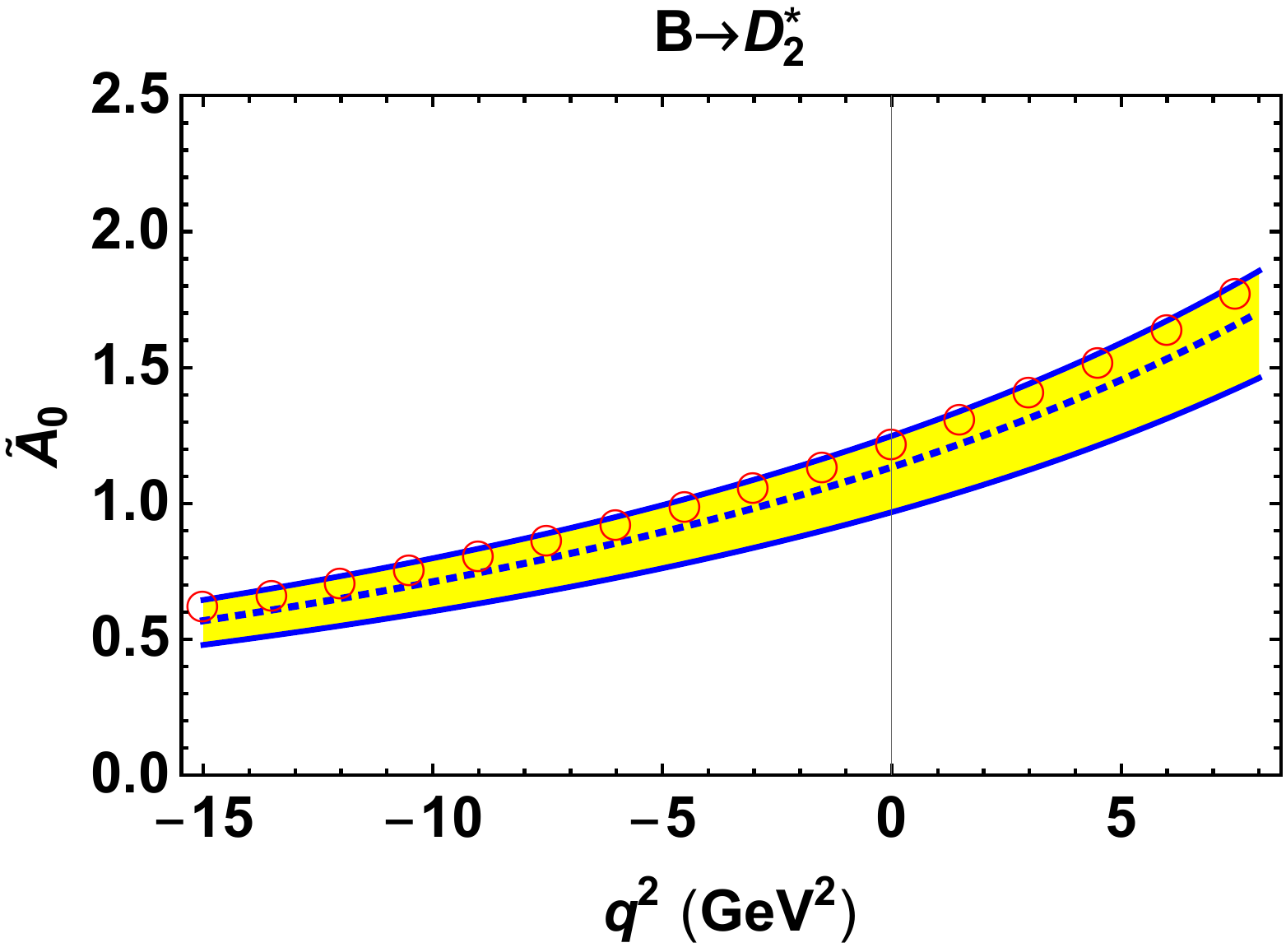} \quad 
\includegraphics[width=0.34\textwidth]{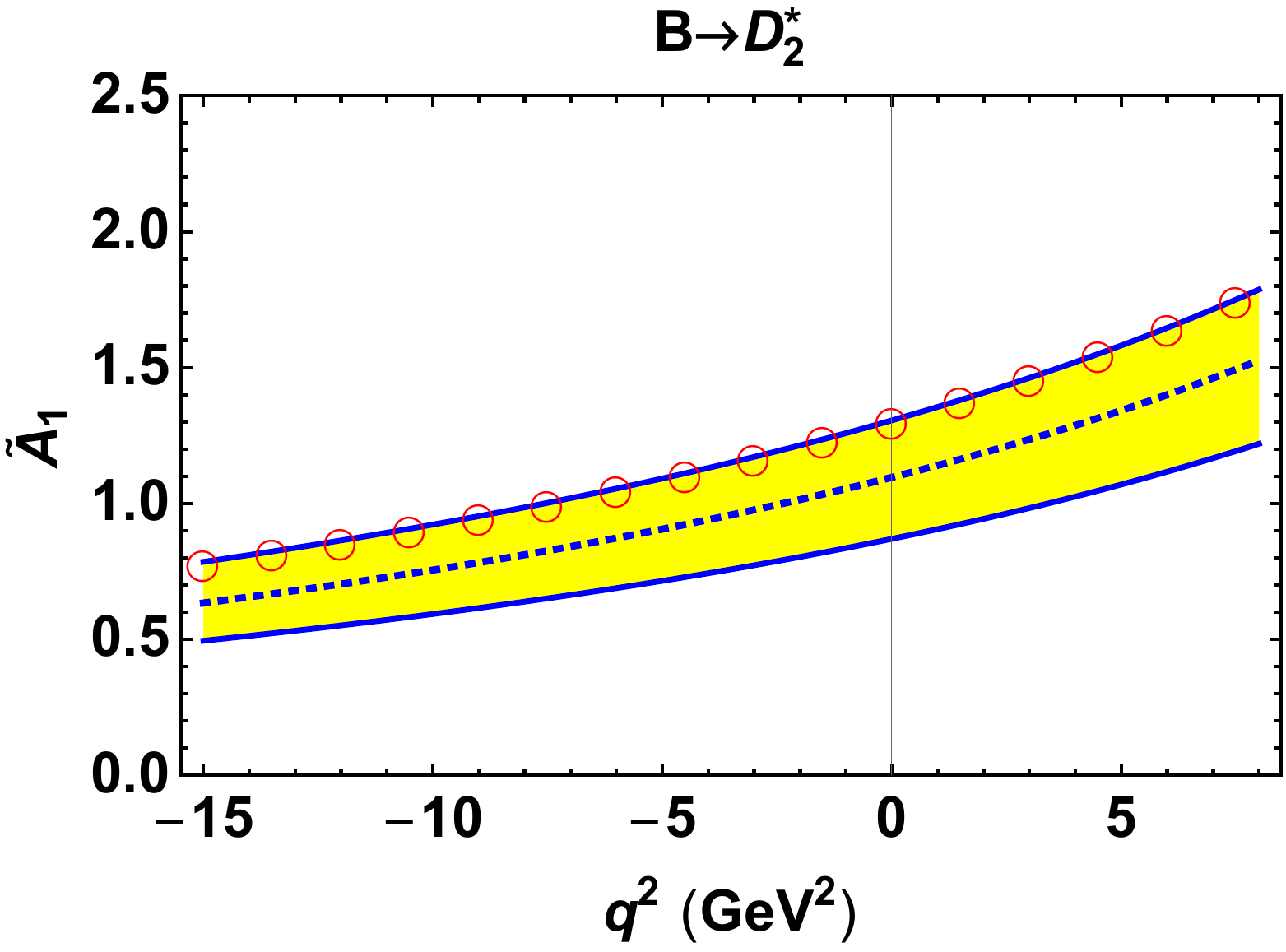} \quad  \\
\includegraphics[width=0.34\textwidth]{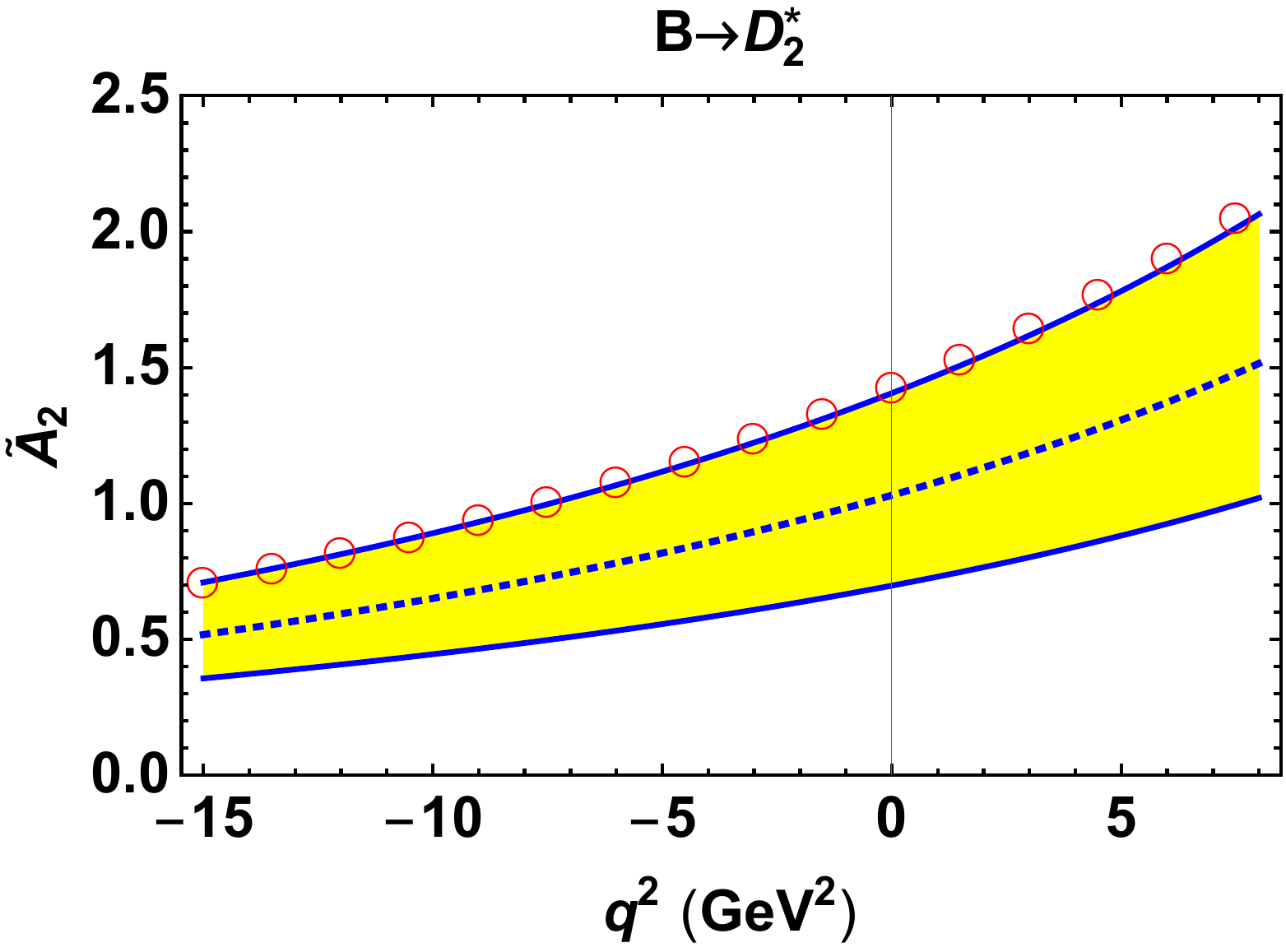} \quad &&
\includegraphics[width=0.34\textwidth]{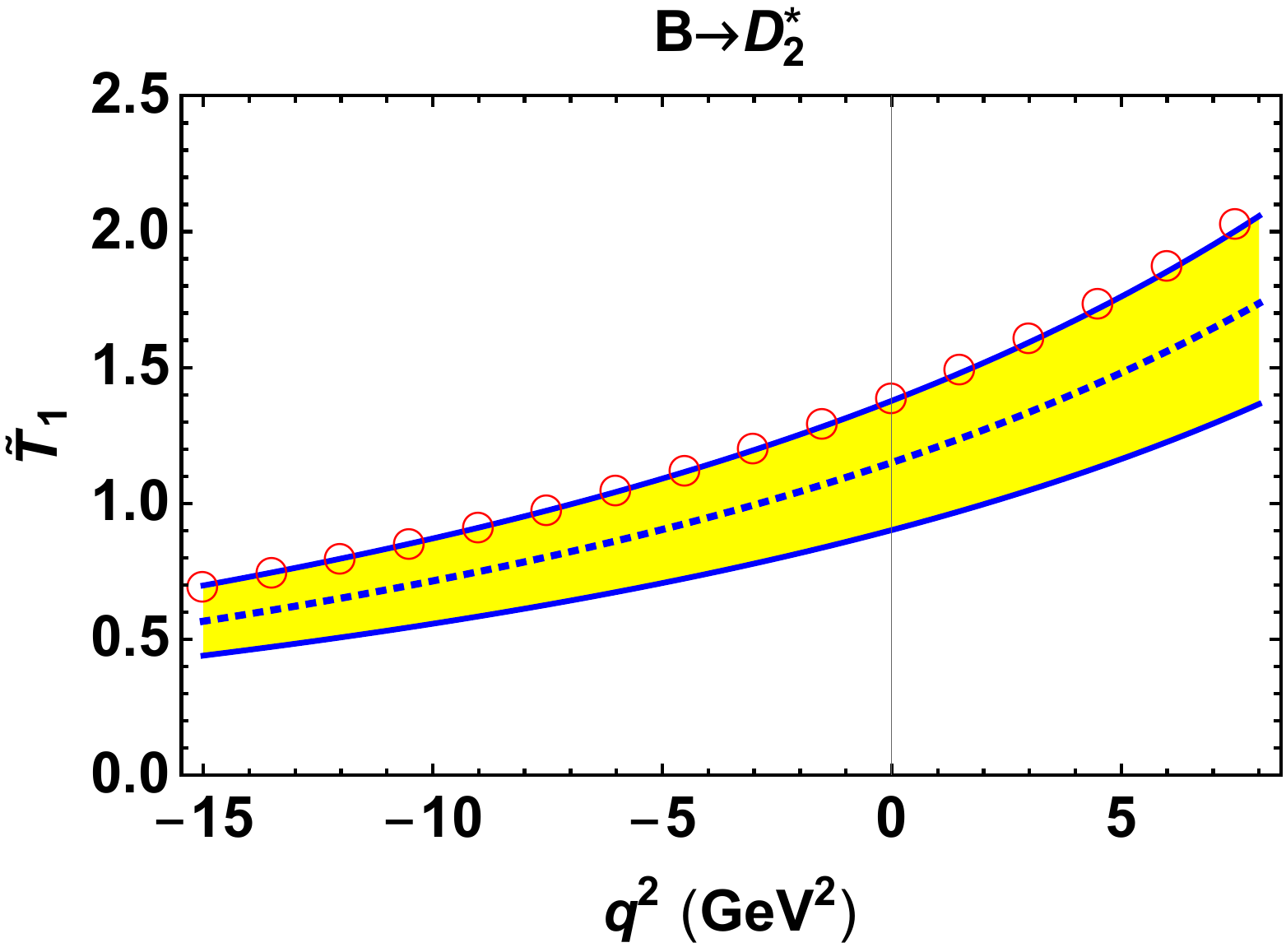} \quad 
\includegraphics[width=0.34\textwidth]{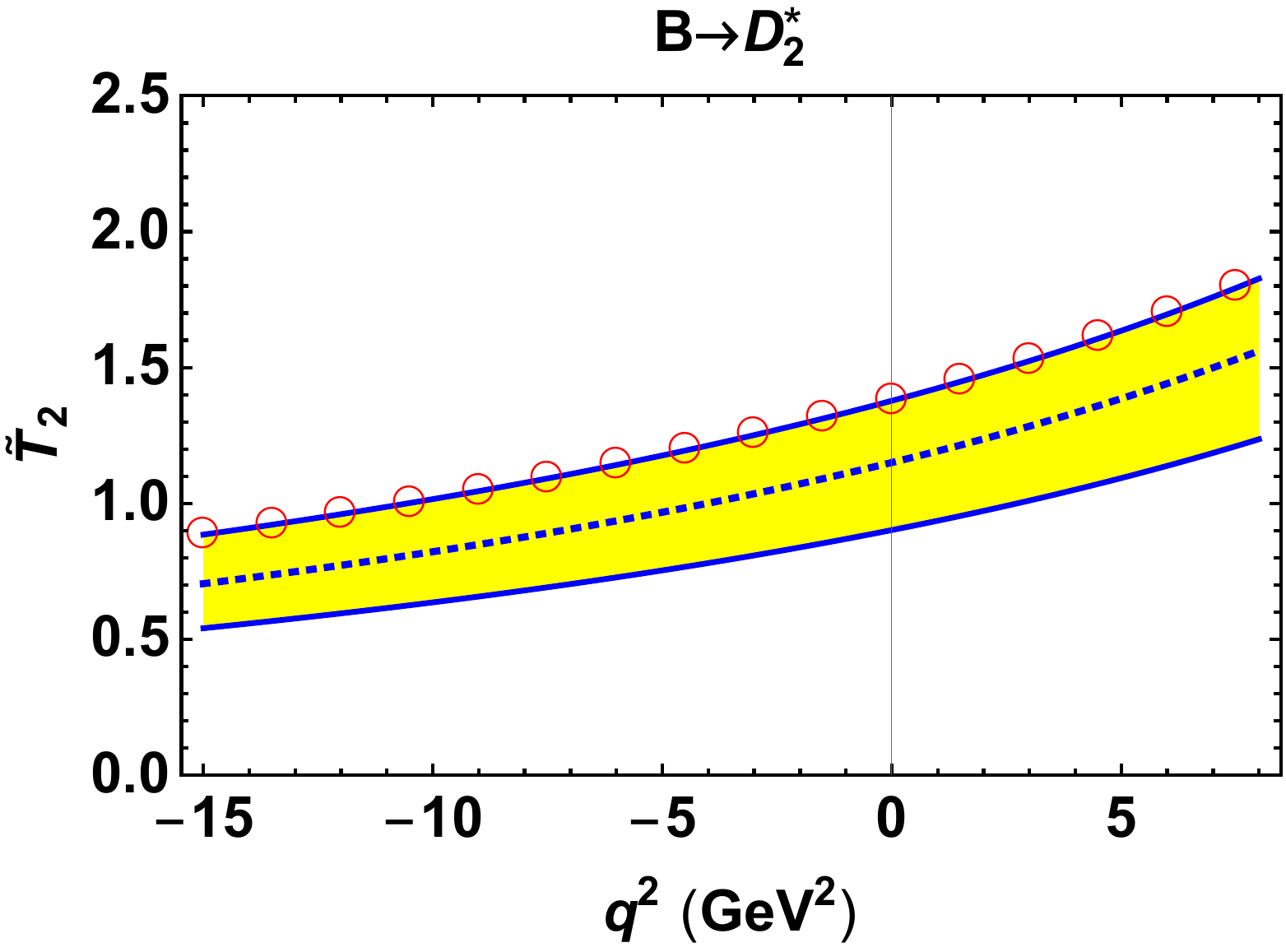} \quad  \\
\includegraphics[width=0.34\textwidth]{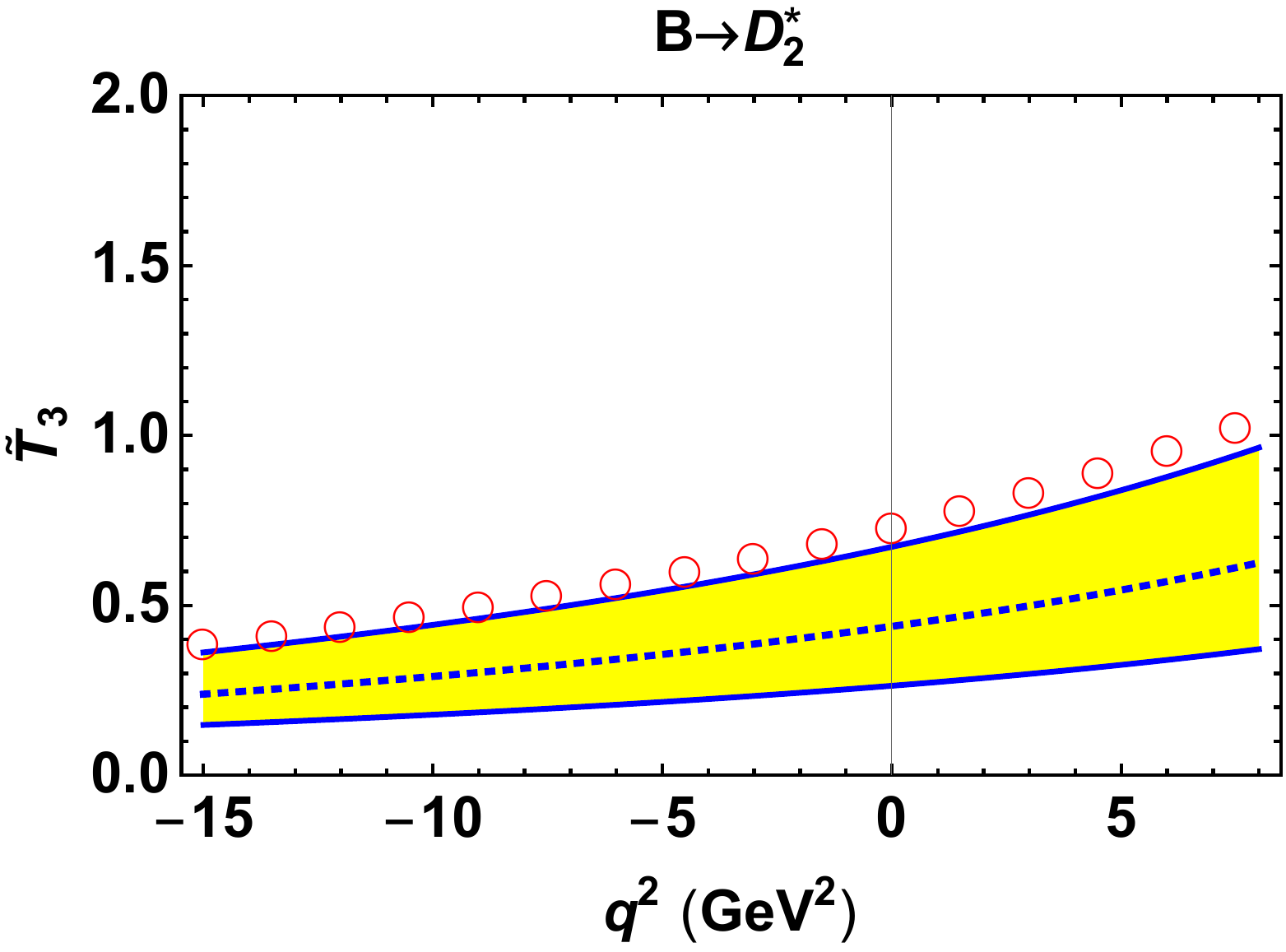}  \quad &&
 \end{tabular}
\caption{$q^2$ dependence of the $B \to D_2^*$ form factors from fits to our LCSR results. The dotted-curves (blue) represent the central values of the form factors as functions of $q^2$ and the shaded areas (yellow) describe the respective error budget on each form factor including the calculated higher twist terms. For comparison purposes, we also show in the same plots the central values of the leading-twist results as empty-circles (red). }
\label{fig:q2DepofBtoD2FFs}
\end{figure}
\end{widetext}
\begin{figure}[htbp]
\centering
\begin{tabular}{c p{0.01\textwidth} c}
\includegraphics[width=0.34\textwidth]{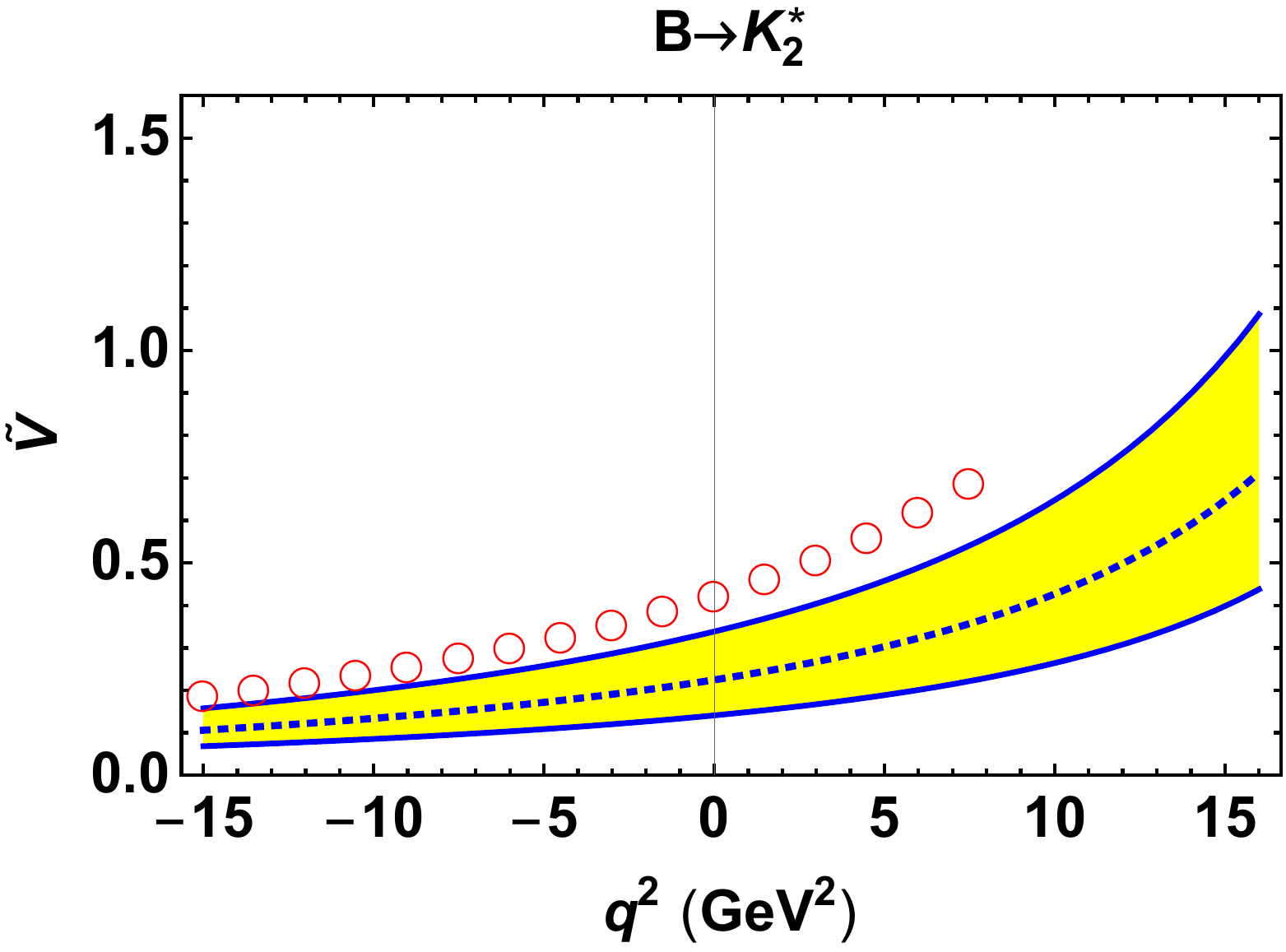}   \quad  &&
\includegraphics[width=0.34\textwidth]{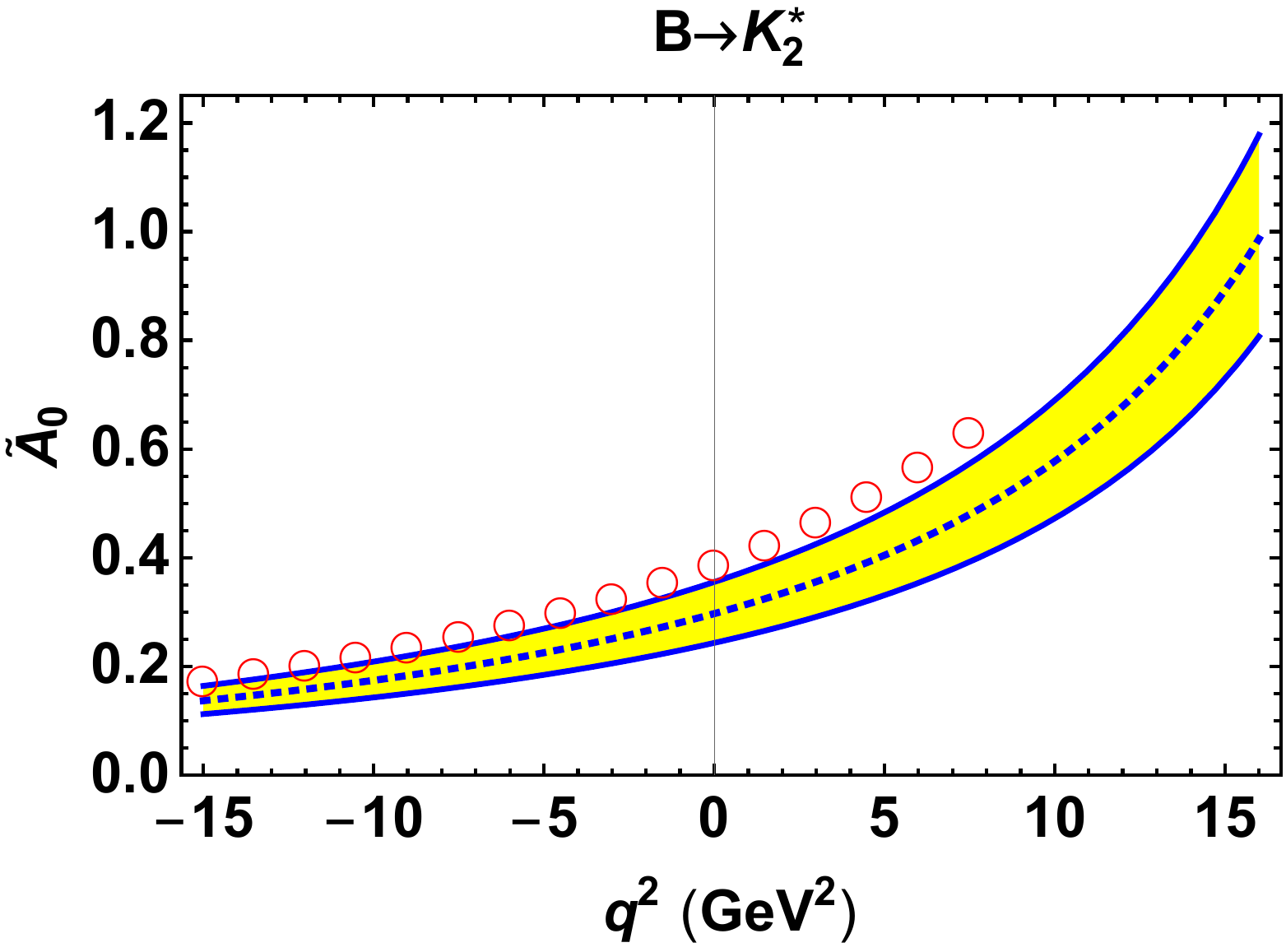} \quad 
\includegraphics[width=0.34\textwidth]{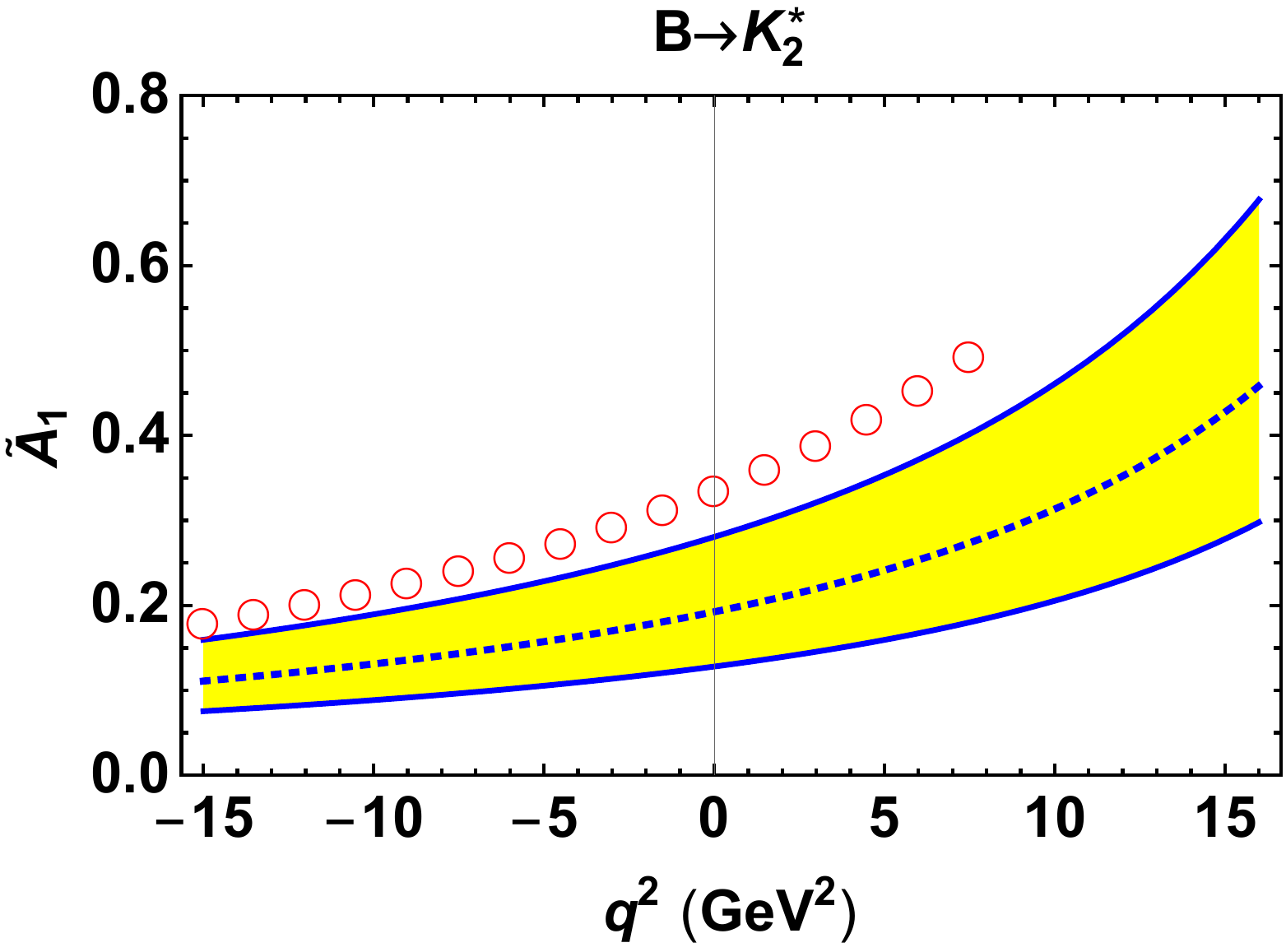} \quad  \\
\includegraphics[width=0.34\textwidth]{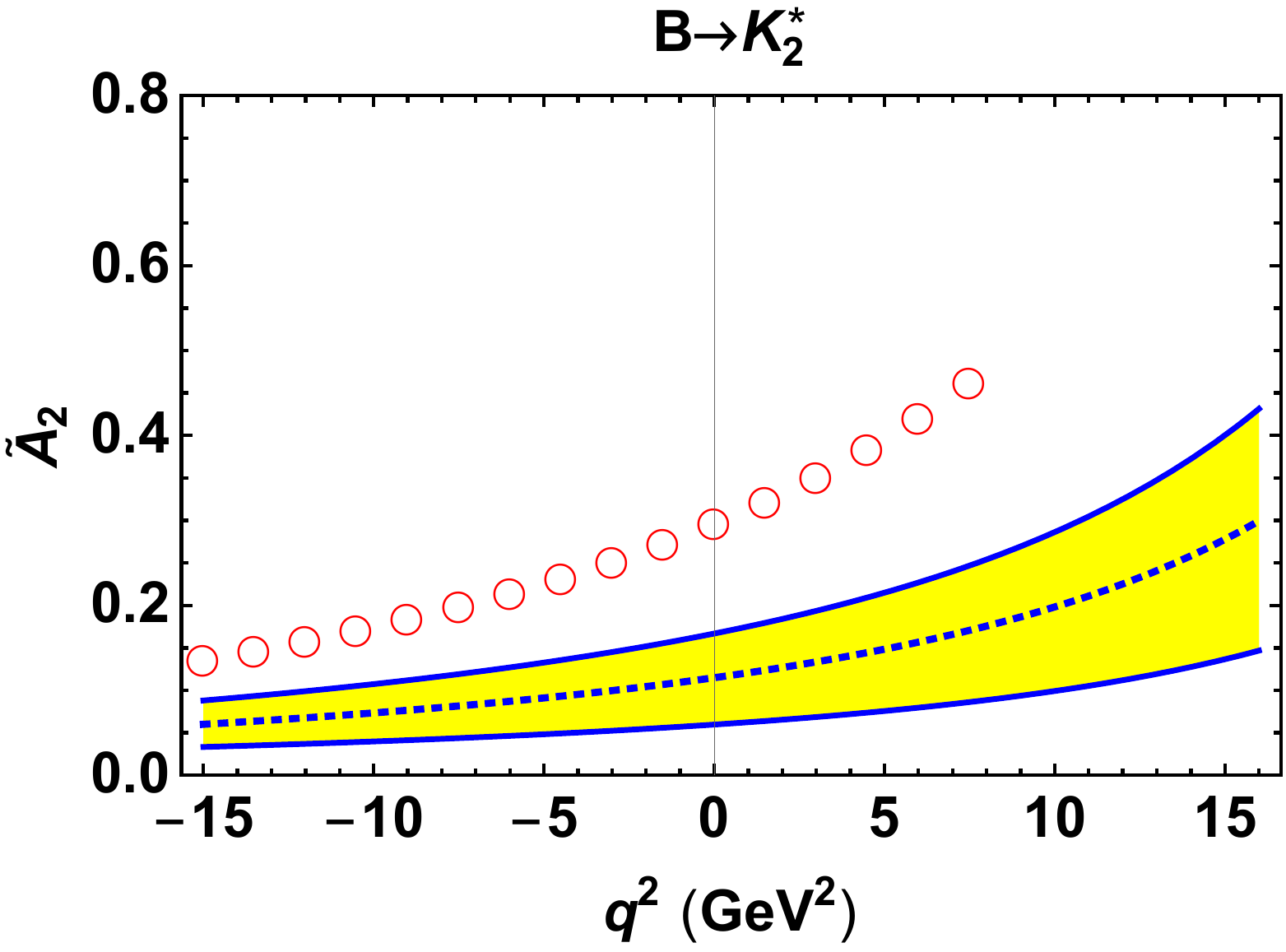} \quad &&
\includegraphics[width=0.34\textwidth]{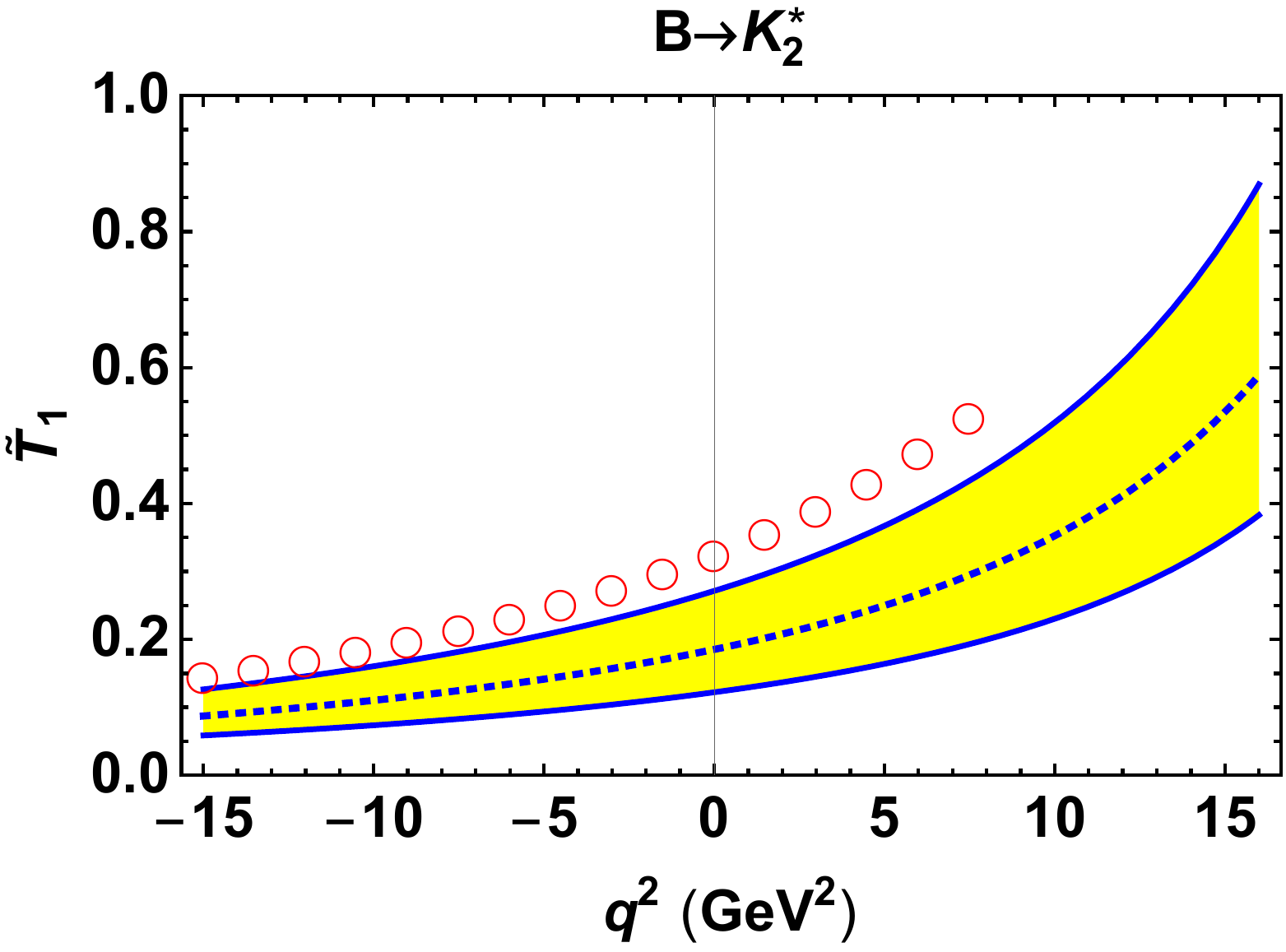} \quad 
\includegraphics[width=0.34\textwidth]{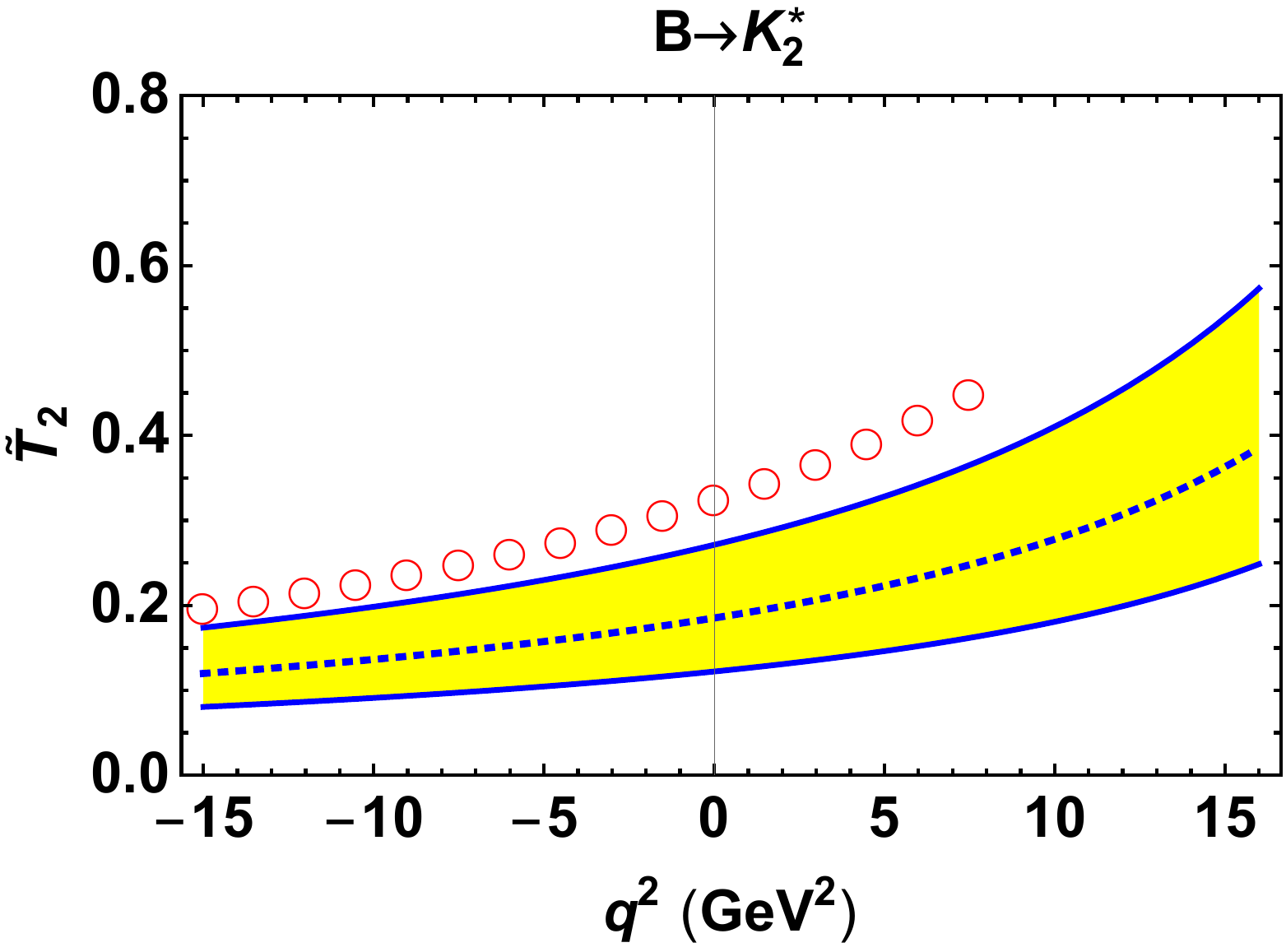} \quad  \\
\includegraphics[width=0.34\textwidth]{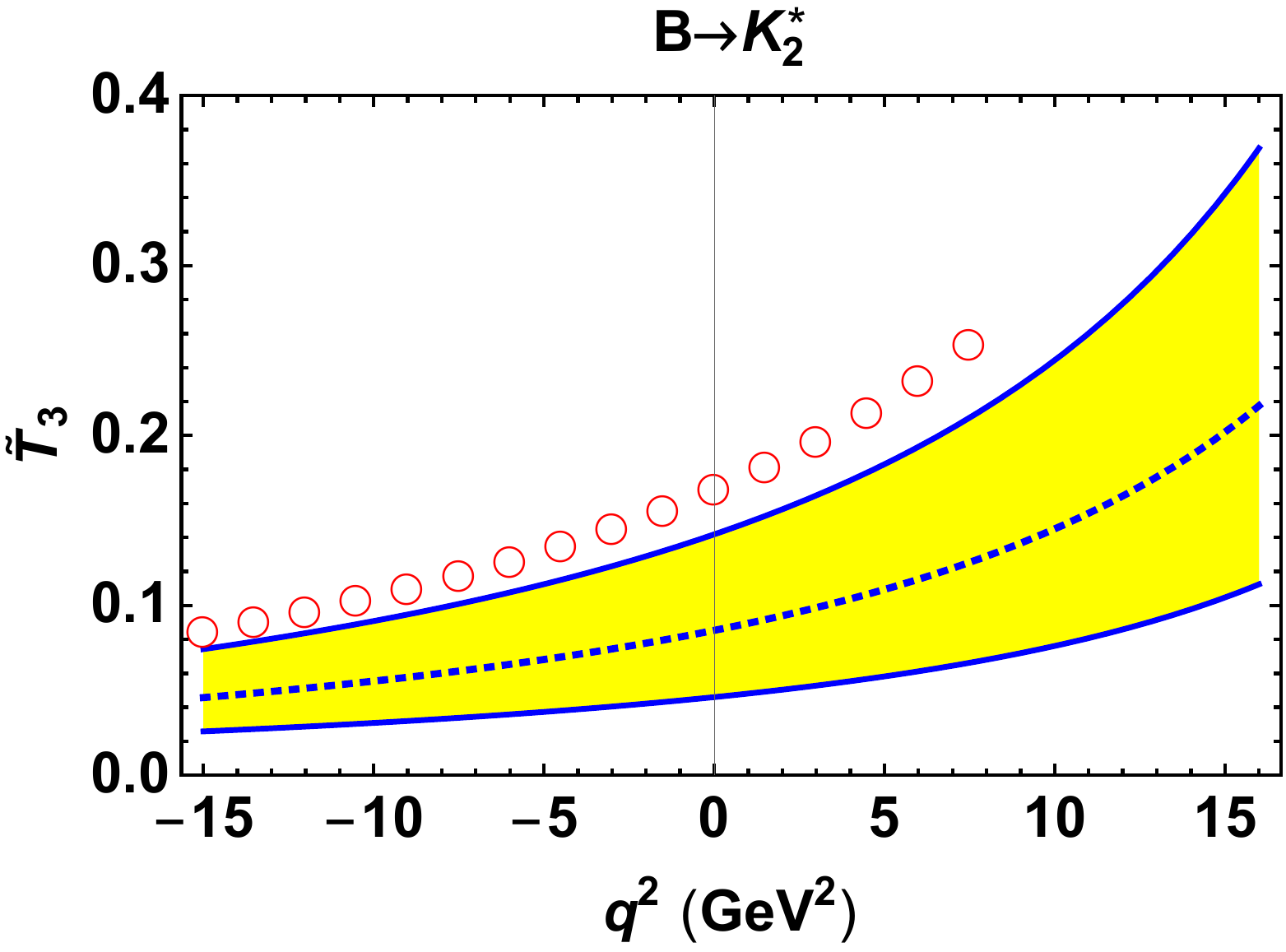}  \quad &&
 \end{tabular}
\caption{$q^2$ dependence of the $B \to K_2^*$ form factors from fits to our LCSR results. For details see \reffig{q2DepofBtoD2FFs}.}\label{fig:q2DepofBtoK2FFs}
\end{figure}
\begin{figure}[htbp]
\centering
\begin{tabular}{c p{0.01\textwidth} c}
\includegraphics[width=0.34\textwidth]{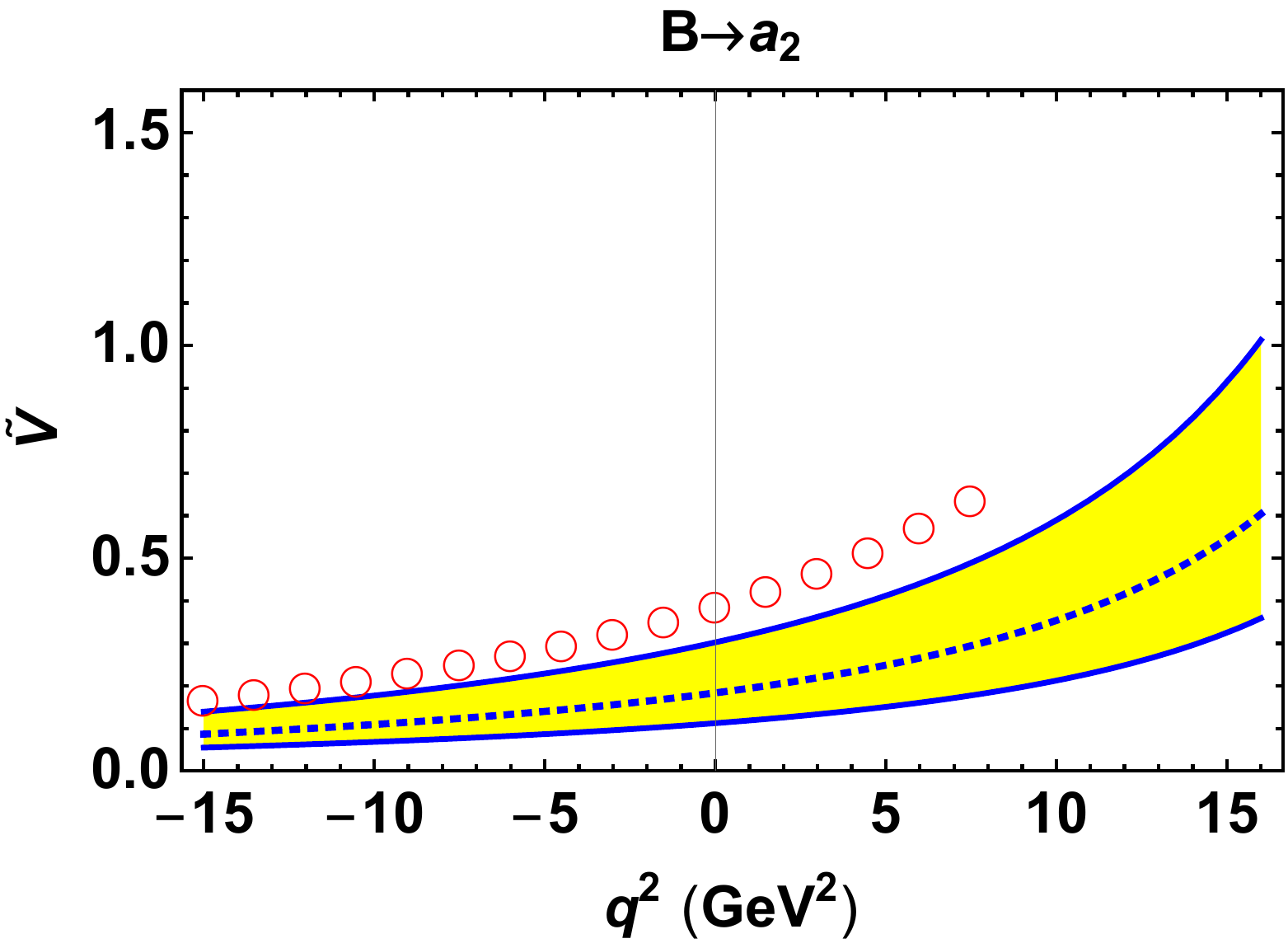}   \quad  &&
\includegraphics[width=0.34\textwidth]{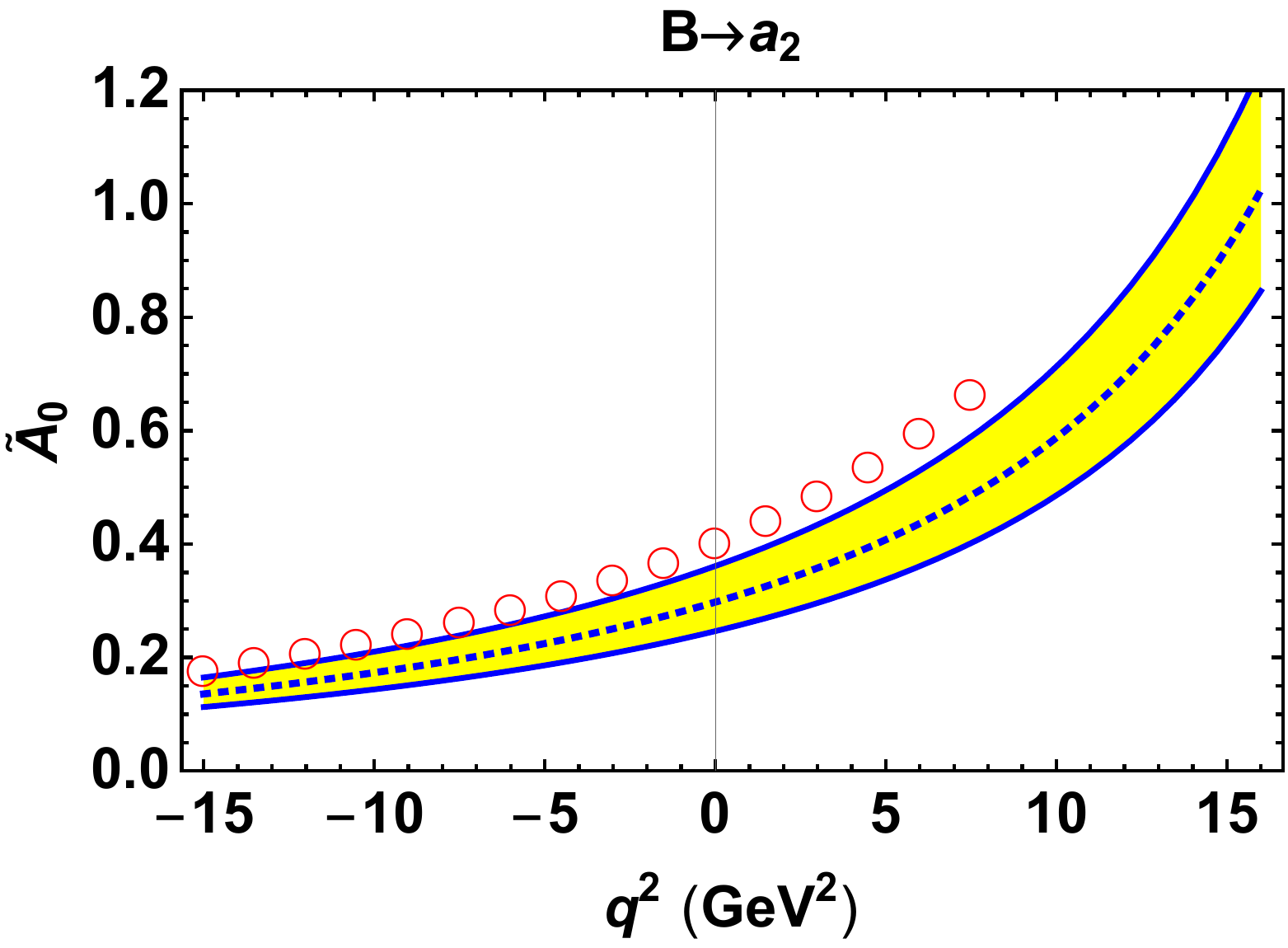} \quad 
\includegraphics[width=0.34\textwidth]{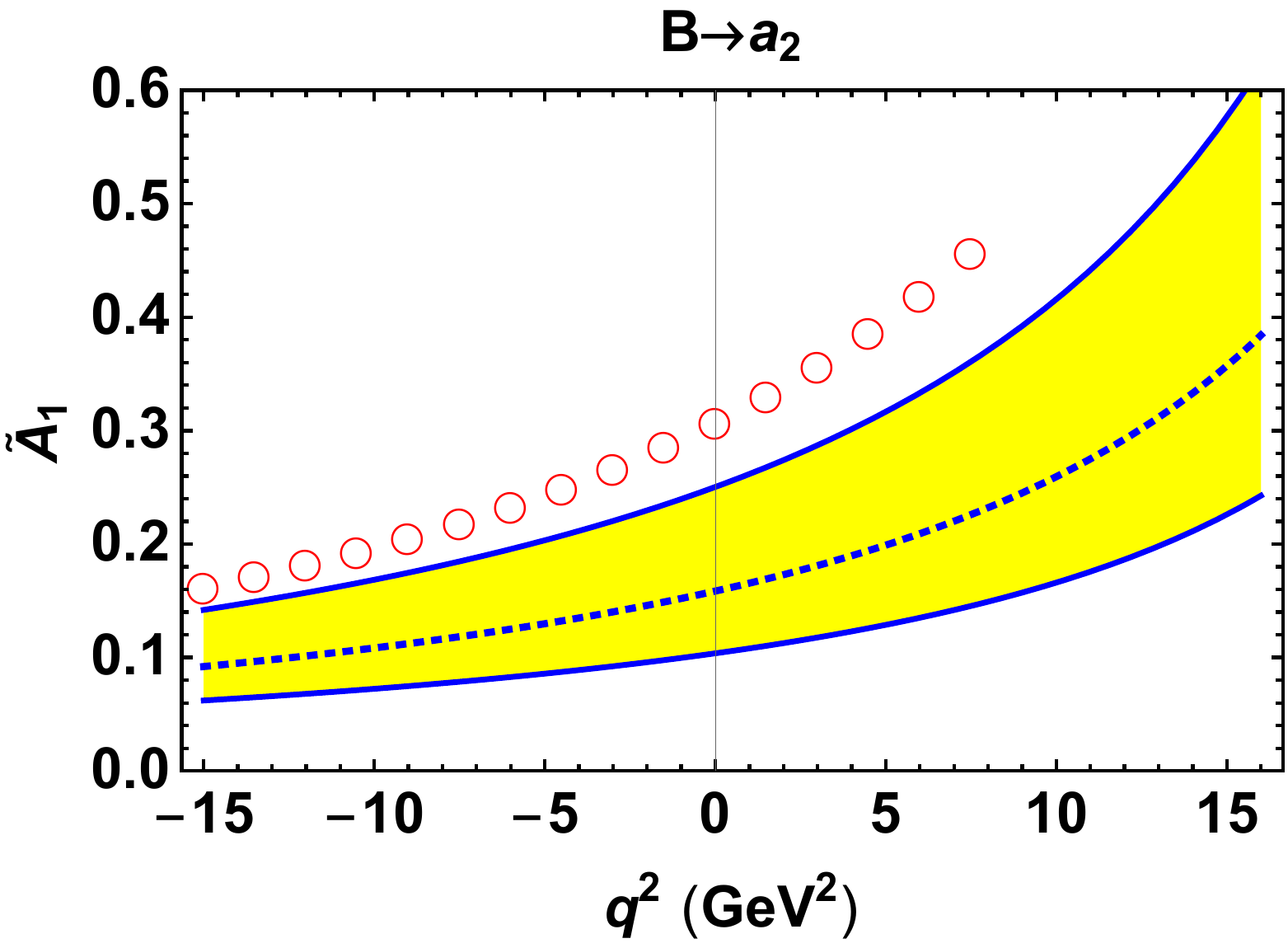} \quad  \\
\includegraphics[width=0.34\textwidth]{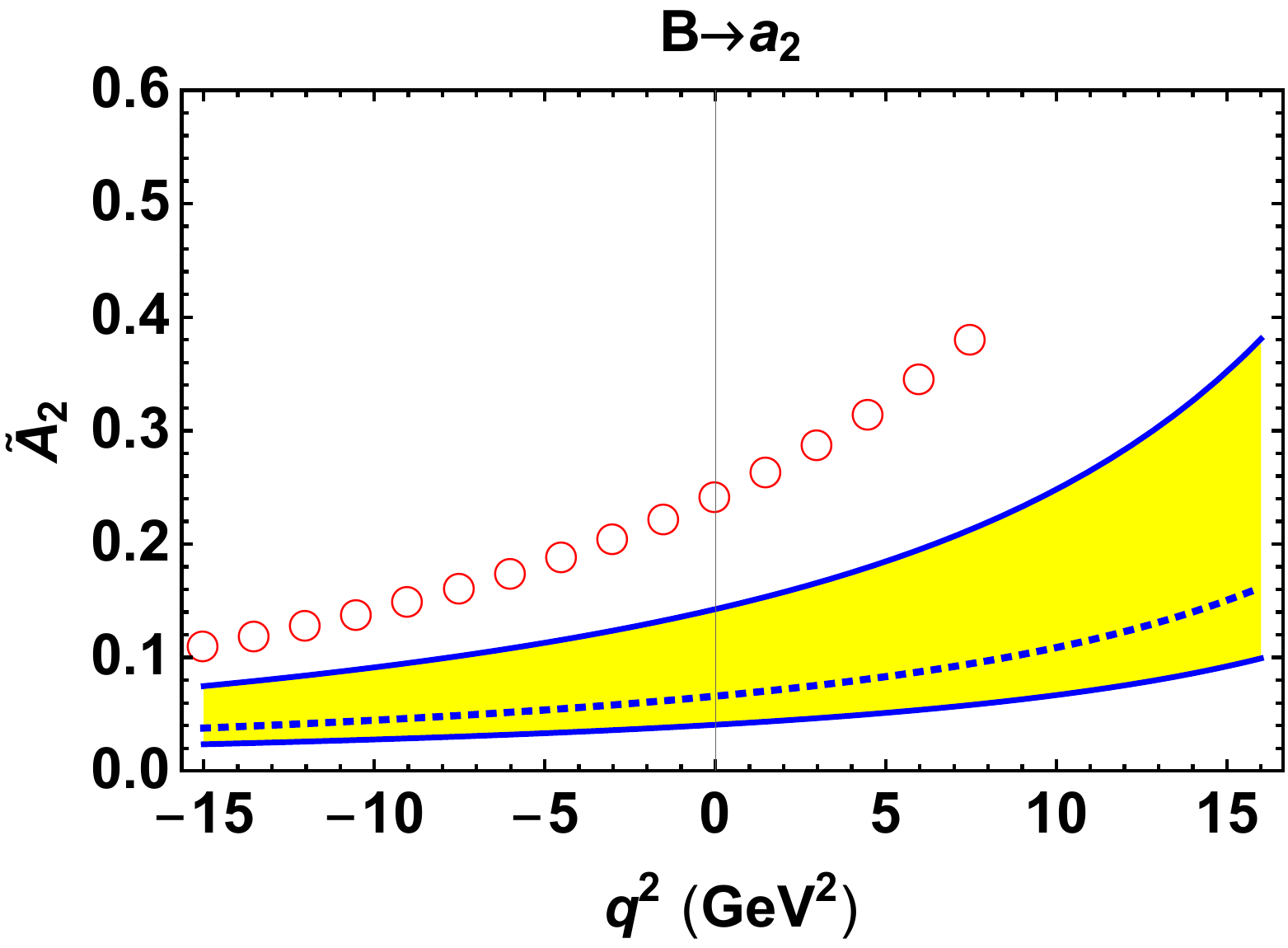} \quad &&
\includegraphics[width=0.34\textwidth]{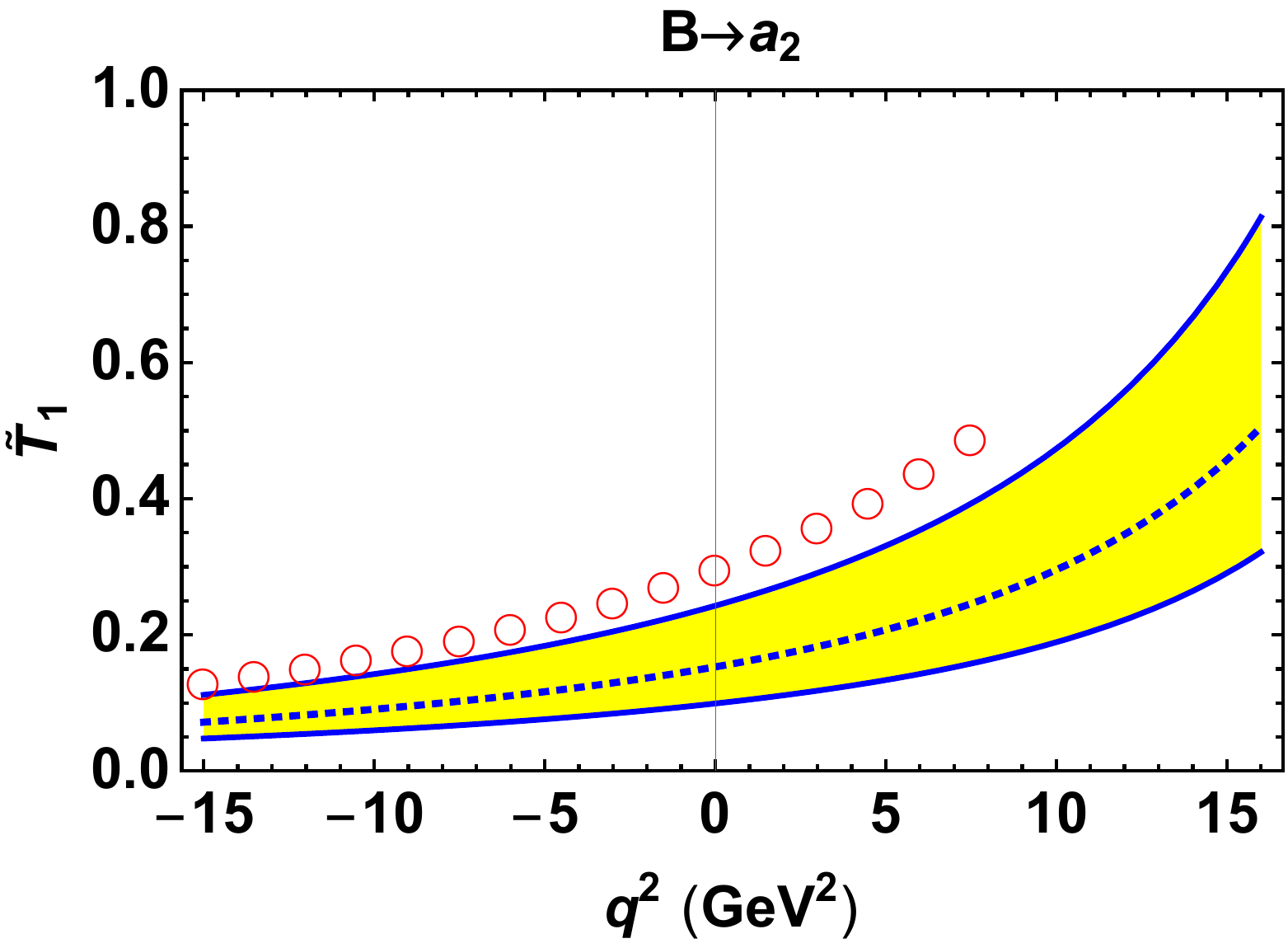} \quad 
\includegraphics[width=0.34\textwidth]{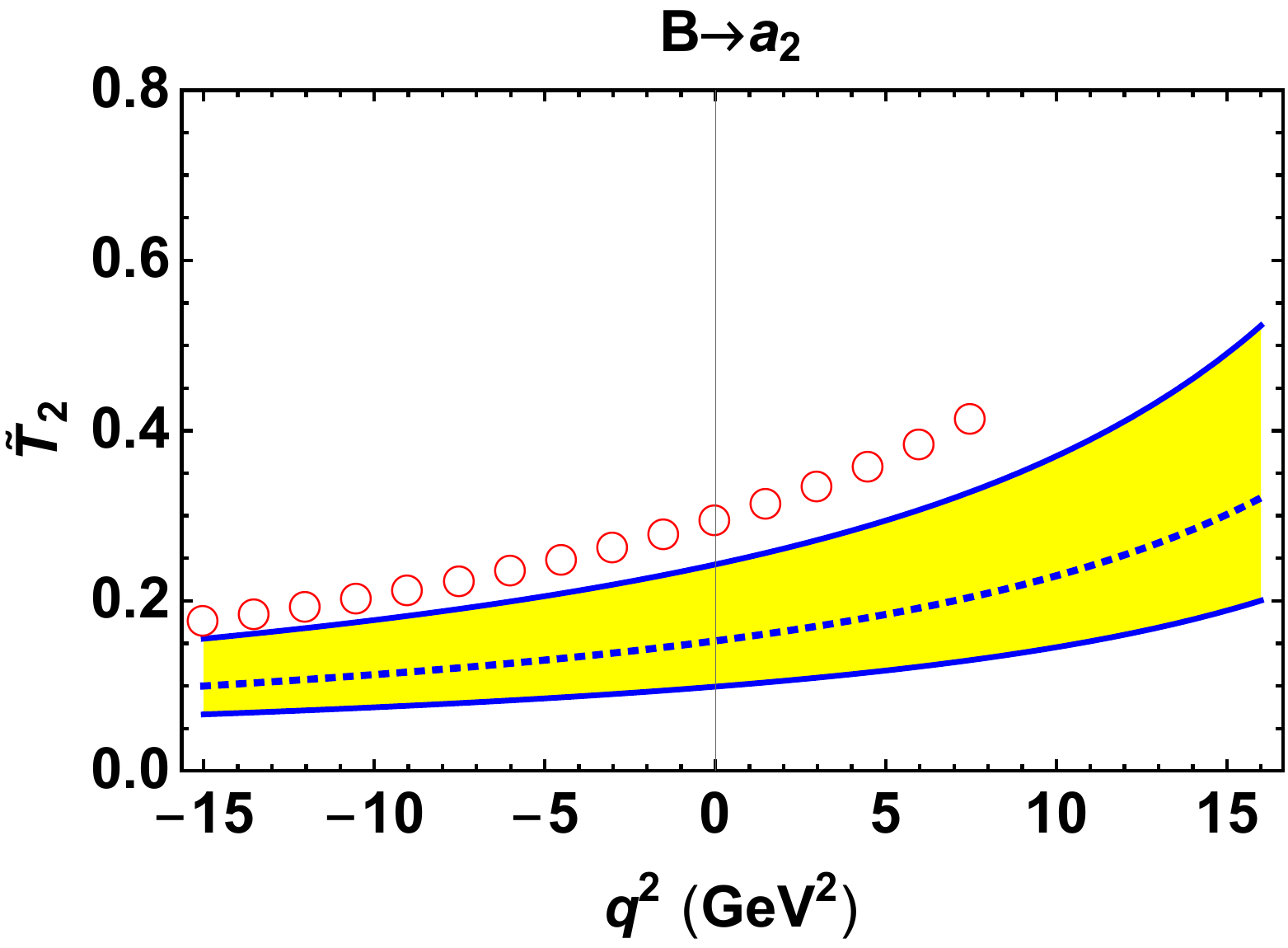} \quad  \\
\includegraphics[width=0.34\textwidth]{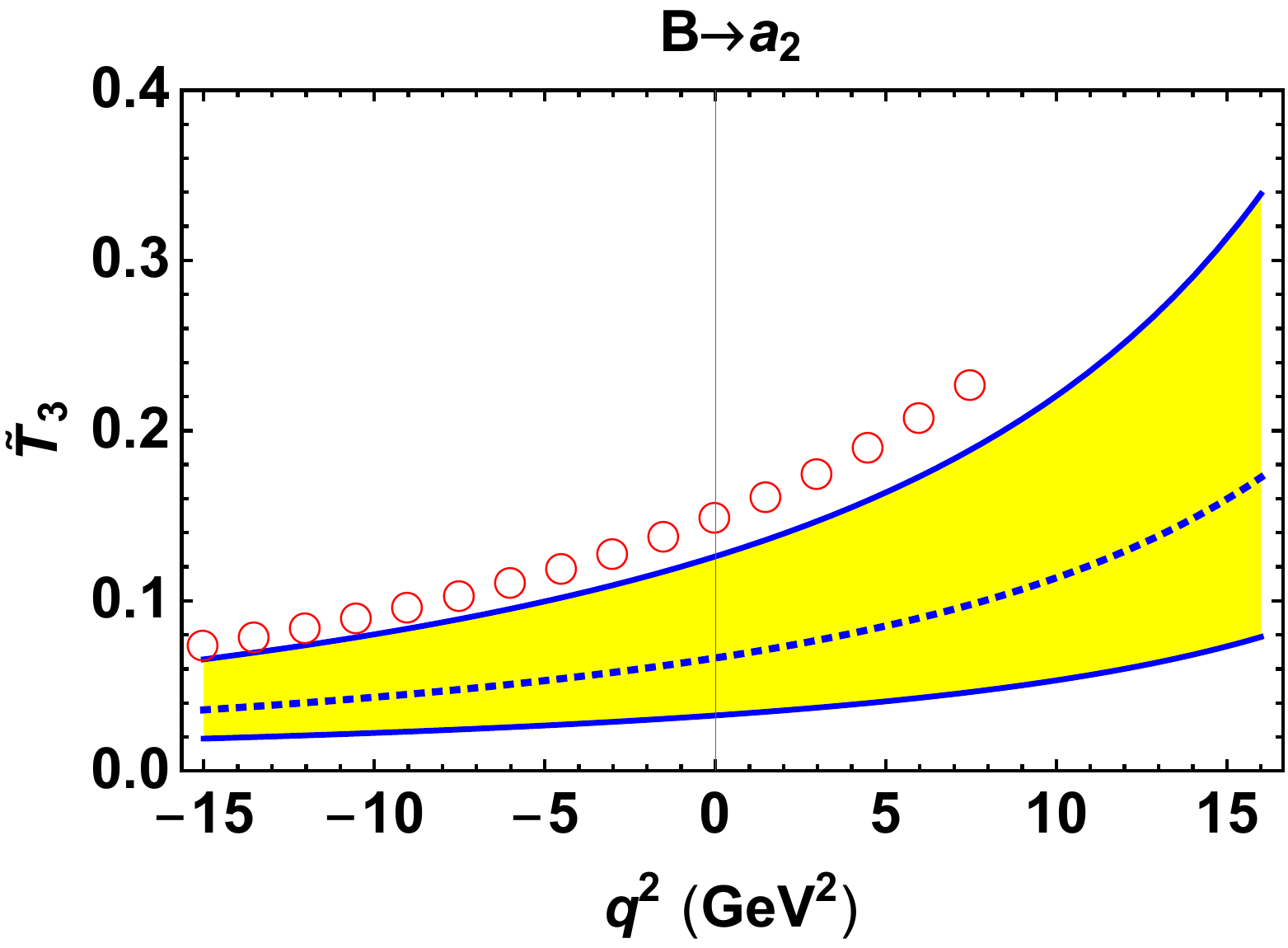}  \quad &&
 \end{tabular}
\caption{$q^2$ dependence of the $B \to a_2^{+}$ form factors from fits to our LCSR results. For details see \reffig{q2DepofBtoD2FFs}.}\label{fig:q2DepofBtoa2FFs}
\end{figure}
\begin{figure}[htbp]
\centering
\begin{tabular}{c p{0.01\textwidth} c}
\includegraphics[width=0.34\textwidth]{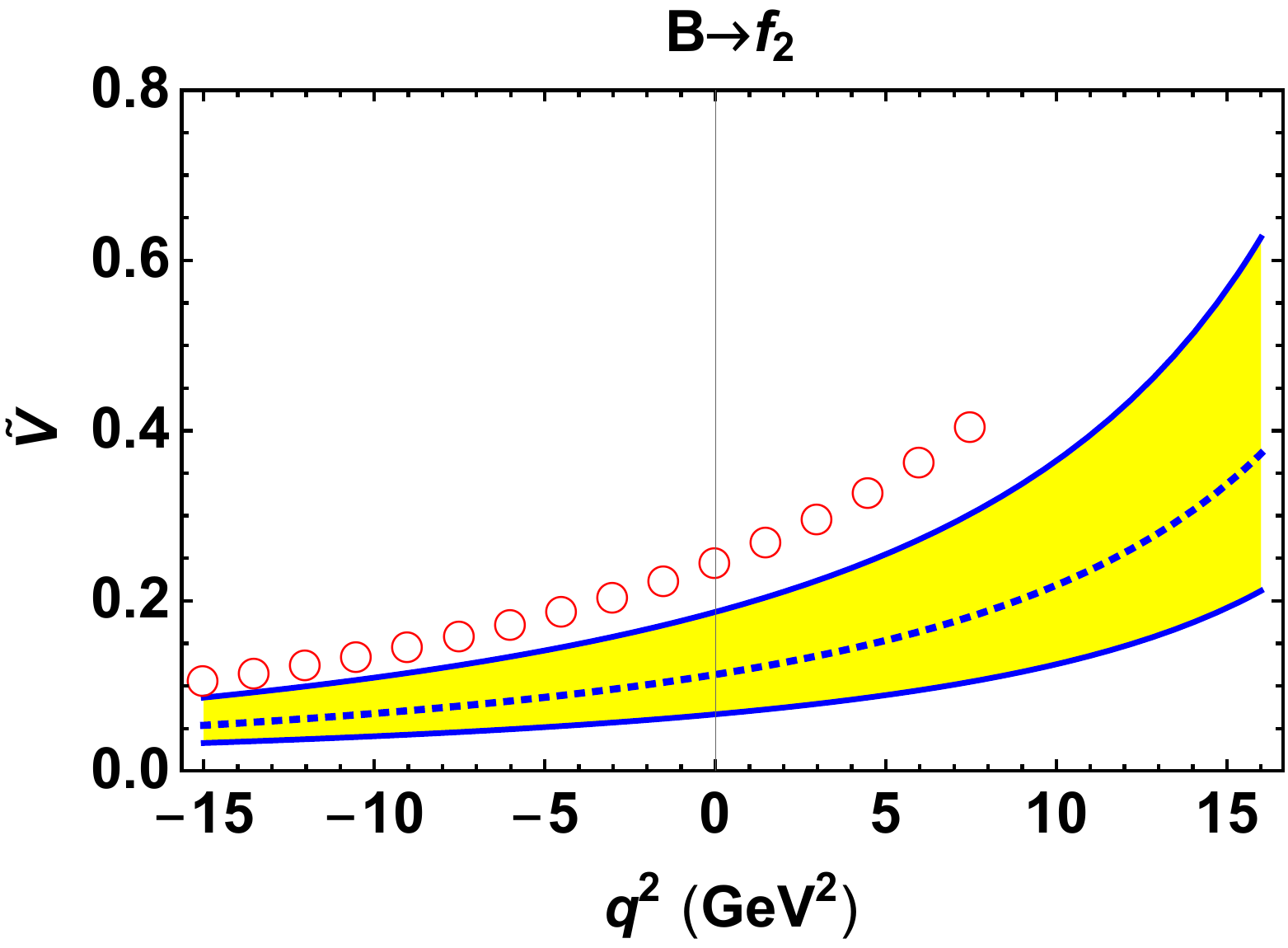}   \quad  &&
\includegraphics[width=0.34\textwidth]{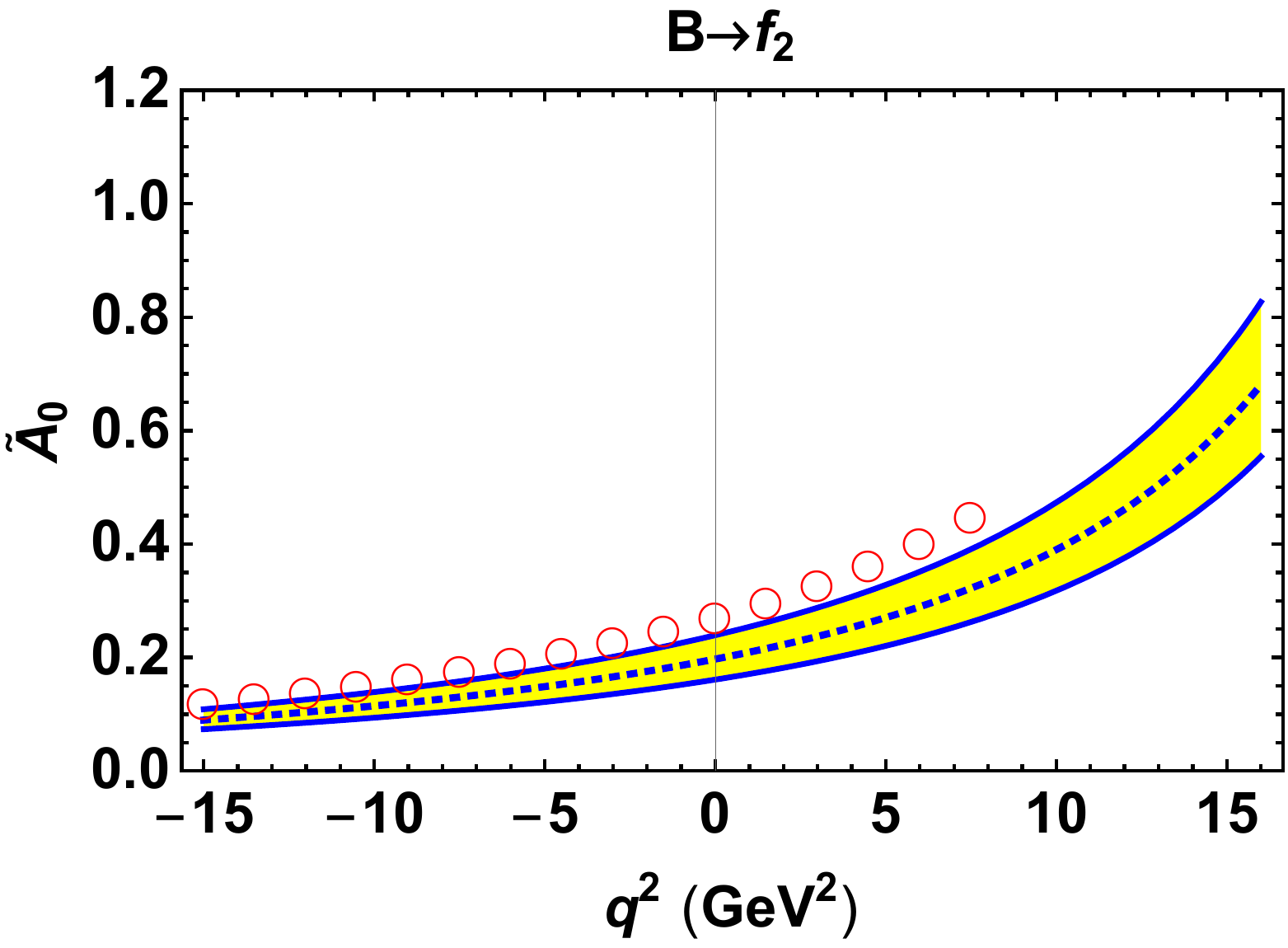} \quad 
\includegraphics[width=0.34\textwidth]{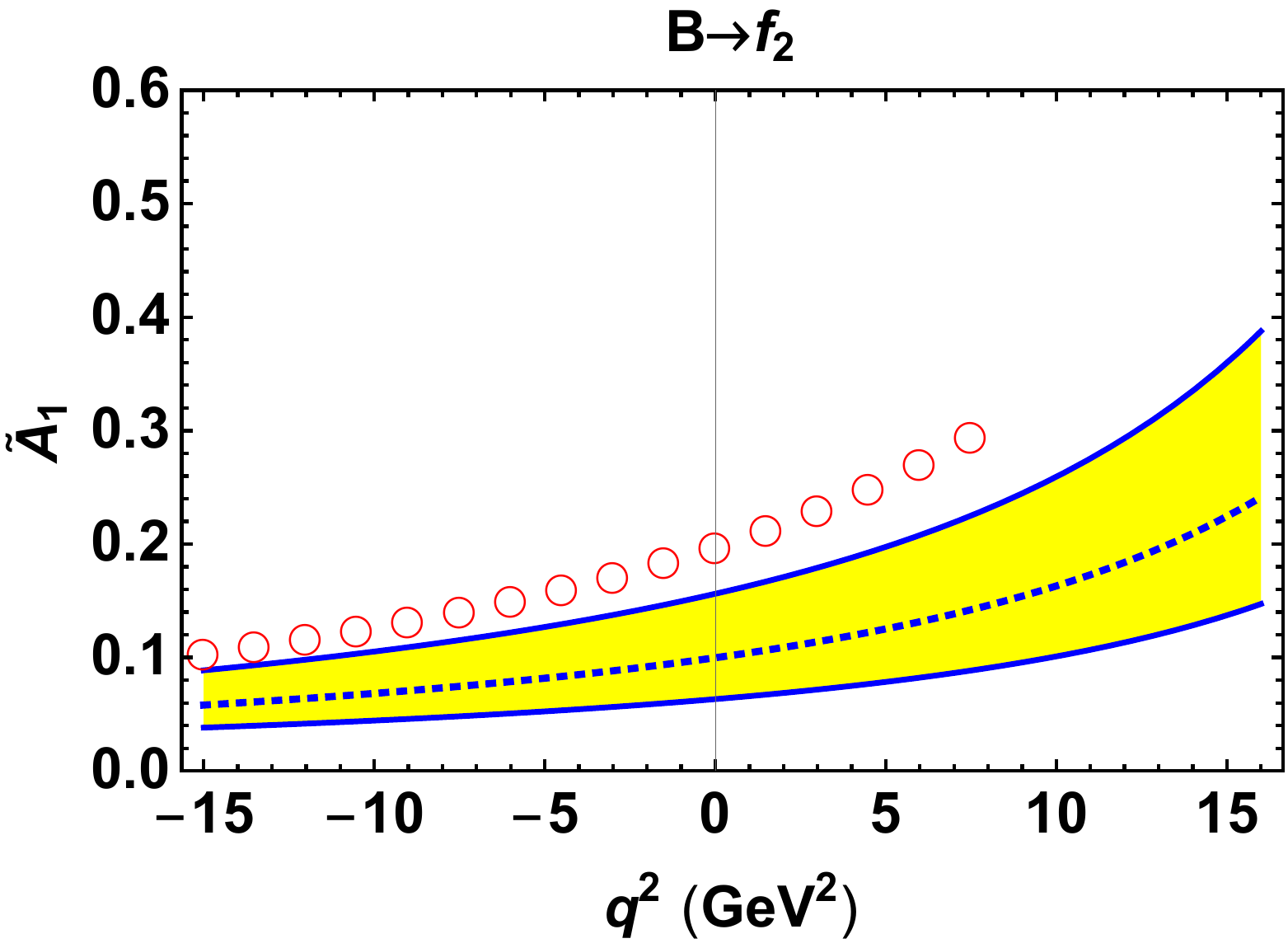} \quad  \\
\includegraphics[width=0.34\textwidth]{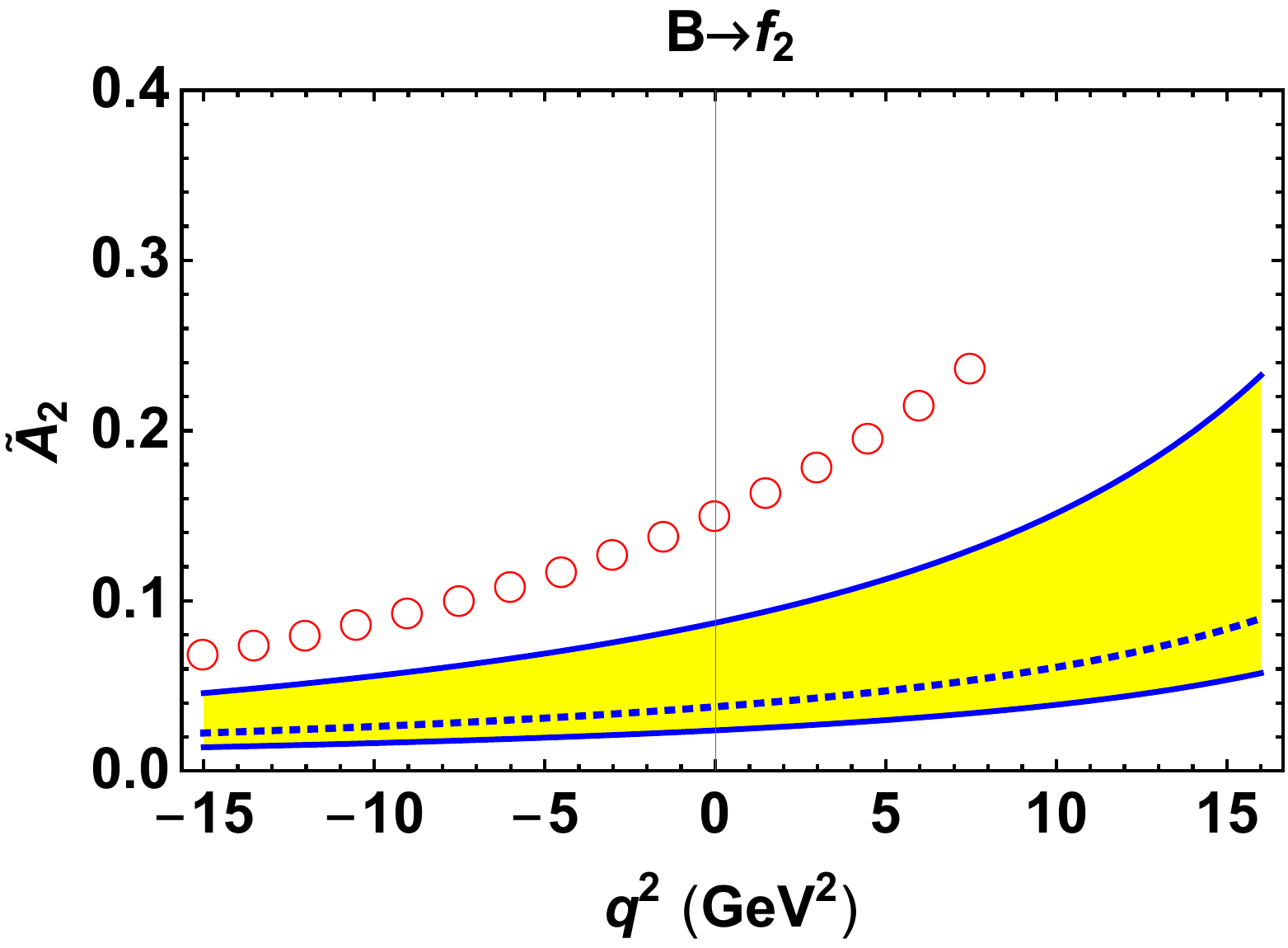} \quad &&
\includegraphics[width=0.34\textwidth]{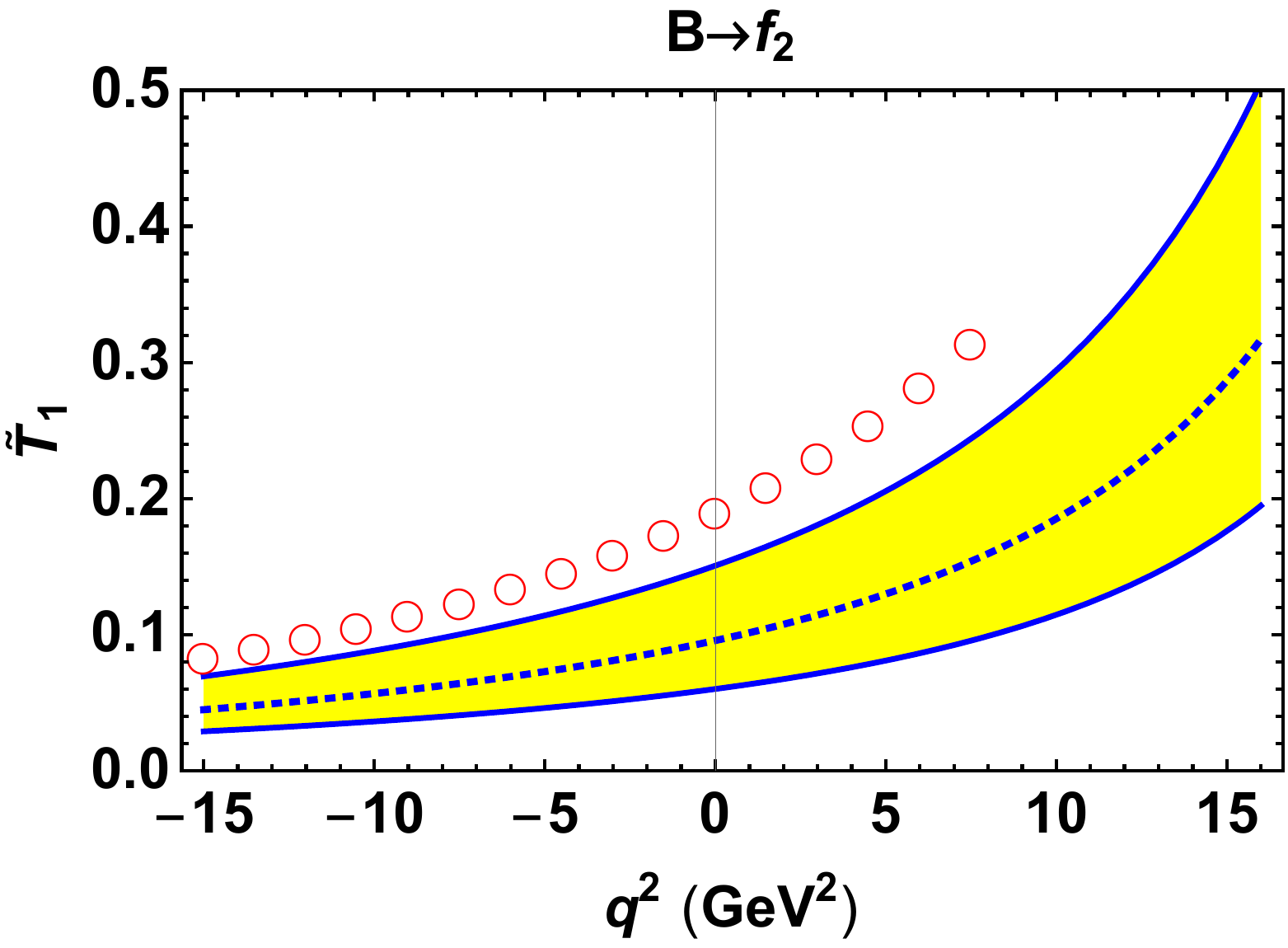} \quad 
\includegraphics[width=0.34\textwidth]{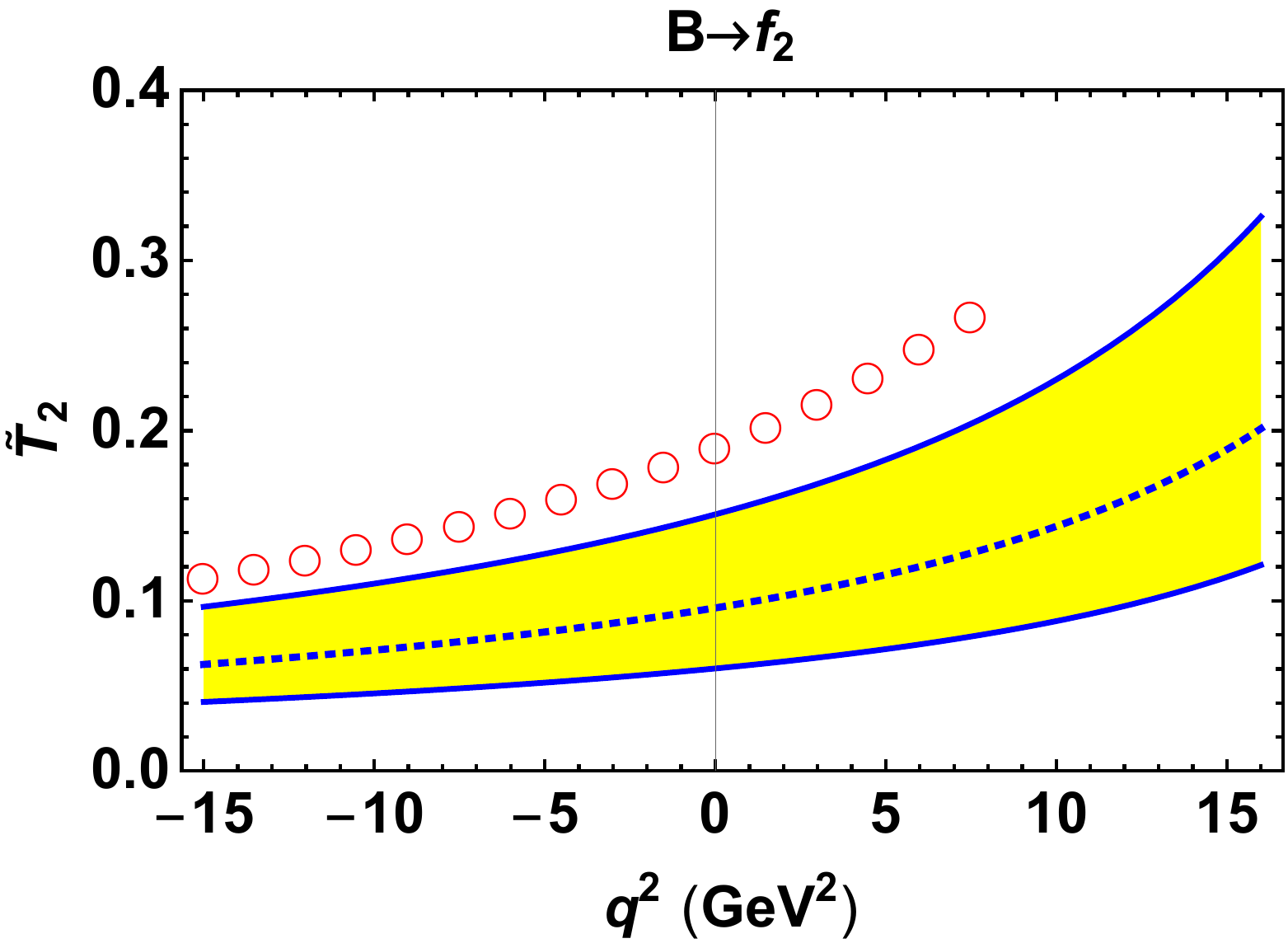} \quad  \\
\includegraphics[width=0.34\textwidth]{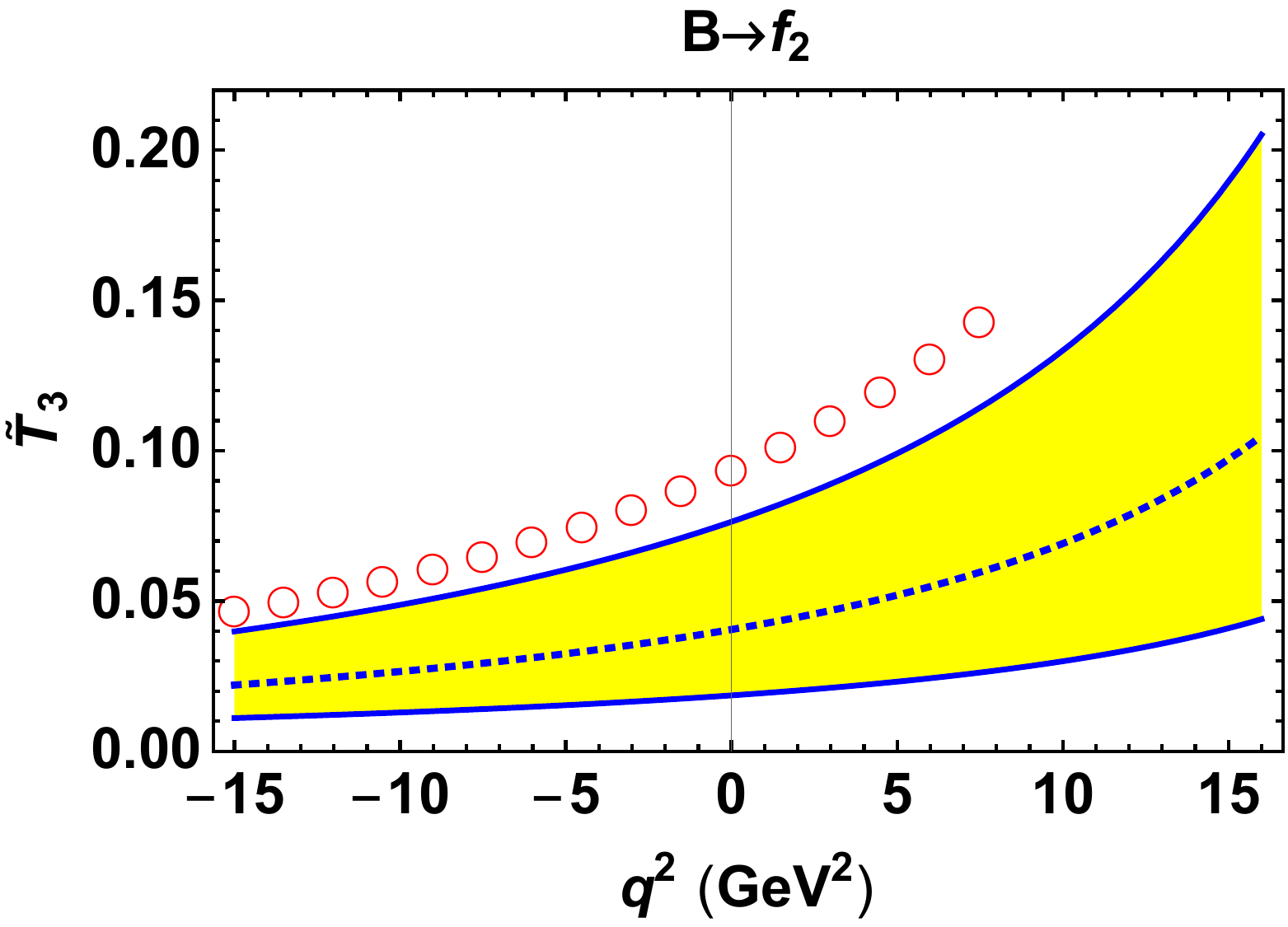}  \quad &&
 \end{tabular}
\caption{$q^2$ dependence of the $B \to f_2$ form factors from fits to our LCSR results. For details see \reffig{q2DepofBtoD2FFs}.}
\label{fig:q2DepofBtof2FFs}
\end{figure}
\FloatBarrier

\section{Phenomenological analyses}
\label{sec:phenoAnalysis}

In this section, using our new results for the relevant form factors we give SM predictions for some selected observables. We considered the decay channels $B\to D_2^{*}\ell\bar\nu$, $B\to K_2^{*}\gamma$ and $B\to K_2^{*}\ell^{+}\ell^{-}$.
\vspace{-0.25cm}

\subsection{SM prediction for $B\to D_2^{*}\ell\bar\nu$}
\label{sec:pheno:D2lnu}

For the $D_2^*(2460)$ mode, currently experimental data only on the decay chain ${\cal{B}} (B\rightarrow
D_2^*\ell\overline{\nu}){\cal{B}} (D_2^*\rightarrow D\pi)$ is available,
\begin{align}
 {\cal{B}} (B\rightarrow
{D}_2^*\ell^{}\overline{\nu}_{\ell}){\cal{B}} ({D}_2^*\rightarrow D\pi)&= 2.2\pm0.3\pm0.4 \quad \quad \mbox{Belle \cite{Liventsev:2007rb}},\nonumber\\
{\cal{B}} (B\rightarrow
{D}_2^*\ell^{}\overline{\nu}_{\ell}){\cal{B}} ({D}_2^*\rightarrow D\pi)&= 1.4\pm0.2\pm0.2  \quad \quad \mbox{BaBar \cite{Aubert:2008ea,Aubert:2008zc}}\, ,
\end{align}
where $\ell=e$ or $\mu$. Despite the current progress in collider physics, there are no data available on $B\to D_2^{*}\ell\bar\nu$ decays yet. On the other side, regarding the vector $D^{*}$ and the pseudoscalar $D$ modes although the most recent measurements for the ratios ${\cal{B}} ({B}\rightarrow D^{(*)}\tau^{}\overline{\nu}_{\tau})/{\cal{B}} ({B}\rightarrow
D^{(*)}\ell^{}\overline{\nu}_{\ell})$ from the Belle collaboration \cite{Abdesselam:2019dgh} alone are compatible with the corresponding SM predictions, when combined with previous experiments the tension between theory and experiment stays around $3.1 \sigma$ \cite{Amhis:2016xyh}, which indicates a violation of lepton flavor universality. 

The size of LFU-violation in ${\cal{B}} ({B}\rightarrow
D_{2}^{*}\tau^{}\overline{\nu}_{\tau})/{\cal{B}} ({B}\rightarrow
D_{2}^{*}\ell^{}\overline{\nu}_{\ell})$ can therefore be further tested for the charmed tensor meson $D_2^*(2460)$ case. Since the matrix elements given in \refeqs{BtoT:vector}{BtoT:axial} are the only ones relevant to $B\to D_2^{*}\ell\bar\nu$ decays in SM, the differential decay widths of these channels are
obtained in terms of $\V{D_2^*} $, $\Azero{D_2^*} $, $\Aone{D_2^*} $ and $\Atwo{D_2^*} $ form factors as \cite{Wang:2009mi,Azizi:2013aua}
\begin{align}
\label{eq:decaywidth}
\frac{d\Gamma}{dq^2} &= \frac{\lambda(m_B^2,m_{D_2^*}^2,q^2)}{4m_{D_2^*}^2}
\Big(\frac{q^2-m_{\ell}^2}{q^2}\Big)^2\frac{\sqrt{\lambda(m_B^2,m_{D_2^*}^2,q^2)}G_F^2
V_{cb}^2}{384m_B^3\pi^3}\Bigg\{\frac{1}{2q^2}\Bigg[\frac{3m_{\ell}^2}{m_B^2}\lambda(m_B^2,m_{D_2^*}^2,q^2)
[{\tilde A}_0(q^2)]^2
\nonumber \\
&+ (m_{\ell}^2+2q^2)\Big|-\frac{1}{2m_{D_2^*}m_B}\Big[(m_B^2-m_{D_2^*}^2-q^2)(m_B+m_{D_2^*}){\tilde A}_1(q^2)
+\frac{\lambda(m_B^2,m_{D_2^*}^2,q^2)}{m_B+m_{D_2^*}}{\tilde A}_2(q^2)\Big]\Big|^2\Bigg]
\nonumber \\
&+ \frac{2}{3}(m_{\ell}^2+2q^2)\lambda(m_B^2,m_{D_2^*}^2,q^2)\Bigg[\Big|
\frac{{\tilde V}(q^2)}{m_B(m_B+m_{D_2^*})}-\frac{(m_B+m_{D_2^*}){\tilde A}_1(q^2)}{m_B\sqrt{\lambda(m_B^2,m_{D_2^*}^2,q^2)}}
\Big|^2
\nonumber \\
&+ \Big|
\frac{{\tilde V}(q^2)}{m_B(m_B+m_{D_2^*})}+\frac{(m_B+m_{D_2^*}){\tilde A}_1(q^2)}{m_B\sqrt{\lambda(m_B^2,m_{D_2^*}^2,q^2)}}
\Big|^2\Bigg]\Bigg\},
\end{align}
where $\lambda(a,b,c)=a^2+b^2+c^2-2ab-2ac-2bc$ is the K\"all\'en function. We presented the $q^2$ dependence of $B\to D_2^{*}\ell\bar\nu$ form factors up to and including twist-four accuracy in Table \ref{tab:Fitparams}. Using these results together with the input parameters $G_F=1.167 \times 10^{-5} \, {\rm GeV}^{-2}$ and $V_{cb}=0.0405$ \cite{Agashe:2014kda}, we obtain the following predictions
\begin{align}
\label{eq:resultsD2}
{\cal{B}} (B\rightarrow D_2^*(2460) e \bar{\nu}_e) & = \begin{cases} (3.80 \pm {0.74})\times10^{-2},   \\
(1.01\pm 0.30)\times 10^{-3}  \quad   \text{\cite{Azizi:2013aua}}\, ,
\end{cases}  \\
{\cal{B}} (B\rightarrow D_2^*(2460) \mu \bar{\nu}_\mu) & = \begin{cases} (3.70 \pm {0.72})\times 10^{-2},  \\
 (1.00\pm 0.29)\times 10^{-3}  \quad  \text{ \cite{Azizi:2013aua}}\, ,  \end{cases} \\
{\cal{B}} (B\rightarrow D_2^*(2460) \tau \bar{\nu}_\tau) & = \begin{cases} (1.50 \pm {0.28})\times 10^{-3}, \\
(0.16\pm 0.03)\times 10^{-3} \quad   \text{\cite{Azizi:2013aua}}\, , \end{cases}
\end{align}
and 
\begin{align}
{\cal{R}}_{D_2^* \tau / \ell }& \equiv \dfrac{{\cal{B}} (B\rightarrow D_2^*(2460) \tau \bar{\nu}_\tau)}{{\cal{B}} (B\rightarrow D_2^*(2460) \ell \bar{\nu}_\ell)} = \begin{cases}  0.041 \pm {0.002},  \\
0.16 \pm0.04  \quad   \text{\cite{Azizi:2013aua}}\, . \end{cases}
\label{eq:resultsD2ratio}
\end{align}

The variance of our predictions from those of Ref.~\cite{Azizi:2013aua} is due to aforementioned discrepancy in the estimation of the form factors (see Table \ref{tab:allFFcomparison}).

\subsection{SM predicition for $B\to K_2^{*}\gamma$}
\label{sec:pheno:K2gamma}

We continue with a phenomenological analysis on exclusive rare radiative decay of $B$ meson to radially excited tensor meson $K_2^*(1430)$. The branching ratio of this radiative mode has been measured by several experiments:
\begin{align}
\label{data:radiativeK2}
 {\cal{B}} (B\rightarrow
K_2^*\gamma)&= (1.66^{+0.59}_{-0.53}\pm0.13)\times 10^{-5} ~~~~~~~~~~~~~\mbox{CLEO \cite{Coan:1999kh}},\nonumber\\
 {\cal{B}} (B\rightarrow
K_2^*\gamma)&= (1.3\pm 0.5 \pm0.1)\times 10^{-5} ~~~~~~~~~~~~~~~~\mbox{Belle \cite{Nishida:2002me}},\nonumber\\
 {\cal{B}} (B\rightarrow
K_2^*\gamma)&= (1.22\pm 0.25 \pm0.10)\times 10^{-5} ~~~~~~~~~~\mbox{BaBar \cite{Aubert:2003zs}},
\end{align}
which gives the PDG average of $(1.24\pm 0.24)\times 10^{-5}$ \cite{Tanabashi:2018oca}. In the SM, $B\to K_2^{*}\gamma$ decay is governed by the electromagnetic dipole operator $\mathcal{O}_7$, and its matrix elements between initial $B$ and final $K_2^*$ states are given in \refeqs{BtoT:tensor}{BtoT:tensor5}. The exclusive decay rate of emission of a real photon ($q^2=0$) depends only on the form factor $\Tone{K_2^*}$ and is given by \cite{Ebert:2000kz} 
\begin{equation}
\label{eq:drateBK2gamma}
\Gamma(B\to K_2^*\gamma)=
\frac{\alpha }{256\pi^4} G_F^2m_b^5|V_{tb}V_{ts}|^2
|C_7(m_b)|^2 {\Tilde T}_1^2(0) \frac{m_B^2}{m_{K_2^*}^2}
\left(1-\frac{m_{K_2^*}^2}{m_B^2}
 \right)^5\left( 1+\frac{m_{K_2^*}^2}{m_B^2}\right),
\end{equation}
where $V_{ij}$ are the CKM matrix elements, $\alpha$ is the fine-structure constant and
$C_7(m_b)$ is the Wilson coefficient associated with $\mathcal{O}_7$. Since the inclusive radiative decay $B\to X_s\gamma$ is accurately measured by several experiments \cite{Koppenburg:2004fz,Aubert:2006gg}, it is more convenient\footnote{Considering this ratio, one avoids most of the parametric uncertainties.} to consider the ratio of exclusive to inclusive branching ratios \cite{Ebert:2000kz}
\begin{align}
\label{eq:ratioBK2gamma}
R_{K_2^*}\equiv
\frac{{\cal{B}}(B\to K_2^*(1430)\gamma)}{{\cal{B}}(B\to X_s\gamma)}=
\frac18 {\Tilde T}_1^2(0)\frac{m_B^2}{m_{K_2^*}^2}
\frac{\left(1-{m_{K_2^*}^2}/{m_B^2}
 \right)^5\left( 1+{m_{K_2^*}^2}/{m_B^2}\right)}{\left(1-{m_s^2}/{m_b^2}
 \right)^3\left( 1+{m_s^2}/{m_b^2}\right)},
\end{align} 
where the world average of the inclusive decay is given by the Heavy Flavour Averaging Group \cite{Amhis:2016xyh} as ${\cal{B}}(B\to X_s\gamma)=(3.32\pm0.16)\times 10^{-4}$, which is compatible with the theoretical estimate \cite{Misiak:2006zs}. 

We determine the experimental ratio $R^{\text{exp}}_{K_2^*}$ by normalizing the experimental world averages of the corresponding decays. Similarly, using the value of the form factor $\Tone{K_2^*}$ from Table \ref{tab:Fitparams} we obtain our SM prediction for $R^{\text{SM}}_{K_2^*}$. They read
\begin{align}
R^{\text{exp}}_{K_2^*} &=  0.037\pm0.007 \, , \nn \\
R^{\text{SM}}_{K_2^*} &=  0.055 \pm 0.023  \, ,
\label{eq:VALUEratioBK2gamma}
\end{align}
which are in agreement within the quoted error budget.

\subsection{SM prediction for $B\to K_2^{*}\ell^+\ell^-$}
\label{sec:pheno:K2ll}
In the standard model, the effective Hamiltonian governing $B\to K_2^{*}\ell^+\ell^-$
decay is
\begin{align}
\mathcal{H}_{eff}=-\frac{4G_{F}}{\sqrt{2}}V_{tb}^{\ast }V_{ts}{\sum\limits_{i=1}^{10}}
C_{i}({\mu }) \mathcal{O}_{i}({\mu }),
\label{eq:effectiveH}
\end{align}
with $\mathcal{O}_{i}({\mu })$ being the effective operators and $%
C_{i}({\mu })$ the respective Wilson coefficients at the renormalization scale $\mu$. Among the ten operators in \refeq{effectiveH}, $\mathcal{O}_{7}$, $\mathcal{O}_{9}$ and $\mathcal{O}_{10}$
\begin{align}
\mathcal{O}_{7} &= \frac{e^{2}}{16\pi ^{2}}m_{b}\left( \bar{s}\sigma _{\mu \nu
}P_{R}b\right) F^{\mu \nu },\,   \notag \\
\mathcal{O}_{9} &= \frac{e^{2}}{16\pi ^{2}}(\bar{s}\gamma _{\mu }P_{L}b)(\bar{\ell}\gamma
^{\mu } \ell),\,   \\
\mathcal{O}_{10} &= \frac{e^{2}}{16\pi ^{2}}(\bar{s}\gamma _{\mu }P_{L}b)(\bar{\ell}
\gamma ^{\mu }\gamma _{5} \ell),   \quad \quad  P_{L,R}=\left( 1\pm \gamma _{5}\right) /2 \, ,\notag
\end{align}
are the only ones contributing to $B\to K_2^{*}\ell^{+}\ell^{-}$. The related Wilson coefficients are discussed thoroughly in the literature (for details, see e.g. \cite{Buras:1994dj,Buras:2011we,Hurth:2010tk} and references therein). In terms of the Wilson coefficients and the form factors defined in \refeqs{BtoT:vector}{BtoT:tensor5}, the general expression of the differential decay width for $B\to K_2^{*}\ell^{+}\ell^{-}$ can be written as \cite{Junaid:2011bh}:
\begin{align}
\frac{d\Gamma }{dq^{2}} &= \frac{G_{F}^{2}\alpha ^{2}}{2^{11}\pi
^{5}m_{B}^{3}}\left\vert V_{tb}V_{ts}^{\ast }\right\vert ^{2} \sqrt{\lambda\left(1-\frac{%
4m_{\ell}^{2}}{q^{2}}\right)} \left(
\frac{\left\vert \mathcal{F}_1 \right\vert ^{2}\left( 2m_{\ell}^{2}+q^{2}\right)
\lambda ^{2}}{6m_{B}^{2}m_{K_{2}^{\ast }}^{2}}\right.  \notag \\
&+ \frac{\left\vert \mathcal{F}_2 %
\right\vert ^{2}m_{B}^{2}\left( 2m_{\ell}^{2}+q^{2}\right) \left(
10q^{2}m_{K_{2}^{\ast }}^{2}+\lambda \right) \lambda }{9m_{K_{2}^{\ast
}}^{4}q^{2}} +\frac{\left\vert \mathcal{F}_3 \right\vert ^{2}\left(
2m_{\ell}^{2}+q^{2}\right) \lambda ^{3}}{9m_{B}^{2}m_{K_{2}^{\ast }}^{4}q^{2}}-%
\frac{\left\vert \mathcal{F}_4 \right\vert ^{2}\left( 4m_{\ell}^{2}-q^{2}\right)
\lambda ^{2}}{6m_{B}^{2}m_{K_{2}^{\ast }}^{2}}\notag \\
&+ \frac{\left\vert \mathcal{F}_5 %
\right\vert ^{2}m_{B}^{2}\left( 2\left( \lambda -20m_{K_{2}^{\ast
}}^{2}q^{2}\right) m_{\ell}^{2}+q^{2}\left( 10q^{2}m_{K_{2}^{\ast }}^{2}+\lambda
\right) \right) \lambda }{9m_{K_{2}^{\ast }}^{4}q^{2}} +\frac{%
2\left\vert \mathcal{F}_7 \right\vert ^{2}m_{\ell}^{2}q^{2}\lambda ^{2}}{%
3m_{B}^{2}m_{K_{2}^{\ast }}^{4}}  \notag \\
&+ \frac{\left\vert \mathcal{F}_6 \right\vert ^{2}\left( 2\left( \left(
m_{B}^{2}-m_{K_{2}^{\ast }}^{2}\right) ^{2}-2q^{4}+4\left(
m_{B}^{2}+m_{K_{2}^{\ast }}^{2}\right) q^{2}\right) m_{\ell}^{2}+q^{2}\lambda
\right) \lambda ^{2}}{9m_{B}^{2}m_{K_{2}^{\ast }}^{4}q^{2}} \notag \\
&+ \frac{2\mbox{Re} (\mathcal{F}_2 \mathcal{F}_3 ^{\ast}) \left( 2m_{\ell}^{2}+q^{2}\right) \left(
-m_{B}^{2}+m_{K_{2}^{\ast }}^{2}+q^{2}\right) \lambda ^{2}}{9m_{K_{2}^{\ast
}}^{4}q^{2}}+\frac{4\mbox{Re} ( \mathcal{F}_6 \mathcal{F}_7 ^{\ast} )m_{\ell}^{2}\left(
m_{B}^{2}-m_{K_{2}^{\ast }}^{2}\right) \lambda ^{2}}{3m_{B}^{2}m_{K_{2}^{%
\ast }}^{4}}  \notag \\
&+ \frac{2\mbox{Re} (  \mathcal{F}_5 \mathcal{F}_6 ^{\ast} )\left( q^{2}\left(
-m_{B}^{2}+m_{K_{2}^{\ast }}^{2}+q^{2}\right) -2m_{\ell}^{2}\left(
m_{B}^{2}-m_{K_{2}^{\ast }}^{2}+2q^{2}\right) \right) \lambda ^{2}}{%
9m_{K_{2}^{\ast }}^{4}q^{2}}\notag \\
& -\left.\frac{4\mbox{Re} (  \mathcal{F}_5 \mathcal{F}_7 ^{\ast}  )m_{\ell}^{2}\lambda ^{2}}{3m_{K_{2}^{\ast }}^{4}}\right),  
\label{eq:BtoK2llDecay}
\end{align}%
where $\lambda \equiv\lambda (m_{B}^{2},m_{K_{2}^{\ast }}^{2},q^{2})$, and the individual quantities $\mathcal{F}_i$ read
\begin{align}
\mathcal{F}_1 &= - C_{9}^{eff}(\mu )\frac{2}{m_{B}+m_{K_{2}^{\ast }}}%
{\tilde V}(q^{2})-C_{7}^{eff}(\mu )\frac{4m_{b}}{q^{2}}{\tilde T}_{1}\left( q^{2}\right)
\notag \\
\mathcal{F}_2 &= \frac{\left( m_{B}+m_{K_{2}^{\ast }}\right) }{m_{B}^{2}}\left[
C_{9}^{eff}(\mu ){\tilde A}_{1}(q^{2})+C_{7}^{eff}(\mu )\frac{2m_{b}(m_{B}-m_{K_{2}^{%
\ast }})}{q^{2}}{\tilde T}_{2}\left( q^{2}\right) \right]  \notag \\
\mathcal{F}_3 &= C_{9}^{eff}(\mu )\frac{1}{m_{B}+m_{K_{2}^{\ast }}}%
{\tilde A}_{2}(q^{2})+C_{7}^{eff}(\mu )\frac{2m_{b}}{q^{2}}\left[ {\tilde T}_{2}\left(
q^{2}\right) +\frac{q^{2}}{m_{B}^{2}-m_{K_{2}^{\ast }}^{2}}{\tilde T}_{3}\left(
q^{2}\right) \right]  \notag \\
\mathcal{F}_4 &= - C_{10}\frac{2}{m_{B}+m_{K_{2}^{\ast }}}{\tilde V}(q^{2})  \notag
\\
\mathcal{F}_5 &= C_{10}\frac{(m_{B}+m_{K_{2}^{\ast }})}{m_{B}^{2}}%
{\tilde A}_{1}(q^{2})  \notag \\
\mathcal{F}_6 &= C_{10}\frac{1}{m_{B}+m_{K_{2}^{\ast }}}{\tilde A}_{2}(q^{2})
\notag \\
\mathcal{F}_7 &= C_{10}\left[ -\frac{2m_{K_{2}^{\ast }}}{q^{2}}%
{\tilde A}_{0}(q^{2}) + \frac{(m_{B}+m_{K_{2}^{\ast }})}{q^{2}}{\tilde A}_{1}(q^{2})-\frac{%
(m_{B}-m_{K_{2}^{\ast }})}{q^{2}}{\tilde A}_{2}(q^{2})\right] .
\label{eq:DW_func}
\end{align}

The new input parameters entering the decay rate prediction here are taken as $V_{tb}=0.77^{+0.18}_{-0.24}$ \cite{Agashe:2014kda}, $V_{ts}=0.0406\pm 0.0027$ \cite{Agashe:2014kda}, $C_{7}^{eff}(m_b)=-0.306$ \cite{Asatrian:2016meg}, $C_{9}^{eff}(m_b)=4.344$ \cite{Buras:1994dj,Dag:2010jr} and $C_{10}=-4.669$ \cite{Buras:1994dj,Dag:2010jr}. Using the calculated LCSR results for the form factors we obtain
\begin{align}
\label{eq:resultsBK2ll}
{\cal{B}} (B\rightarrow
K_2^*e^+e^-)& = (7.72 \pm 4.28)\times 10^{-7},\\
{\cal{B}} (B\rightarrow
K_2^*\mu^+\mu^-)& = \begin{cases} 
(6.05 \pm 3.81)\times 10^{-7},  \\
(2.43_{-0.5}^{+0.6})\times 10^{-7}  &\text{\cite{Junaid:2011bh}}, \\
(2.5_{-1.1}^{+1.5})\times 10^{-7}     &\text{\cite{Li:2010ra}},
                  \end{cases} \\
                 {\cal{B}} (B\rightarrow
K_2^*\tau^+\tau^-)& = \begin{cases} 
(1.12 \pm 0.59)\times 10^{-9}, \\
(2.74_{-0.9}^{+0.9})\times 10^{-10}   &\text{\cite{Junaid:2011bh}} ,\\
(9.6_{-4.5}^{+6.1})\times 10^{-10}     &\text{\cite{Li:2010ra}}.
                 \end{cases} 
\end{align}

Our predictions are compatible with the references given within the error budget. Furthermore, in analogy to \refeq{resultsD2ratio} we also give our prediction for the LFU ratio:
\begin{align}
{\cal{R}}_{K_2^* \tau / \mu}& \equiv \dfrac{{\cal{B}} (B\rightarrow K_2^* \tau^+ \tau^-)}{{\cal{B}} (B\rightarrow K_2^* \mu^+ \mu^-)} =  0.0020 \pm  0.0004 \, .
\label{eq:resultK2ratio}
\end{align}

As a final remark before summary, we would like to stress that the results presented in this work include only factorizable contributions and non-factorizable (non-local $c\bar{c}$-loop) effects are not taken into account in this work. Analysis of such non-factorizable contributions lies beyond the scope of this paper and we plan to come back to discuss this point separately in the future.

\section{Conclusion}
\label{sec:summary}

The study of semileptonic $B$-meson decays involving tensor mesons can provide additional information on physics beyond the Standard Model due to the rich polarization structure of the tensor mesons. In connection to that we calculated the $B\to D_2^*,K_2^*,a_2,f_2$ ($J^{P}=2^{+}$) transition form factors within
light-cone sum rules using $B$-meson distribution amplitudes, including the twist-four terms. We find that the calculated higher-twist terms have a noticeable impact on the sum rules. Using the obtained results for the form factors we estimate the decay rates of $B\to D_2^{*}\ell\bar\nu$, $B\to K_2^{*}\gamma$ and $B\to K_2^{*}\ell^{+}\ell^{-}$ in the SM. Our results indicate that these decays can be within reach for LHCb and Belle experiments in the near future.

\acknowledgments

The work of A.K. is supported by the DFG within the Emmy Noether Programme under Grant No. DY-130/1-1
and the DFG Collaborative Research Center 110 "Symmetries and the emergence of structure in QCD.'' H.D. acknowledges support
through the Scientific and Technological Research Council of Turkey (TUBITAK)
Grant No. BIDEP-2219. A.O. thanks the Physics Department at the Technical University of Munich for the hospitality at the early stage of this work. A.K. is thankful to Danny van Dyk and Javier Virto for helpful discussions on the topic.

\appendix

\section{Distribution Amplitudes of the $B$-meson}
\label{app:LCDAs}

The two-particle momentum-space projector can be expressed in terms of $B$-LCDAs (up to twist-four) as
\begin{align}
\bra{0} \bar{q}_{2}^{\alpha}(x) h_{v}^{\beta}(0) \ket{\bar{B}_{q_2}(v)} =
    &-\frac{i f_B m_B}{4} \int^\infty_0 \text{d}\omega e^{-i\omega v\cdot x} \bigg\{
        (1 + \slashed{v}) \bigg[
            \phi_+(\omega) -g_+(\omega) \partial_\sigma \partial^\sigma
            \nn \\&+\left(\frac{\bar{\phi}(\omega)}{2}
            -\frac{\bar{g}(\omega)}{2} \partial_\sigma \partial^\sigma\right) 
        \gamma^\mu \partial_\mu
        \bigg] \gamma_5
    \bigg\}^{\beta\alpha},
    \label{eq:BLCDAs2pt}
\end{align}
where $v_\mu$ is the four-velocity of the $B$-meson, and $\partial_\mu \equiv \partial / \partial l^\mu$ with $l^\mu=\omega v^\mu$ in the two-particle case. The above momentum-space derivatives are understood to act on the hard-scattering kernel of \refeq{correlatorOPE2pt}. Moreover, we abbreviate
\begin{equation}
\label{eq:def:barred-LCDAs}
\begin{aligned}
    \bar{\phi}(\omega) & \equiv \int_0^{\omega} \text{d}\xi\, \left(\phi_+(\xi) - \phi_-(\xi)\right)\,,\\
    \bar{g}(\omega) & \equiv \int_0^{\omega} \text{d}\xi\, \left(g_+(\xi) - g_-(\xi)\right)\,.\\
\end{aligned}
\end{equation}

In our numerical estimates for the form factors we follow the {\it local duality} model\footnote{The model we employ in this work corresponds to model II A of Ref.~\cite{Braun:2017liq}.} proposed in Ref.~\cite{Braun:2017liq} for the two-particle $B$-LCDAs $\phi_+$, $\phi_-$, and $g_+$. The explicit expressions for $\phi_+$, $\phi_-$, and $g_+$ in this model are given in Eqs.~5.22--5.23 of Ref.~\cite{Braun:2017liq}.

For $g_-$ no model expression is available yet; we therefore use the Wandzura-Wilczek (WW) approximation
\begin{align}
\label{eq:gmWW}
 g_-^{WW}(\omega)
        & =  +\frac{1}{4} \int_0^{\omega} \mathrm{d}\eta_2 \, \int_0^{\eta_2} \mathrm{d}\eta_1 \, \left[ \phi_+ (\eta_1) - \phi_-^{WW} (\eta_1) \right]  - \frac{1}{2} \int_0^{\omega} \mathrm{d}\eta_1 \, (\eta_1 - \bar{\Lambda}) \phi_-^{WW} (\eta_1)   \,,  \\ 
 \phi_-^{WW} (\omega)& =  \int_{\omega}^{\infty} \mathrm{d}\eta_1 \, \frac{\phi_+ (\eta_1)}{\eta_1} \, .  \nn 
\end{align}
In the local duality model considered in this work, \refeq{gmWW} explicitly yields:
\be
 g_-^{WW}(\omega) =  \frac{\omega (3\lambda_B-\omega)^3 }{48 \lambda_B^3}\, \theta(3\lambda_B-\omega) \, ,
\label{eq:gmWWexplicit}
\ee
where $\theta(x)$ is the heavy-side step function.

The parameters $\lambda _E^2, \lambda _H^2$ and $\lambda _B$ appearing in the explicit expressions of $B$-LCDAs are provided as input in \refsec{results:input}.

\section{$B\to T$ coefficients of  \refeq{masterformula} from two-particle contributions}
\label{app:coefficients}

\subsection{$K^{(F)}$ factors of \refeq{masterformula}}

The pre-factors appearing in \refeq{masterformula} read:
\begin{equation}
\begin{aligned}
    &K^{( \V{T})}= {-} \frac{f_T m_T^3}{m_B(m_B+m_T)},&
    &K^{( \Aone{T})}=  - \frac{f_T m_T^3 (m_B+m_T)}{2 m_B},&        \\
    &K^{(\Atwo{T} )}=   \frac{2 f_T m_T^3}{m_B (m_B+m_T)},&
    &K^{(\Athreezero{T} )}=   {-f_T m_T^4}/m_B  ,&              \\
    &K^{(\Tone{T} )} = {-f_T m_T^3}/m_B \, , K^{(\TtwothreeA{T} )} = K^{(\TtwothreeB{T} )}= {-2 f_T m_T^3}/m_B  \, .&&&
\end{aligned}
\end{equation}

\subsection{$C^{(F,\psi_\text{2p})}_n$ coefficients of \refeq{CoeffFuncs2pt}}

We collect here the (non-vanishing) coefficients appearing in \refeq{CoeffFuncs2pt}.

For $ \V{T}$ we obtain: 
\begin{equation}
\begin{aligned}
C^{( \V{T},\phi_+)}_1     & =  \frac{\sigma}{2}  \,, &  \\
C^{( \V{T},\bar{\phi})}_2 & =   \frac{m_{q_1} \sigma}{2}  \,, \\
C^{( \V{T},g_+)}_2        & =  4 \sigma , &
C^{( \V{T},g_+)}_3        & =  -4 m_{q_1}^2 \sigma \,, \\
C^{( \V{T},\bar{g})}_3    & =   4 m_{q_1} \sigma   \, ,  &
C^{( \V{T},\bar{g})}_4    & = - 12 m_{q_1}^3 \sigma \,.
\end{aligned}
\end{equation}

For $\Aone{T}$ we obtain:  
\begin{equation}
\begin{aligned}
C^{(\Aone{T},\phi_+)}_1     & =  -\frac{\sigma \left( q^2-(m_B\bar{\sigma}+m_{q_1})^2 \right) }{4\bar{\sigma}}  \,, &  \\
C^{(\Aone{T},\bar{\phi})}_1 & =  \frac{\sigma m_{q_1} }{4\bar{\sigma}}   ,&
C^{(\Aone{T},\bar{\phi})}_2& = - \frac{\sigma m_{q_1} \left( q^2-(m_{q_1} + m_B\bar{\sigma})^2 \right) }{4\bar{\sigma}}  \,, \\
C^{(\Aone{T},g_+)}_1        & = \frac{2\sigma}{\bar{\sigma}}   , &
C^{(\Aone{T},g_+)}_2       & =  -\frac{2 \sigma (q^2 - m_B \bar{\sigma} (m_{q_1} + m_B \bar{\sigma}))}{\bar{\sigma}}    \,, \\
C^{(\Aone{T},g_+)}_3      & =  \frac{2\sigma m_{q_1}^2 \left( q^2-(m_{q_1} + m_B\bar{\sigma})^2 \right) }{\bar{\sigma}}  \,, \\
C^{(\Aone{T},\bar{g})}_2    & = \frac{2\sigma (m_{q_1}+2 m_B \bar{\sigma})}{\bar{\sigma}}  \,, &
C^{(\Aone{T},\bar{g})}_3    & = - \frac{2  m_{q_1}\sigma (2m_{q_1}^2 + q^2 - m_B^2 \bar{\sigma}^2)}{\bar{\sigma}}  \,, \\
C^{(\Aone{T},\bar{g})}_4    & =  \frac{ 6 m_{q_1}^3 \sigma (q^2- (m_{q_1}+m_B\bar{\sigma})^2 )}{\bar{\sigma}}  \,.
\end{aligned}
\end{equation}

For $\Atwo{T}$ we obtain:  
\begin{equation}
\begin{aligned}
C^{(\Atwo{T},\phi_+)}_1     & =  \sigma (1-2 \bar{\sigma})   \,, &  \\
C^{(\Atwo{T},\bar{\phi})}_2 & =  -\sigma (m_{q_1}-2m_B\bar{\sigma}+2m_B\bar{\sigma}^2 )  \,,\\
C^{(\Atwo{T},g_+)}_2        & =  8 \sigma  (1-2 \bar{\sigma}) \,, &
C^{(\Atwo{T},g_+)}_3        & =  -8m_{q_1}^2 \sigma  (1-2 \bar{\sigma})  \,, \\
C^{(\Atwo{T},\bar{g})}_3    & =  - 8 \sigma (m_{q_1}-4m_B\sigma\bar{\sigma} ) \,, &
C^{(\Atwo{T},\bar{g})}_4    & =  24 m_{q_1}^2 \sigma (m_{q_1}-2m_B\sigma\bar{\sigma} ) \,.
\end{aligned}
\end{equation}

For $\Athreezero{T}$ we obtain:  
\begin{equation}
\begin{aligned}
C^{(\Athreezero{T},\phi_+)}_1     & =  \frac{\sigma q^2 (2 {\bar \sigma}-3)}{4}  \,, &  \\
C^{(\Athreezero{T},\bar{\phi})}_2 & =   \frac{ \sigma q^2 (2\sigma (1+\sigma)m_B- m_{q_1}) }{4}  \,,\\
C^{(\Athreezero{T},g_+)}_2        & =  -2 q^2 \sigma (1+2 \sigma )   \,, &
C^{(\Athreezero{T},g_+)}_3        & =   2q^2 m_{q_1}^2 \sigma  (1+2 \sigma ) \,, \\
C^{(\Athreezero{T},\bar{g})}_3    & = {2\sigma q^2 (  4 m_B \sigma (1+\sigma) -m_{q_1})} \,, &
C^{(\Athreezero{T},\bar{g})}_4    & =  {6 m_{q_1}^2 q^2 \sigma (m_{q_1} -2m_B \sigma (1+\sigma))}  \,.
\end{aligned}
\end{equation}

For $\Tone{T}$ we obtain: 
\begin{equation}
\begin{aligned}
C^{(\Tone{T},\phi_+)}_1     & = \sigma (m_B \bar{\sigma}+m_{q_1})/2    \,, &  \\
C^{(\Tone{T},\bar{\phi})}_2 & =  \sigma m_{q_1}(m_B \bar{\sigma}+m_{q_1})/2  \,,\\
C^{(\Tone{T},g_+)}_2        & = 2 \sigma (2 m_B \bar{\sigma}+m_{q_1})    \,, &
C^{(\Tone{T},g_+)}_3        & =  -4 \sigma m_{q_1}^2 (m_B \bar{\sigma}+m_{q_1})     \,, \\
C^{(\Tone{T},\bar{g})}_2    & = 4 \sigma  \,, &
C^{(\Tone{T},\bar{g})}_3    & = 4 \sigma \bar{\sigma}m_B m_{q_1}     \,,\\
C^{(\Tone{T},\bar{g})}_4    & =  -12 \sigma m_{q_1}^3 (m_B \bar{\sigma}+m_{q_1})       \,. &
\end{aligned}
\end{equation}

For $\TtwothreeA{T}$ we obtain:
\begin{equation}
\begin{aligned}
C^{(\TtwothreeA{T},\phi_+)}_1     & = \sigma (m_B \bar{\sigma}+m_{q_1})   \,, &  \\
C^{(\TtwothreeA{T},\bar{\phi})}_2 & =  \sigma (m_{q_1}(m_B \bar{\sigma}+m_{q_1})-2 \sigma q^2)     \,,\\
C^{(\TtwothreeA{T},g_+)}_2        & = 4 \sigma(m_{q_1}+2m_B \bar{\sigma})      \,, &
C^{(\TtwothreeA{T},g_+)}_3        & =  -8 m_{q_1}^2\sigma(m_{q_1}+m_B \bar{\sigma})          \,, \\
C^{(\TtwothreeA{T},\bar{g})}_2    & = 8 \sigma     \,, &
C^{(\TtwothreeA{T},\bar{g})}_3    & =    8 \sigma (\bar{\sigma} m_B m_{q_1} -4\sigma q^2)    \,,\\
C^{(\TtwothreeA{T},\bar{g})}_4    & = -24 \sigma m_{q_1}^2 (\bar{\sigma} m_B m_{q_1} +m_{q_1}^2-2\sigma q^2)     \,. &
\end{aligned}
\end{equation}

For $\TtwothreeB{T}$ we obtain: 
\begin{equation}
\begin{aligned}
C^{(\TtwothreeB{T},\phi_+)}_1     & = -\sigma(\sigma m_B -m_{q_1})      \,, & \\
C^{(\TtwothreeB{T},\bar{\phi})}_1 & =   -\frac{\sigma^2}{\bar{\sigma}}    \,, &
C^{(\TtwothreeB{T},\bar{\phi})}_2 & =  \frac{\sigma\hat{\sigma}_1}{\bar{\sigma}}          \,,\\
C^{(\TtwothreeB{T},g_+)}_2        & = 4 \sigma(-2\sigma m_B +m_{q_1})    \,, &
C^{(\TtwothreeB{T},g_+)}_3        & =  8 \sigma m_{q_1}^2 (\sigma m_B -m_{q_1})  \,, \\
C^{(\TtwothreeB{T},\bar{g})}_2    & =    \frac{8 \sigma (1 -3 \sigma)}{\bar{\sigma}}       \,, &
C^{(\TtwothreeB{T},\bar{g})}_3    & =    \frac{8 \sigma^2 \hat{\sigma}_2 }{\bar{\sigma}}   \,,\\
C^{(\TtwothreeB{T},\bar{g})}_4    & =-\frac{24 m_{q_1}^2 \sigma\hat{\sigma}_1}{\bar{\sigma}}       \,,&\hspace{3.45cm}
\end{aligned}
\end{equation}
where shorthand notations $\hat{\sigma}_1=m_{q_1}^2(1-2\sigma)-m_B m_{q_1} \sigma \bar{\sigma}+\sigma(m_B^2\bar{\sigma}^2+q^2(2\sigma-1))$ and $\hat{\sigma}_2=2\bar{\sigma}^2 m_B^2-\bar{\sigma}m_B m_{q_1}-2q^2+4\sigma q^2+m_{q_1}^2$ are introduced for simplicity. The $\TtwothreeA{T}$ and $\TtwothreeB{T}$ form factors are defined via \refeq{defTA} and \refeq{defTB}.

\bibliographystyle{apsrev4-1}
\bibliography{references}

\end{document}